\pdfoutput=1  

\documentclass[aps,groupedaddress,floatfix,nofootinbib,preprintnumbers,eqsecnum,twocolumn]{revtex4}
\usepackage{graphicx}
\usepackage{amssymb,amsmath,multirow,dsfont}
\usepackage{dcolumn}   
\usepackage{bm}        

\usepackage{hyperref}
\usepackage{xcolor}
\hypersetup
{
  colorlinks,
  linkcolor={blue!80!black},
  citecolor={blue!70!black},
  urlcolor={blue!70!black}
}
\usepackage{comment,color,placeins,amsmath,amsfonts,enumitem}

\makeatletter
\renewcommand*\env@matrix[1][\arraystretch]{%
  \edef\arraystretch{#1}%
  \hskip -\arraycolsep
  \let\@ifnextchar\new@ifnextchar
  \array{*\c@MaxMatrixCols c}}
\makeatother

\newcommand{\newc}{\newcommand}
\newc{\gsim}{\lower.7ex\hbox{$\;\stackrel{\textstyle>}{\sim}\;$}}
\newc{\lsim}{\lower.7ex\hbox{$\;\stackrel{\textstyle<}{\sim}\;$}}
\makeatletter
\newcommand{\biggg}{\bBigg@{3}}
\newcommand{\Biggg}{\bBigg@{4}}
\makeatother

\def\beq{\begin{equation}}
\def\eeq{\end{equation}}
\def\beqn{\begin{eqnarray}}
\def\eeqn{\end{eqnarray}}
\def\calM{{\cal M}}

\def\half{{\textstyle{1\over 2}}}

\def\ie{{\it i.e.}\/}
\def\eg{{\it e.g.}\/}

\def\mbar{{ \overline{m} }}

\def\mbar{{\overline{m}}}
\def\thetabar{{\overline{\theta}}}
\def\rhobar{{\overline{\rho}}}
\def\xibar{{\overline{\xi}}}
\def\lambdabar{{\overline{\lambda}}}
\def\etabar{{\overline{\eta}}}
\def\xbar{{\overline{x}}}

\def\IR{\relax{\rm I\kern-.18em R}}
 \font\cmss=cmss10 \font\cmsss=cmss10 at 7pt
\def\IQ{\relax{\rm I\kern-.18em Q}}
\def\IZ{\relax\ifmmode\mathchoice
 {\hbox{\cmss Z\kern-.4em Z}}{\hbox{\cmss Z\kern-.4em Z}}
 {\lower.9pt\hbox{\cmsss Z\kern-.4em Z}}
 {\lower1.2pt\hbox{\cmsss Z\kern-.4em Z}}\else{\cmss Z\kern-.4em Z}\fi}

\newcommand{\alignedfrac}[2]{%
    \setbox0\hbox{$#1$}        
    \dimen0=\wd0               
    \setbox1\hbox{$#2$}        
    \dimen1=\wd1               
    \ifdim\wd0<\wd1            
        \dfrac{#1\hfill}{#2}   
    \else                      
        \dfrac{#1}{#2\hfill}   
    \fi
}

\newcommand{\nn}{\nonumber}
\newcommand{\abs}[1]{\left| #1 \right|}

\newcommand{\be}{\begin{equation}}
\newcommand{\ee}{\end{equation}}
\newcommand{\ba}{\begin{align}}
\newcommand{\ea}{\end{align}}

\def\ie{{\it i.e.}\/}
\def\eg{{\it e.g.}\/}

\begin{document}
\title{A Tale of Two Timescales:\\  Mixing, Mass Generation, and Phase Transitions in the Early Universe}
\author{Keith R. Dienes$^{1,2}$\footnote{E-mail address:  {\tt dienes@email.arizona.edu}},
      Jeff Kost$^{1}$\footnote{E-mail address:  {\tt jkost@email.arizona.edu}},  
      Brooks Thomas$^{3,4}$\footnote{E-mail address:  {\tt bthomas@ColoradoCollege.edu}}}
\affiliation{
     $^1$ Department of Physics, University of Arizona, Tucson, AZ  85721  USA\\
     $^2$ Department of Physics, University of Maryland, College Park, MD  20742  USA\\
     $^3$ Department of Physics, Reed College, Portland, OR 97202 USA\\
     $^4$ Department of Physics, Colorado College, Colorado Springs, CO  80903 USA}

\begin{abstract}
Light scalar fields such as axions and string moduli can play an important role in early-universe cosmology.  However, many factors can significantly impact their late-time cosmological abundances. For example, in cases where the potentials for these fields are generated dynamically --- such as during cosmological mass-generating phase transitions --- the duration of the time interval required for these potentials to fully develop can have significant repercussions.  Likewise, in scenarios with multiple scalars, mixing amongst the fields can also give rise to an effective timescale that modifies the resulting late-time abundances.  Previous studies have focused on the effects of either the first or the second timescale in isolation.  In this paper, by contrast, we examine the new features that arise from the interplay between these two timescales when both mixing and time-dependent phase transitions are introduced together.  First, we find that the effects of these timescales can conspire to alter not only the total late-time abundance of the system --- often by many orders of magnitude --- but also its distribution across the different fields.  Second, we find that these effects can produce large parametric resonances which render the energy densities of the fields highly sensitive to the degree of mixing as well as the duration of the time interval over which the phase transition unfolds.  Finally, we find that these effects can even give rise to a ``re-overdamping'' phenomenon which causes the total energy density of the system to behave in novel ways that differ from those exhibited by pure dark matter or vacuum energy.  All of these features therefore give rise to new possibilities for early-universe phenomenology and cosmological evolution.  They also highlight the importance of taking into account the time dependence associated with phase transitions in cosmological settings.
\end{abstract}

\maketitle





\section{Introduction\label{sec:Intro}}


Light scalar fields are a common feature in many cosmological scenarios and
thus frequently play an important role in early-universe cosmology.
Such scalars include, for example, 
the QCD axion~\cite{Peccei:1977hh,Peccei:1977ur,Weinberg:1977ma,Wilczek:1977pj}
and other axion-like particles~\cite{Jaeckel:2010ni},
supersymmetric partner particles
such as sneutrinos and staus~\cite{ Binetruy:1983jf, Griest:1990kh, Jungman:1995df, Feng:2003zu},
string moduli such as the dilaton and other geometric 
moduli~\cite{green1988superstring,polchinski2001string,becker2006string},
Q-balls~\cite{Coleman:1985ki,Kusenko:1997zq},
quintessence fields~\cite{Ratra:1987rm}
and chameleons~\cite{Khoury:2003aq,Khoury:2003rn,Brax:2004qh},
familons~\cite{Wilczek:1982rv} and  
Majorons~\cite{Chikashige:1980ui, Gelmini:1980re},
certain degrees of freedom within inert Higgs-doublet models~\cite{Deshpande:1977rw},
additional scalars present in little-Higgs theories~\cite{ArkaniHamed:2001nc},
and others~\cite{Caldwell:1997ii,Boehm:2002yz,Boehm:2003hm,Kusenko:2012ch}.
Some of these fields may even serve as candidates for dark matter or as
significant contributors to dark energy.
Moreover, the number of such scalars can be relatively large;
indeed, the recent Dynamical Dark Matter framework~\cite{Dienes:2011ja,Dienes:2011sa}
posits a dark sector containing an entire ensemble of such fields which together
conspire to produce a number of phenomenological, astrophysical, and cosmological effects
which differ markedly from those arising from more traditional dark sectors.
Likewise, string-inspired models typically involve significant numbers 
of moduli  --- and
frequently a large number of light axions and axion-like fields as 
well~\cite{Witten:1984dg,Conlon:2006tq,Svrcek:2006yi, Arvanitaki:2009fg}.
  
When a scalar field is light, the reason is usually that a mass 
term for this field is forbidden by a symmetry of the full, high-temperature theory. 
Such a mass term must therefore be generated at a lower temperature scale through  
some dynamics which violates these symmetries --- dynamics typically 
associated with a cosmological phase transition.  For example, masses for axions 
or axion-like particles can be generated by non-perturbative instanton effects
which become significant at temperatures near the confining temperature of the 
corresponding gauge group.  Indeed, dynamical mass generation 
through cosmological phase transitions provides a natural mechanism for engineering 
mass-scale hierarchies.

However, dynamical mass generation can also have 
potentially significant ramifications for
the cosmology of the particles involved.  
First, cosmological phase transitions are never instantaneous.  
In the case of a second-order phase transition, the
transition between the high- and low-temperature phases is smooth but nevertheless time-dependent.  
Even in the case of a first-order transition, a rapid epoch of bubble nucleation and expansion 
interpolates between the two phases.
As a result, scalar masses which are dynamically generated during either kind of 
phase transition do not ``turn on'' instantaneously everywhere in the 
universe;  rather, these masses depend non-trivially on time and 
evolve continuously over the course of the phase transition towards
whatever asymptotic values they will have at late times. 
It turns out that this time-dependence ---  and the existence of an associated 
timescale or ``width'' over which the phase transition unfolds --- 
can have a significant impact 
on the cosmology of light scalars.  For example, in the case of an axion receiving its
mass via non-perturbative instanton effects,
the non-trivial consequences of this 
time-dependence on the resulting axion cosmological abundance 
were assessed in Ref.~\cite{Turner:1985si}.

Another important consequence of dynamical mass generation is that it can give rise 
to mixing amongst scalars sharing the same quantum numbers.  
Indeed, if multiple such scalars couple to the same source of dynamical mass 
generation, there is no guarantee that the mass eigenstates of the theory will
be the same before and after the corresponding phase transition.  The
phenomenological consequences of such mixing were investigated in 
Refs.~\cite{Dienes:1999gw,Dienes:2011sa,Dienes:2012jb} for an ensemble of light 
axion-like fields.   Mixing can also be induced amongst light scalar fields at low energies
through a variety of other dynamical processes, including the integration out of massive fields
to which our light scalars are coupled.  
For example, the phenomenological implications of such mixings amongst axion-like fields have
been studied in Ref.~\cite{Babu:1994id}.  
Regardless of the source, however, we find that dynamically-induced mixing generally results 
in a non-trivial modification of the coupling structure of the fields involved and
can thus have a significant impact on the resulting energy densities of the fields at late times.

The effects of mixing and of a non-trivial time-dependence for scalar mass generation
have historically been studied separately. 
In this paper, by contrast, we shall examine the effects that arise when the mass matrix
for a set of multiple light scalar fields includes {\it both}\/ mixing {\it and}\/ a non-trivial 
time-dependence.
As we shall demonstrate, 
the interplay between the different timescales associated with these effects
can, in the presence of a time-dependent background cosmology,
give rise to a number of qualitatively new features which are not seen when 
these effects are each considered in isolation.

These new features generally concern the late-time cosmological
abundances (energy densities) associated with the scalar fields in our coupled multi-scalar system.
For example, we shall demonstrate that 
the total late-time energy density of our system can be altered --- often by many orders of magnitude --- 
compared with traditional expectations.  
This includes situations in which the late-time energy density is enhanced to greater values
as well as situations in which it is strongly suppressed.
Second, we shall demonstrate that not only can the total energy density be altered, but so can 
the distribution of this total energy density across the different fields in our system --- often
in dramatic ways.
Indeed, these effects can even completely
change which field carries the largest abundance.
Third, we shall demonstrate that under certain circumstances, 
the combined effects of our two timescales can give rise to 
large parametric resonances which render the energy densities of the fields highly sensitive to 
the degree of mixing as well as the duration of the time interval over which the phase transition unfolds. 
As a result of this heightened sensitivity, any effects which 
modify the mixing slightly 
or which change the rate at which our mass-generating phase transition unfolds
can have a huge influence on the corresponding late-time cosmological abundances.
Finally, we shall demonstrate that mixing in conjunction with a time-dependent phase
transition can even give rise to a ``re-overdamping'' phenomenon which causes the field values and
total energy density of the system to behave in novel ways which differ from those normally associated with
pure dark matter or vacuum energy.  

Taken together, we see that all of these features give rise to new possibilities for 
the phenomenology of the early universe and cosmological evolution.
In particular, many new possibilities for model-building emerge.
However, our results can also serve as warning:  
in the presence of multiple mixed scalar fields,
it is essential to treat mass-generating phase transitions
rigorously, with the proper time-dependence 
included.   Indeed, approximations in which 
such phase transitions are treated as instantaneous can produce late-time 
energy densities which differ from their true values by many orders of magnitude.
This warning is particularly relevant in the case of axions,
where the mass-generating phase transition is nothing but the instanton-induced QCD phase transition.
This phase transition unfolds over a calculable non-zero timescale,
and we shall demonstrate explicitly that ignoring this timescale 
(by setting it to be either zero or effectively infinite)
can lead to highly inaccurate late-time axion abundances.

This paper is devoted to a general study of the new features discussed above.
Because these features are common to many systems of scalars which 
exhibit mixing in conjunction with
a time-dependent mass-generating phase transition, 
in this paper we shall work within the context of a general model 
of scalars $\phi_i$ whose identities remain unspecified.
We shall likewise make no assumptions about the nature of the mass-generating phase transition
except that it unfolds over a certain timescale.
Moreover, quite remarkably, we shall find that a simple toy model consisting of only {\it two}\/ scalar 
fields is sufficient to illustrate all of the features outlined above.
This paper will therefore focus on an analysis of this two-component 
toy model, and we shall defer a study of more complex scenarios to future work~\cite{toappear}.    

This paper is organized as follows.
In Sect.~\ref{sec:AToyModel}, we introduce the two-component toy model
which forms the basis for all subsequent discussions in this paper.
As outlined above, this toy model contains both a mixing between our two fields
as well as a time-dependent mass-generating phase transition.
Along the way we also introduce all needed definitions, conventions, and notation.
We also discuss some of the basic properties of this model.
Then, in Sect.~\ref{sec:LateTimeEnergyDensity},
we study the behavior of the total late-time energy    
density of our two-component system, while in Sect.~\ref{sec:indiv}
we study how this total late-time energy density is
distributed between the two fields of our model.
In Sect.~\ref{sec:TheParametricResonance} we 
discuss the parametric resonance which appears in the energy
density, while in Sect.~\ref{sec:ReOverDamping} we turn our attention
to the ``re-overdamping'' phenomenon.

Although our analysis up to this point is completely general, 
our results have many immediate phenomenological applications.
In Sect.~\ref{sec:Axion}, we therefore focus on a special case
by considering the effects on the standard QCD axion
that emerge when a second axion is incorporated into the theory.
In Sect.~\ref{sec:BeyondTwoFields} we then sketch some of the additional 
features that emerge when more than two fields are considered.
These features will ultimately be explored more fully in Refs.~\cite{toappear}.
Finally, in Sect.~\ref{sec:Conclusions}, we conclude with a discussion of our
main results and possible avenues for extension and generalization.

This paper also contains two Appendices.
Although most of the results of this paper are obtained through numerical
analysis, Appendix~\ref{notation} provides exact analytical results in certain tractable 
special cases.   Appendix~\ref{alternative} then discusses    
an alternative approach towards analyzing some of 
the features presented in this paper.

\FloatBarrier

\section{A toy model\label{sec:AToyModel}}


In this section we delineate the 
toy model which shall be the basis of our analysis in this paper.
This section will also serve to establish our notation and conventions.


\subsection{Motivation:  A one-component warm-up\label{sec:TheSingleField}}


As a prelude to the presentation of our model, we begin with a short
discussion of the one-component case.
Along the way we will also review
some basic facts about
scalar fields in an FRW cosmology and
provide motivation for the particular construction of our eventual
multi-component toy model.

Towards this end, let us consider a single scalar field $\phi$ of mass $m$ evolving in a flat FRW universe.
If all spatial variations in $\phi$ are assumed to be negligible,
such a field evolves according to the standard equation of motion
\beq
\frac{\partial^2\phi}{\partial t^2} + 3H(t)\frac{\partial\phi}{\partial t} + m^2\phi ~=~ 0 ~
\label{eq:singlephieqofmotion}
\eeq
where the Hubble parameter $H(t)$ scales as \mbox{$3H(t)\approx \kappa/t$} with \mbox{$\kappa=2$}
(respectively \mbox{$\kappa=3/2$}) during a matter- (radiation-) dominated epoch.
As a result, the field behaves as a damped oscillator with a 
time-dependent damping ratio \mbox{$\zeta(t) \equiv 3H(t)/2m$}.
Since $H(t)$ falls with time,
a field which is initially overdamped will inevitably become underdamped and experience decaying oscillations.

For simplicity, the energy-momentum tensor for such a field can be modeled as a perfect fluid
with energy density $\rho$ and pressure $p$ given by
\beqn
  \rho &=&  \frac{1}{2}\left\lbrack \left(\partial \phi/\partial t\right)^2 + m^2\phi^2\right\rbrack ~, \nn \\
   p   &=&  \frac{1}{2}\left\lbrack \left(\partial \phi/\partial t\right)^2 - m^2\phi^2\right\rbrack ~.
\label{eq:singlefieldrho}
\eeqn
If $\phi$ is initially in an overdamped phase with \mbox{$\partial \phi/\partial t=0$}, then
\mbox{$\rho= -p$}:  the energy density $\rho$
associated with such a field in this phase behaves as vacuum energy.
By contrast, after $\phi$ transitions to an underdamped phase, the field eventually experiences oscillations
which are approximately virialized:
in this phase \mbox{$p=0$} and the corresponding energy density $\rho$ can be associated with massive matter.
Of course, there is also an intermediate time interval {\it during}\/ the transition from the overdamped to 
underdamped phase within which 
the behavior of the field exhibits transient features that eventually dissipate.

\begin{figure}[t]
\centering
\hspace*{-0.4cm}\includegraphics[width=0.45\textwidth,keepaspectratio]{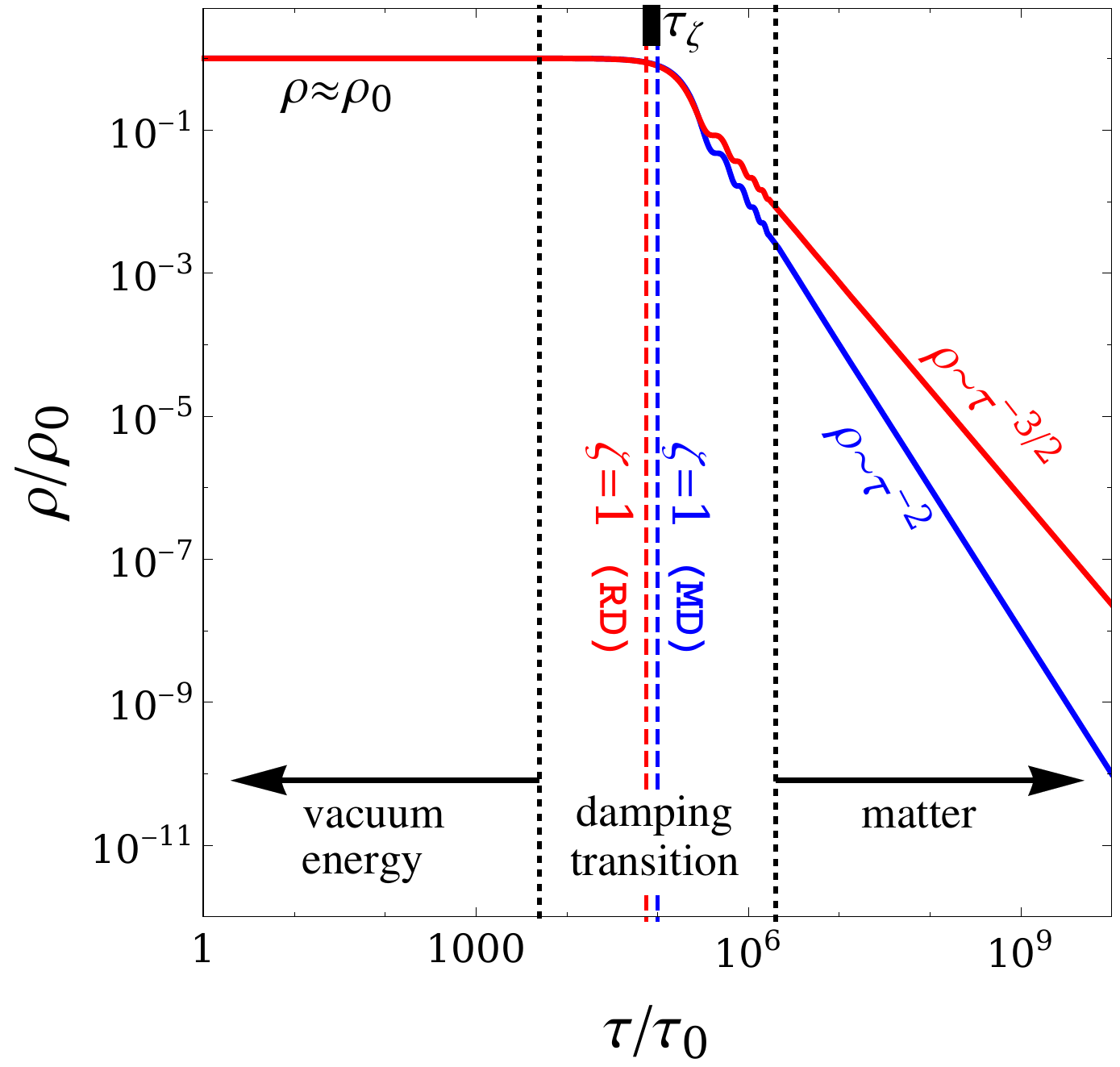}
\caption{The energy density $\rho$, normalized to its initial value $\rho_0$, for 
a solution to Eq.~\eqref{eq:singlephieqofmotion} in a radiation-dominated (RD) universe, \ie, 
\mbox{$\kappa=3/2$} (red), or a matter-dominated (MD) universe, \ie, \mbox{$\kappa=2$} (blue).
In the overdamped phase, the energy density remains nearly 
constant at \mbox{$\rho=-p>0$}, behaving as vacuum energy. 
By contrast,
once the critical damping \mbox{$\zeta= 1$} threshold is crossed at $\tau_\zeta$
(a critical time which is slightly different for RD and MD universes),
the energy density begins to dissipate, asymptotically exhibiting
the simple power-law scaling behavior \mbox{$\rho\sim \tau^{-\kappa}$} expected for matter.}
\label{fig:rhosinglefield}
\end{figure}

The equation of motion \eqref{eq:singlephieqofmotion} can be simplified by defining a dimensionless time variable
\mbox{$\tau\equiv mt$}.
The solutions can then be expressed analytically in terms of Bessel functions of the first and second kind.
The exact evolution of the corresponding energy density $\rho$ is shown in Fig.~\ref{fig:rhosinglefield}, where
$\tau_\zeta$ denotes the critical damping time at which \mbox{$\zeta=1$}.
The different phases of the system are clearly evident in Fig.~\ref{fig:rhosinglefield}.
For early times \mbox{$\tau \ll \tau_\zeta$}, the system is overdamped and the energy density
$\rho$ is essentially constant:  \mbox{$\rho\sim \rho_0$}.
For late times \mbox{$\tau\gg \tau_\zeta$}, by contrast,
the system is underdamped and 
the energy density $\rho$ scales as \mbox{$\rho\sim \tau^{-\kappa}$}.
Finally, during the intermediate times \mbox{$\tau\sim \tau_\zeta$},
the energy density $\rho$ exhibits a corresponding transition
between the two limiting behaviors above, punctuated by small transient oscillations.
Although the specific dynamics of the field during this transitional
period does not affect the late-time asymptotic scaling of $\rho$ with time,
these transients are nevertheless important in determining the overall scale
of the late-time energy density.

In this example, we have taken the mass $m$ to be a constant, non-zero for all time.
However, in many cosmological situations, masses are generated by phase transitions. 
In such cases, the masses of such fields can be time-dependent.
A well-motivated example of this is case of the QCD axion: 
the axion potential is flat (\ie, \mbox{$m=0$}) at temperatures \mbox{$T\gg\Lambda_{\text{QCD}}$},
but this flat potential is modified at lower temperatures by instanton effects which generate an 
effective mass.

\begin{figure}[t]
\centering
\includegraphics[width=0.41\textwidth, keepaspectratio]{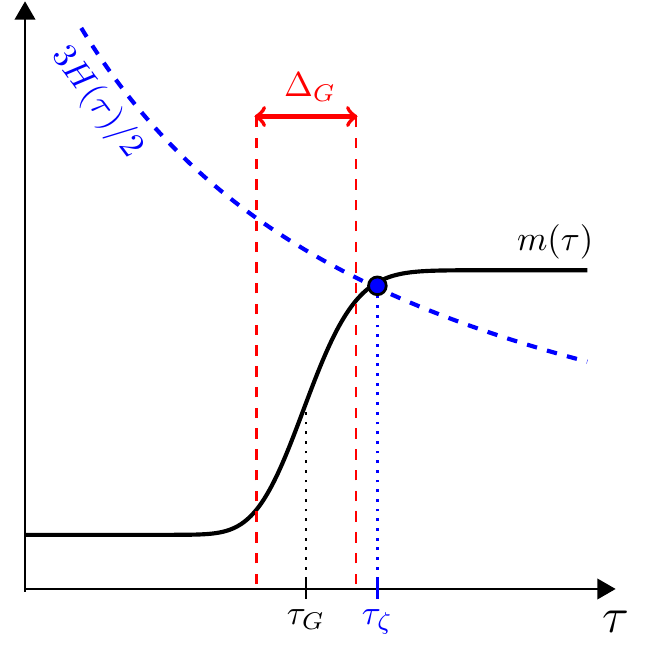}
\caption{Two time-dependent mass scales:  
a growing mass function $m(\tau)$ and a falling Hubble friction scale $3H(\tau)/2$.
Their intersection determines the transition time $\tau_\zeta$.
We assume that the growth of $m(\tau)$ takes place during an interval
of approximate duration $\Delta_G$ centered at $\tau_G$.
As we shall see, the introduction of a non-zero $\Delta_G$
can have many dramatic effects on the late-time energy densities of
those fields which (either directly or indirectly) are sensitive to this change in mass.
For example, one new possibility 
is that 
the transition between overdamped and underdamped phases can 
occur {\it during}\/ the mass-generating phase transition.}
\label{fig:3Hmfig}
\end{figure}

Time-dependent masses $m(\tau)$ can significantly modify the evolution of such scalar fields.
In this paper, we are primarily interested in cases in which at least a portion of the contributions to
the scalar masses are generated as the result
of phase transitions.   We shall therefore generally consider cases in which 
$m(\tau)$ has one (smaller) value at early times,
a second (larger) value at asymptotically late times,
and a smooth time-dependent transition between the two.
Indeed, we shall let $\tau_G$ represent the ``central'' time at which
this mass-generating phase transition occurs,
and imagine that this transition unfolds over a time interval
of duration or width $\Delta_G$.
We then obtain the situation illustrated schematically in Fig.~\ref{fig:3Hmfig}.

The introduction of a non-zero time interval $\Delta_G$ 
during which our phase transition unfolds
can have a significant effect on the time-evolution of $\phi$.
When \mbox{$\Delta_G=0$}, there are essentially two regimes 
which one can consider, each with its own distinctive 
phenomenology:  \mbox{$\tau_\zeta< \tau_G$} and \mbox{$\tau_\zeta>\tau_G$}.
However, a non-zero width $\Delta_G$ opens up a third possibility, with \mbox{$\tau_\zeta\sim\tau_G$}.
Indeed, more generally, the introduction of a non-zero $\Delta_G$ introduces
a new timescale into the problem, and we shall see that this can have dramatic
effects.
We shall explore the phenomenology of this region as part of our larger general study.
We emphasize, however, that phenomenon of having  \mbox{$\tau_\zeta\sim\tau_G$} is completely different from having $\tau_G$ 
during the ``damping transition'' region of 
Fig.~\ref{fig:rhosinglefield}.  Indeed, while
$\Delta_G$ may be considered to represent a width around $\tau_G$, 
the ``damping region'' of 
Fig.~\ref{fig:rhosinglefield} can instead be considered a
width around $\tau_\zeta$, a width which exists --- as shown in Fig.~\ref{fig:rhosinglefield} ---
 {\it even if the mass is constant}\/.


\subsection{Defining our toy model}


The introduction of non-zero time interval $\Delta_G$ for our phase transition
is just one feature we wish to study in this paper.
The other concerns the possibility of {\it mixing}\/ between different
field components $\phi_i$, \mbox{$i=1,\ldots,N$}.    
The phenomenological effects stemming 
from each can certainly be studied in isolation. 
However, as we shall find, these two features can conspire to produce a number
of remarkable effects that transcend what is possible with either alone.
For this reason, we shall study the implications of both features
together.  The effects of either feature in isolation can then 
be extracted through the limiting cases in which the 
effects due to the other feature are gradually turned off.

In this paper, we shall explore these effects
within the context of a simple toy model
involving only two components, $\phi_0$ and $\phi_1$.
As we shall see, our toy model is simple enough to be tractable, 
yet rich enough to incorporate all of the phenomena of interest.

Our model consists of two real scalar fields, $\phi_0$ and $\phi_1$.
If we again assume
that the spatial variations in these fields are negligible,
their equations of motion in a flat FRW universe take the form
\beq
{\partial^2 \phi_i \over \partial t^2} + 3H(t) {\partial\phi_i \over \partial t}  +
        \sum_{j}\mathcal{M}^2_{ij}\,\phi_{j} ~=~ 0 \
\label{eq:doublefieldequationsofmotion}
\eeq
where $\mathcal{M}^2$ is the corresponding squared-mass matrix.
At times \mbox{$t\ll t_G$}, long before mass generation occurs, we shall take 
$\mathcal{M}^2$ constant and diagonal. 
In fact, for simplicity, we shall further assume that $\phi_0$ is massless at such early times.
Thus, at early times, we shall assume
\beq
\mathcal{M}^2 ~\xrightarrow[~t \ll t_G~]{}~ \begin{bmatrix}[1.2] {}^{}\ 0 &  0 \\ \ 0 & M^2\end{bmatrix} \ 
\label{eq:massmatrixhighT}
\eeq
where \mbox{$M\not=0$} is a general unfixed mass parameter.
By contrast, 
long after the phase transition has occurred,
we assume that new components $\overline{m}_{ij}^2$ will have been generated in the squared-mass matrix:
\beq
\mathcal{M}^2 ~\xrightarrow[~t\gg t_G~]{}~ \begin{bmatrix}[1.2] \ 0 & 0\\ \ 0 & M^2\end{bmatrix} + \begin{bmatrix}[1.2] \overline{m}_{00}^2 & \overline{m}_{01}^2\\ \overline{m}_{01}^2 & \overline{m}_{11}^2\end{bmatrix} \ ,
\label{eq:massmatrixlowT}
\eeq
In order to connect these two asymptotic extremes, 
we shall let $m_{ij}^2(t)$ denote 
time-dependent elements of the mass matrix 
which interpolate between zero at early times and $\overline{m}_{ij}$ at late times.
We can then write
\beq
\mathcal{M}^2(t) ~=~ \begin{bmatrix}[1.2] {}^{}\ 0 &  0 \\ \ 0 & M^2\end{bmatrix} + \begin{bmatrix}[1.2] m_{00}^2(t) & m_{01}^2(t) \\ m_{01}^2(t) & m_{11}^2(t)\end{bmatrix} \ .
\label{eq:fullM2}
\eeq

Having specified the mass matrix, we can now introduce a dimensionless time variable
\mbox{$\tau\equiv M t$}, as in the single-component case.
Our equations of motion then take the form
\beq
   \ddot{\phi}_i + 3H(\tau) \dot{\phi}_i + \sum_{j}\mathcal{M}^2_{ij}(\tau)\,\phi_{j} ~=~ 0 \ 
\label{eq:newerdoublefieldequationsofmotion}
\eeq
where the dots indicate $\partial/ \partial \tau$ 
and where our mass matrix is also dimensionless and takes the form
\beq
   \calM^2(\tau) ~=~    \begin{bmatrix}[1.2] m_{00}^2(\tau) & m_{01}^2(\tau) \\ m_{01}^2(\tau) & 1+ m_{11}^2(\tau)
       \end{bmatrix}
\label{massmatrix}
\eeq
with each dimensionless $m_{ij}$ now understood to be a fraction of $M$.
 {\it We shall adopt these conventions throughout the rest of this paper.}

In general, there is no reason to expect that the mass generation occurs in the same basis 
as in Eq.~\eqref{eq:massmatrixhighT}.  
We shall therefore allow for the possibility that \mbox{$m_{01}^2\not= 0$} --- \ie,
the possibility 
that our phase transition induces a mixing between our primordial field components $\phi_0$ and $\phi_1$.  
However, we shall nevertheless make the simplifying assumption that the time-dependence of each component is identical,
allowing us to focus on those effects that come from a single common timescale for mass generation.
Indeed, since the mass matrix is nothing but the curvature matrix 
associated with the potential $V(\phi_0,\phi_1)$ induced by the phase transition, this assumption 
is tantamount to assuming a single time-dependence for the potential as a whole.
As a result, we can write each of the individual mass components in a factorized form
\beq
          m_{ij}(\tau) ~=~ \overline{m}_{ij}\cdot  h(\tau) \ ,
\label{eq:doubleTDfactoring}
\eeq
where $h(\tau)$ is a smooth function of time which describes the time-development of the phase transition.

Our next step in specifying our toy model is to choose a suitable function $h(\tau)$.
As indicated above, we require that \mbox{$h(\tau)\to 0$} as \mbox{$\tau\to 0$} and \mbox{$h(\tau)\to 1$} as \mbox{$\tau\to\infty$}.
However, because one of our main interests in this paper 
concerns the timescale associated with the mass-generating phase transition,
we would also like $h(\tau)$ to incorporate a dimensionless parameter $\sigma$ 
which controls how abruptly the phase transition occurs. 
The limit \mbox{$\sigma \to 0$} might then correspond to a phase transition which is effectively instantaneous, 
while non-zero values of $\sigma$
correspond to phase transitions which occur increasingly slowly.
It is also desirable, regardless of the width of the transition, 
that the midpoint at $\tau_G$ be a fixed point of reference. 
We therefore include in our construction the requirement that \mbox{$h(\tau_G)=1/2$} for all $\sigma$. 
Indeed, this may be taken as a definition of $\tau_G$.

\begin{figure}[t]
\centering
\hspace*{-0.4cm}\includegraphics[width=0.45\textwidth, keepaspectratio]{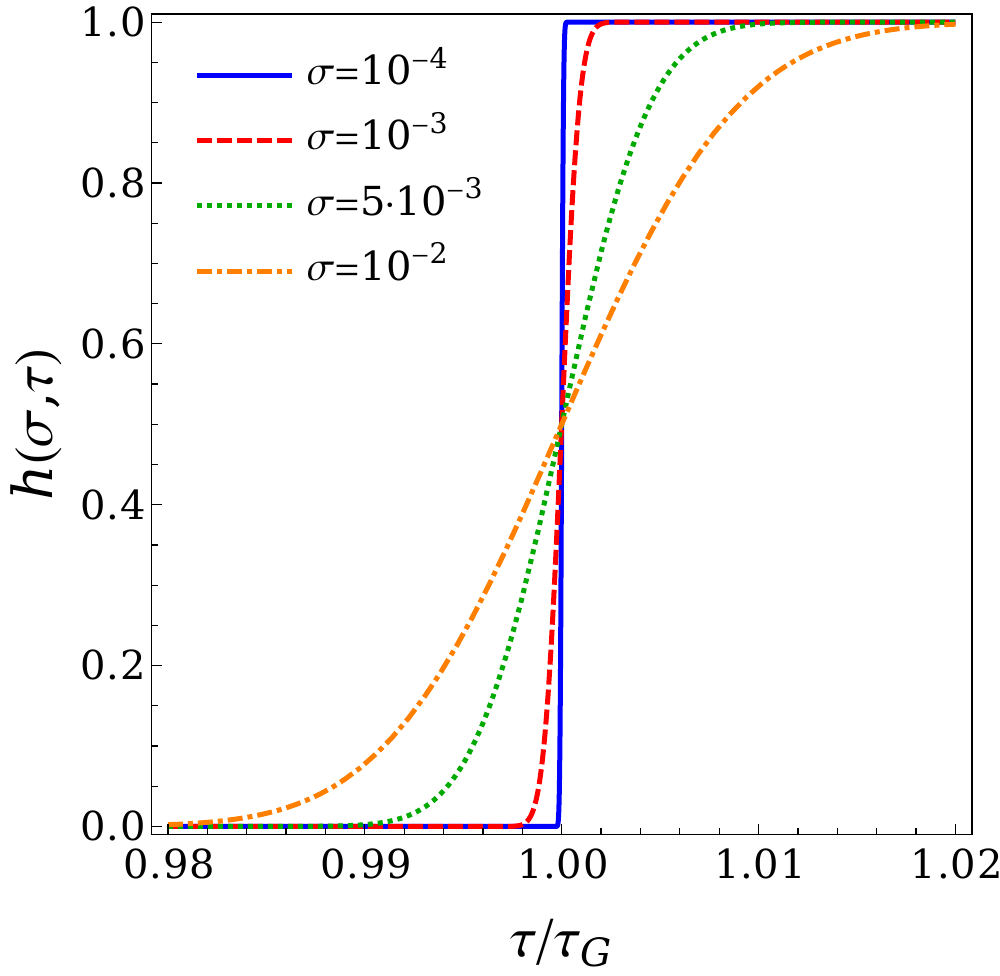}
\caption{The general form of the $h$-function in Eq.~(\protect\ref{eq:hdefinition}), 
plotted for several choices of the parameter $\sigma$. 
In all cases we see that $h(\tau)$ rises smoothly from \mbox{$h=0$} to \mbox{$h=1$} and crosses \mbox{$h=1/2$} 
at \mbox{$\tau=\tau_G$}.
Large values of $\sigma$ correspond to more gradual transitions, while for \mbox{$\sigma\to 0$} 
we find that $h$ approaches the Heaviside step function \mbox{$\Theta(\tau-\tau_G)$}.}
\label{fig:hsamples}
\end{figure}

Beyond these constraints, the choice of $h(\tau)$ is completely arbitrary, and many functions may be chosen.
For concreteness, however, we shall take
\beq
         h(\tau) ~=~ \frac{1}{2}\left\{1+\text{erf}\left[\frac{1}{\sigma}\log\left(\frac{\tau}{\tau_G}\right)\right]\right\} \ ,
\label{eq:hdefinition}
\eeq
where the error function ${\rm erf}(z)$ is given by 
\beq
     \text{erf}(z) ~\equiv~ \frac{2}{\sqrt{\pi}}\int_0^z e^{-x^2}dx~.
\eeq
As illustrated in Fig.~\ref{fig:hsamples}, this function satisfies all of our requirements.
Of course, many other choices for $h(\tau)$ are possible.
However, none of the qualitative results of this paper will ultimately depend on the specific choice for $h(\tau)$.
Thus, any smooth, monotonic function $h(\tau)$ satisfying the above constraints will 
lead to similar results.

Corresponding to each non-zero value of the parameter $\sigma$ there exists a  non-zero timescale $\Delta_G$ over which the phase transition 
occurs.  In general, we may define $\Delta_G$ in terms of the slope of the $h(\tau)$ function at its midpoint $\tau_G$: 
\beq
          h(\tau_G + \delta \tau) ~\approx~ \frac{1}{2} + \frac{\delta \tau}{\Delta_G} ~ ~~~~ 
                    {\rm for}~\delta \tau\ll \Delta_G~.
\label{eq:hwidthderivation}
\eeq
Adopting this definition for $\Delta_G$, we then find for our $h(\tau)$ function 
that
\beq
            \Delta_G ~\equiv~ \sqrt{\pi}\sigma\tau_G ~.
\label{eq:sigDelta}
\eeq
We thus see that \mbox{$\Delta_G\to 0$} as \mbox{$\sigma\to 0$}, as expected.
Indeed, this limit corresponds to the case of an instantaneous phase transition, with the $h(\tau)$
taking the form of the Heaviside step function $\Theta(\tau-\tau_G)$.

In this paper, we shall consider the effects that arise when $\sigma$ is non-zero.
There is, however, a critical value $\sigma^\ast$  above which the behavior of $h(\tau)$ 
changes in an important way.
To see this, let us first consider the limit in which \mbox{$\sigma \to \infty$}.
In this limit, we find that $h(\tau)$ is essentially constant and $\tau_G$-independent:
\mbox{$h(\tau)\approx \half \Theta(\tau)$} for all $\tau$.
This limit may therefore be interpreted as one in which 
our original phase transition at $\tau_G$ effectively disappears and
is replaced by a new, infinitely sharp ``phase transition'' at \mbox{$\tau=0$}.
Note that this latter transition is nothing but an artifact of our boundary condition
that \mbox{$h=0$} at \mbox{$\tau=0$}.  
As a result, dialing $\sigma$ from $0$ to $\infty$ has the effect of slowing 
our original phase transition near $\tau_G$ while simultaneously building up 
an artificial phase transition near \mbox{$\tau=0$}.  
Indeed, if $\sigma$ grows too large,
the most rapid changes in the mass parameters of our system will no longer be associated with our original phase
transition near \mbox{$\tau=\tau_G$}, but with the artificial one near \mbox{$\tau=0$}.

In this paper, we wish to maintain the notion that increasing the value of $\sigma$
corresponds to slowing the time-development 
of our mass matrix.
Even more importantly, we 
also wish to ensure that the time period exhibiting the most rapid time-development 
of our mass matrix is still associated with our
original phase transition near $\tau_G$ and not the artifact near \mbox{$\tau=0$}.
Therefore, in this paper, we shall always restrict our attention 
to values \mbox{$0\leq \sigma \leq \sigma^\ast$} where
$\sigma^\ast$ is that value of $\sigma$ for which 
         $\max_\tau (dh/d\tau)$
is minimized.
For the $h(\tau)$ function given in Eq.~(\ref{eq:hdefinition}), we find that \mbox{$\sigma^\ast =\sqrt{2}$}
for all $\tau_G$.
As a result, in what follows we 
shall only consider values of $\Delta_G$ in the range
\beq
                0 ~\leq~ \Delta_G ~\leq \sqrt{2\pi} \,\tau_G~.
\eeq

\begin{figure}[t]
\centering
\hspace*{-0.4cm}\includegraphics[width=0.45\textwidth, keepaspectratio]{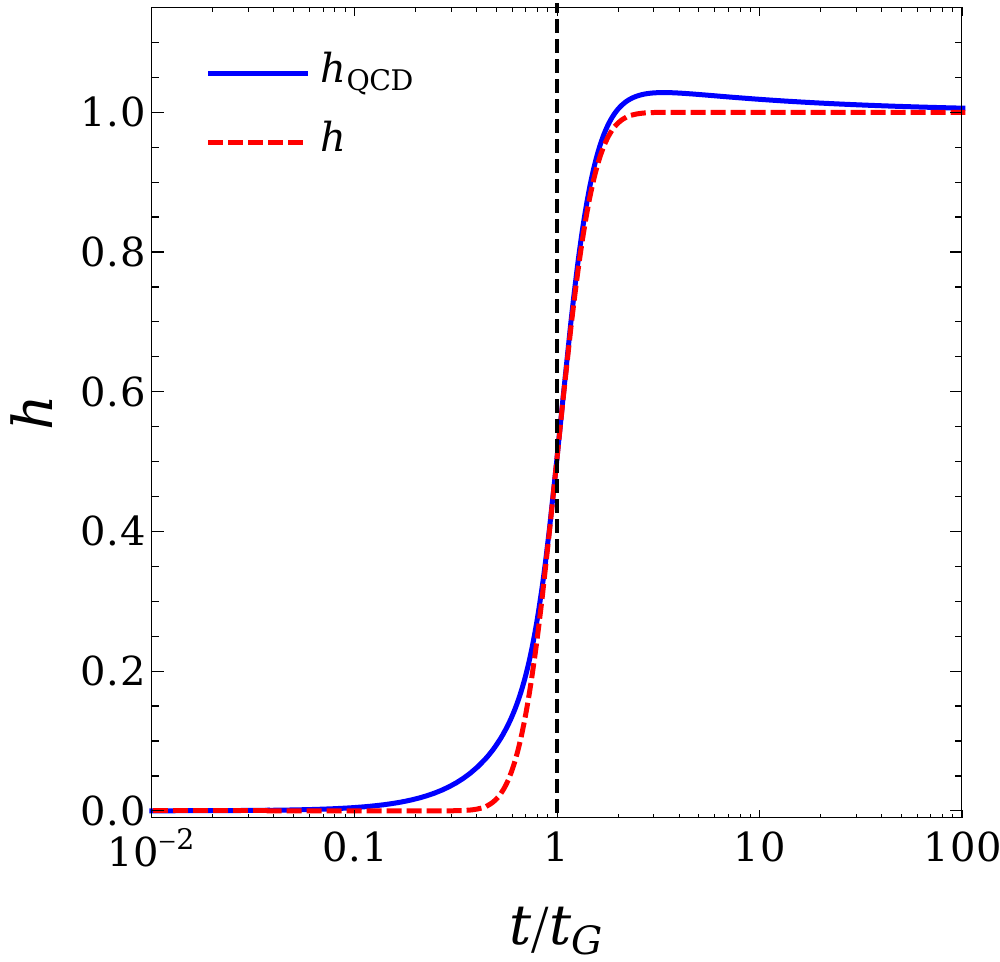}
\caption{The function $h_{\text{QCD}}$ describing the instanton-induced
time-dependent mass of the axion (solid blue line),
plotted as a function of $t/t_G$.
This curve is generated for \mbox{$\Lambda_{\rm QCD}=200$~MeV}, with 
$t_G$ defined as that location where the curve crosses \mbox{$h=1/2$}.
For comparison purposes, our $h$-function in Eq.~(\protect\ref{eq:hdefinition}) 
with \mbox{$\sigma\approx 0.47$} is superimposed (dashed red line).
We see that our $h$-function matches $h_{\text{QCD}}$ surprisingly well throughout the entire
range of values of plotted, and fits particularly well near the ``center'' of the phase
transition at \mbox{$t=t_G$.}}
\label{fig:axioncompare}
\end{figure}

In order to gain insight into the values of $\sigma$ that we might expect for a well-motivated phase transition,
let us consider the case of the instanton-induced phase transition
during which a mass is generated for the QCD axion.
Detailed lattice studies~\cite{Wantz:2009mi,Wantz:2009it} have yielded approximate expressions
for the mass of the QCD axion as a function of temperature $T$.
However, the cosmological time-temperature relation
appropriate for a radiation-dominated epoch is given by
\beq
    T(t) ~=~ \left(\frac{5}{2g_*}\right)^{1/4}\left(\frac{3M_P}{\pi t}\right)^{1/2}~,
\eeq
where $M_P$ is the Planck mass and 
where \mbox{$g_*$} is the temperature-dependent effective number of relativistic degrees of freedom.
Use of this relation then 
allows us to determine the axion mass 
as a function of cosmological time $t$,
with the results shown in Fig.~\ref{fig:axioncompare}.
For comparison, our $h$-function with $\sigma\approx 0.47$
is also shown in Fig.~\ref{fig:axioncompare}.
It is clear that these two functions match quite well over the entire range of times shown.

The only remaining ingredient to be specified as part of our toy model is 
a set of initial conditions to be imposed at some early time $\tau_0$.  For the differential equations
in Eq.~(\ref{eq:newerdoublefieldequationsofmotion}),
this means specifying initial values for our two fields $\phi_{0,1}$ and their
first derivatives $\dot{\phi}_{0,1}$.
While many possibilities exist, one particularly natural choice is to 
consider a mere displacement for the $\phi_0$ field:
\beq
\begin{pmatrix}[1.2]
   \phi_{0} \\
   \phi_{1}
   \end{pmatrix}_{\tau = \tau_0}  = \begin{pmatrix}[1.2] A_0 \\ 0\end{pmatrix} ~,~~~~~ 
\begin{pmatrix}[1.2]
    \dot{\phi}_{0}\\
    \dot{\phi}_{1}
    \end{pmatrix}_{\tau=\tau_0}  = 
    \begin{pmatrix}[1.2]
    \ 0 \ \ \\  \ 0 \ \ \end{pmatrix} ~.~~
\label{eq:initialconditions}
\eeq
There are three special properties associated with this initial configuration which
make this choice especially appealing.
First, as long as \mbox{$\tau_0 \ll \tau_G$}, 
this configuration does not introduce any initial energy into the system.
Thus, all of the energy that our system accrues will be solely that injected through the
phase transition.
Second, as long as \mbox{$\tau_0 \ll \tau_G$}, the specific choice of the initial time $\tau_0$
will be irrelevant.   In other words, given our other assumptions, 
the system is essentially time-independent for all \mbox{$\tau\ll \tau_G$}
and thus completely insensitive to variations of $\tau_0$.  
As a result, we shall never need to refer to $\tau_0$ again, understanding implicitly
throughout this paper that $\tau_0$ has always been chosen to be sufficiently early that our 
results are independent of $\tau_0$.
Finally, the choice of $A_0$ in Eq.~(\ref{eq:initialconditions}) merely serves
to set an overall mass scale for our system.   However, since we shall always 
be considering {\it ratios}\/ of field values or energy densities in this paper,
our results will also be insensitive to $A_0$.

This completes the specification of our toy model. 
It contains only five free parameters:  
two (namely $\tau_G$ and \mbox{$\Delta_G\sim \sigma$}) which describe the time-development of the phase transition,
and three (namely $m_{00}^2$, $m_{11}^2$, and $m_{01}^2$) which describe the effects of the phase
transition and the mixing it induces.
Despite its simplicity, however, we shall
find that this toy model is not only of sufficient generality to accommodate a wide variety of
physical systems but also of sufficient richness 
to give rise to a number of surprising phenomena.
 

\subsection{Preliminary analysis:  Constraints on mixing}


In subsequent sections, we shall analyze the behavior of our fields and of their corresponding energy 
densities within different regions of our five-dimensional parameter space.  
However, even before proceeding, there are a number of preliminary
observations that hold more generally and which serve 
to significantly constrain the allowed parameter space.

These constraints all ultimately stem from the observation that
not every choice of the masses $m_{ij}^2$ can be made independently.
Requiring that the eigenvalues of $\mathcal{M}^2$ be real is tantamount
to demanding that $\calM^2$ be Hermitian.  However we must further
demand that these eigenvalues be non-negative, which requires that
$\calM^2$ be positive-semidefinite.
This requires that all $m_{ij}^2$ be real and satisfy the three constraints
\beq
\begin{cases}
     \bullet  &  m_{00}^2 \geq 0~\cr
     \bullet  &  m_{11}^2 \geq -1~ \cr
     \bullet  &  m_{00}^2 (1+ m_{11}^2) \geq m_{01}^4~.\cr
\end{cases}
\label{constraints}
\eeq

These constraints provide a set of intrinsic limits on the 
magnitude of the mixing that may occur within our toy model. 
Indeed, these constraints imply that 
\beq
   \left| m_{01}^2 \right|  ~\leq~ 
   [m_{01}^2]_{\rm max} ~\equiv~ \sqrt{  m_{00}^2 (1+m_{11}^2) }~.
\label{mixingmax1}
\eeq
However, for many purposes it will prove  
useful to introduce the variables
\begin{align}
    m_{\text{sum}}^2 &~\equiv~ \mathcal{M}^2_{11} + \mathcal{M}^2_{00} ~=~ 1+ m_{00}^2 + m_{11}^2 \nn \\
    \Delta m^2 &~\equiv~ \mathcal{M}_{11}^2 - \mathcal{M}_{00}^2 ~=~ 1 - m_{00}^2 + m_{11}^2   
\label{eq:ms2Dm2defs}
\end{align}
in terms of which our squared-mass matrix $\calM^2$ in Eq.~(\ref{massmatrix}) takes the more symmetric form
\beq
\mathcal{M}^2 = \begin{bmatrix}[1.2] \frac{1}{2}\left(m_{\text{sum}}^2 - \Delta m^2\right) & m_{01}^2 \\ m_{01}^2 & \frac{1}{2}\left(m_{\text{sum}}^2 + \Delta m^2\right)\end{bmatrix} \ .
\label{eq:redefinedM2}
\eeq
Note while the individual $m_{ij}^2$ quantities each carry the same time-dependence [proportional to $h(\tau)^2$],
the same is no longer true for
$m_{\rm sum}^2$ and $\Delta m^2$.
In terms of these variables, the constraints in Eq.~(\ref{constraints})
then take the form
\beq
\begin{cases}
     \bullet  & m_{\rm sum}^2 \geq 0 \cr   
     \bullet  & |\Delta m^2| \leq m_{\rm sum}^2 \cr   
     \bullet  & | m_{01}^2 | \leq {1\over 2} \sqrt{ (m_{\rm sum}^2)^2 - (\Delta m^2)^2}~, \cr  
\end{cases}
\eeq
whereupon we find from Eq.~(\ref{mixingmax1}) that
\beq
   [m_{01}^2]_{\rm max} ~=~ \frac{1}{2}\sqrt{(m_{\text{sum}}^2)^2 - \left(\Delta m^2\right)^2} ~.
\label{mixingmax2}
\eeq

Finally, perhaps the most useful way to parametrize the mixing in our toy model
is in terms of a rotation angle $\theta$ which relates the mass eigenstates
$\phi_{\lambda_0}$ and $\phi_{\lambda_1}$
at any instant of time to the original mass eigenstates prior to the onset of the phase transition:
\beq
  \begin{pmatrix}
      \phi_{\lambda_0}\\ \phi_{\lambda_1}
     \end{pmatrix} ~=~ 
   \begin{pmatrix}
       \cos\theta & -\sin\theta\\ 
       \sin\theta & \phantom{-}\cos\theta
        \end{pmatrix}
     \begin{pmatrix}  \phi_0 \\ \phi_1\end{pmatrix} ~,
\label{eq:Udefinition}
\eeq
from which it follows that
\beq
    \tan(2\theta) ~=~ \frac{2m_{01}^2}{\Delta m^2} ~ .
\label{eq:mixingrelation}
\eeq
Indeed, while $\theta$ generally populates the range \mbox{$-\pi \leq \theta \leq \pi$} in 
Eq.~(\ref{eq:Udefinition}), 
we see from Eq.~(\ref{eq:mixingrelation}) 
that it is sufficient to focus on \mbox{$-\pi/2\leq \theta\leq \pi/2$}
for the purpose of calculating mixing angles,
since these are invariant under the mapping
\mbox{$(\phi_{\lambda_0},\phi_{\lambda_1})\to
-(\phi_{\lambda_0},\phi_{\lambda_1})$}.
The same will also be true for calculating energy densities ---
our main interest in this paper ---
because these energy densities will depend only quadratically
on the fields in Eq.~(\ref{eq:Udefinition}). 
However, we further observe that
in the $\lbrace \phi_{\lambda_0},\phi_{\lambda_1}\rbrace$
basis introduced in Eq.~(\ref{eq:Udefinition}),
the equations of motion in Eq.~(\ref{eq:newerdoublefieldequationsofmotion})
now take the form
\begin{align}
\ddot{\phi}_{\lambda_0} + 3H\dot{\phi}_{\lambda_0} + &\left(\lambda_0^2 - \dot{\theta}^2\right)\phi_{\lambda_0} \nn \\
&~=~ -2\dot{\theta}\dot{\phi}_{\lambda_1} -\left(\ddot{\theta} + 3H\dot{\theta}\right)\phi_{\lambda_1} \nn \\
\ddot{\phi}_{\lambda_1} + 3H\dot{\phi}_{\lambda_1} + &\left(\lambda_1^2 - \dot{\theta}^2\right)\phi_{\lambda_1} \nn \\
&~=~ +2\dot{\theta}\dot{\phi}_{\lambda_0} +\left(\ddot{\theta} + 3H\dot{\theta}\right)\phi_{\lambda_0}~.
\label{eq:massbasiseqnsofmotion}
\end{align}
These equations of motion exhibit an invariance 
under the simultaneous correlated 
transformations \mbox{$(\phi_{\lambda_0}, \phi_{\lambda_1}) \to (\phi_{\lambda_0},-\phi_{\lambda_1})$}
and \mbox{$\theta \to -\theta$}.
Since this is also an invariance of our initial conditions 
in Eq.~(\ref{eq:initialconditions}), 
our toy model will 
exhibit this invariance 
at all points during its time-evolution.
Note that 
this invariance is also tantamount to 
the simultaneous correlated transformations
\mbox{$(\phi_{0}, \phi_{1}) \to (\phi_{0},-\phi_{1})$}
and \mbox{$m_{01}^2 \to -m_{01}^2$}.
We can henceforth truncate our attention to 
\mbox{$m_{01}^2 \geq 0$} and \mbox{$0\leq \theta\leq \pi/2$} 
without loss of generality,
and we shall do so throughout the rest of this paper.
The region \mbox{$0\leq \theta <\pi/4$} then corresponds to \mbox{$\Delta m^2 >0$},
while the region 
\mbox{$\pi/4 < \theta \leq \pi/2$} corresponds to \mbox{$\Delta m^2 <0$}.

\begin{figure}[t]
\centering
\hspace*{-0.4cm}\includegraphics[width=0.45\textwidth,keepaspectratio]{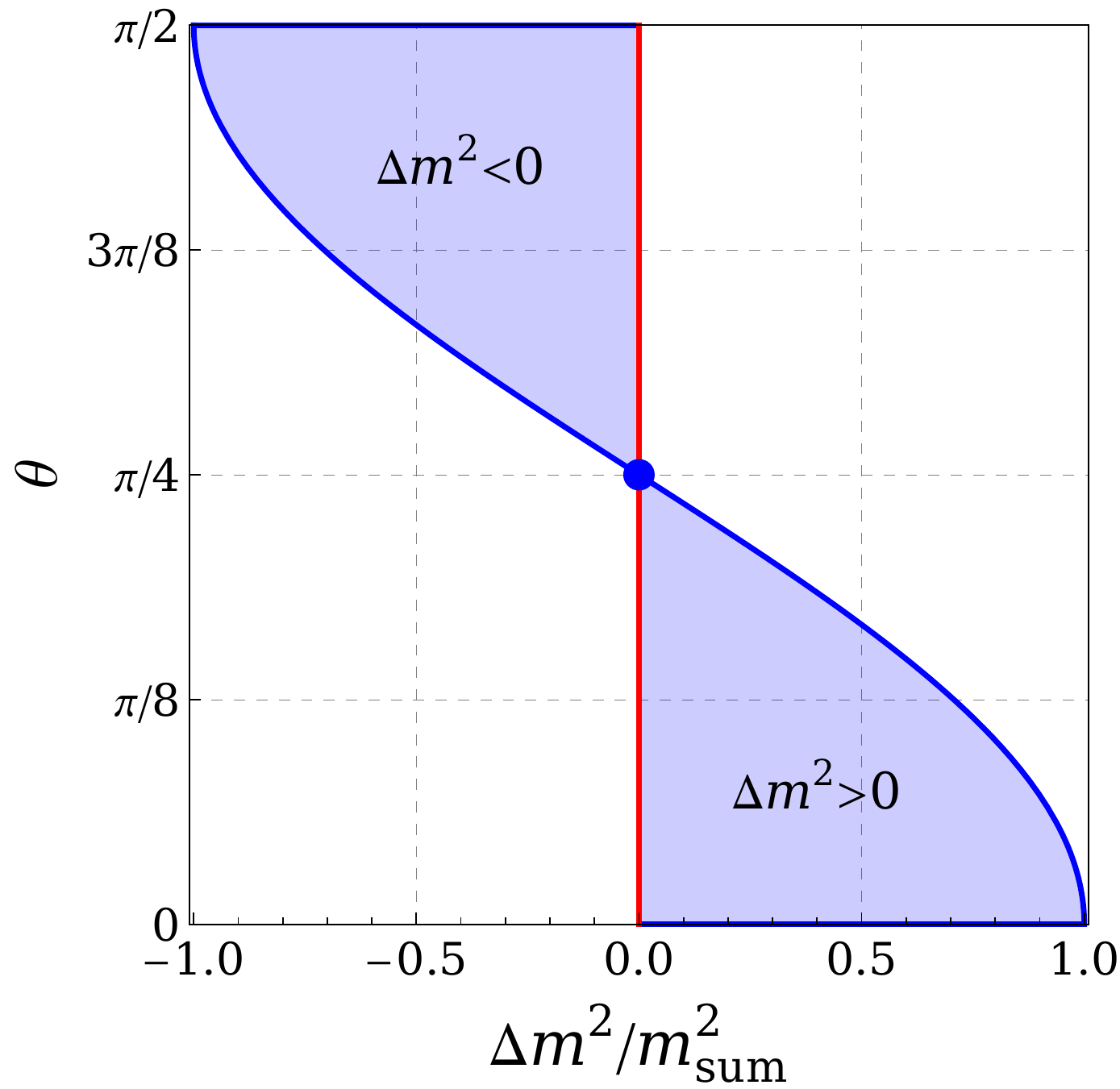}
\caption{Allowed ranges for the mixing angle $\theta$ (blue shaded region), 
plotted as a function of $\Delta m^2/m_{\rm sum}^2$.
Note that 
$\theta$ is restricted to $0$ or $\pi/2$ when \mbox{$\Delta m^2 = \pm m_{\rm sum}^2$} ---
\ie, when \mbox{$m_{00}^2=0$} or \mbox{$m_{11}^2=-1$}, respectively.
By contrast, when \mbox{$\Delta m^2 =0$}, the allowed range for $\theta$ depends on the value
of $m^2_{01}$:  for \mbox{$m^2_{01}\not=0$}, the only allowed $\theta$-value 
is \mbox{$\theta=\pi/4$} (blue dot), while for
\mbox{$m^2_{01}=0$} we find that $\tan(2\theta)$ is indeterminate and any value for $\theta$ is 
allowed (red line), depending on how relevant limits are taken.}
\label{fig:thetamaxfig}
\end{figure}

It then follows from Eq.~(\ref{eq:mixingrelation})
that for any values of $m_{\rm sum}^2$ and $\Delta m^2$, the allowed 
mixing angles $\theta$ are those for which
\beq
    |\tan (2\theta)| ~\leq~ 
    |\tan (2\theta_{\rm max})| ~\equiv~
    \sqrt{  \left( {m_{\rm sum}^2 \over \Delta m^2 }\right)^2 -1}~
\label{thetamaxdeff}
\eeq
where $\theta_{\max}$ is defined as the value of $\theta$ which maximizes $|\tan(2\theta)|$. 
Thus \mbox{$\theta\leq \theta_{\rm max}$} for \mbox{$\theta_{\rm max}\leq \pi/4$},
while \mbox{$\theta\geq \theta_{\rm max}$} for \mbox{$\theta_{\rm max}\geq \pi/4$}.
These angles are illustrated in Fig.~\ref{fig:thetamaxfig}.
Note that for \mbox{$\Delta m^2 = \pm m_{\rm sum}^2$}
(corresponding to 
\mbox{$m_{00}^2=0$} or \mbox{$m_{11}^2=-1$}, respectively),
the corresponding values of $\theta$ are restricted to $0$ or $\pi/2$.
Indeed, any other values would lead to a
squared-mass matrix $\calM^2$ with at least one negative eigenvalue, corresponding
to a tachyonic mode. 
Finally, for \mbox{$\Delta m^2=0$}, we have \mbox{$\theta=\pi/4$} for all \mbox{$m_{01}^2>0$},
while the angle $\theta$ is in principle undetermined for  \mbox{$m_{01}^2=0$}. 

We conclude this section with some further definitions that shall also prove useful
in what follows.
Quite often, we shall be evaluating a quantity $X$ which is a function of the time $\tau$.
For example, such quantities might include 
the mass parameters $m_{ij}(\tau)$, 
the mixing angle  $\theta(\tau)$,
the field values $\phi_{\lambda}(\tau)$,  
or the total energy density $\rho(\tau)$.
For any such quantity $X(\tau)$, we shall define
\beq
\overline{X} ~\equiv~ \lim_{\tau\rightarrow \infty} X(\tau)~.
\label{overbardefns}
\eeq
In other words, 
$\overline{X}$ shall denote the asymptotic {\it late-time}\/
value of $X(\tau)$,
where in practical terms the notion of a ``late'' time
can be taken as referring to a time 
at which both fields $\phi_{\lambda}$ have reached 
the asymptotic underdamped regime, with corresponding energy densities $\rho_{\lambda}$
exhibiting the virialized damped scaling behavior \mbox{$\rho_{\lambda}\sim \tau^{-\kappa}$} shown in Fig.~\ref{fig:rhosinglefield}.~ 
Finally, we shall also find that a crucial measure for the degree of mixing that might be present in a given
system is not the absolute value of the
mixing parameter $m^2_{01}$ (or $\theta$),
but rather the value of this parameter {\it as a fraction of the total degree of mixing that would have been allowed 
for that system}, given the constraints discussed above.
The same is often true for the splitting parameter $\Delta m^2$.
Towards this end, we define the {\it mixing saturation}\/
\beq
   \xi ~\equiv~ \frac{m_{01}^2}{\left[m_{01}^2\right]_{\text{max}}} ~=~ 
         \frac{\abs{\tan(2\theta)}}{\abs{\tan(2\theta_{\rm max})}}~=~
         \frac{m_{01}^2}{\sqrt{ m_{00}^2 (1 + m_{11}^2 ) }}       
\label{eq:xidefinition}
\eeq
as well as the analogous {\it splitting saturation}\/
\beq
   \eta ~\equiv~ \frac{ \Delta m^2 }{ [\Delta m^2]_{\rm max} } ~=~ 
         \frac{ \Delta m^2 } { m_{\rm sum}^2} ~=~
         \frac{ 1 - m_{00}^2 + m_{11}^2} 
              { 1 + m_{00}^2 + m_{11}^2}~.
\label{eq:eteadefinition}
\eeq
Note that in terms of these variables\footnote{
      We caution the reader that unlike the other variables we have thus far introduced,
      neither $\xi$ nor $\eta$ has a unique time-dependence.
      In other words, knowledge of the value of either $\xi$ or $\eta$ at a given time
      does not fix the value of $\xi$ or $\eta$ at other times without
      knowledge of the values of the more primordial $m_{ij}^2$ variables 
      from which $\xi$ and $\eta$ are derived.
      Thus, when we write a relation such as that in Eq.~(\ref{tanrelation}),
      we are illustrating the functional dependence of $\theta$ on $\xi$ and $\eta$
      and asserting that $\theta$ does not depend on the more primordial 
      variables except in these combinations.   However,  the time-dependence of
      $\theta$ cannot be determined from such an expression and must 
      be calculated directly in terms of the primordial variables themselves.}
we have
\beq
       \tan (2\theta) ~=~ \frac{\xi}{\eta}\,  \sqrt{1-\eta^2}~.
\label{tanrelation}
\eeq
We emphasize, however, that a maximally saturated mixing configuration with \mbox{$\xi=1$} does \emph{not}\/ necessarily 
imply that the mixing itself is maximal or even large on an absolute scale.
For this reason, 
we shall usually specify the degree of mixing in a given configuration
by quoting both $\xi$ and $\theta_{\rm max}$.
Indeed, for many quantities, it will be $\xi$ rather than $\theta$ or $\theta_{\rm max}$ which 
characterizes the behavior of interest and which allows us to compare across systems with
different values of $\theta$ or $\theta_{\rm max}$.

Note that for \mbox{$\Delta m^2=0$}, ambiguities can arise when defining $\xi$.
In particular, the middle expression in Eq.~(\ref{eq:xidefinition}) 
is formally indeterminate in such situations.
In this paper, for \mbox{$\Delta m^2=0$}, we shall therefore define $\xi$ 
through the final expression in Eq.~(\ref{eq:xidefinition}).
Thus $\xi$ can vary even though $\theta$ may be fixed at $\pi/4$.
Moreover, again following 
the final expression in Eq.~(\ref{eq:xidefinition}),
in this paper we shall define
\beq
    X\big|_{\substack{
     \Delta m^2=0 \\
      ~~~~\xi =0 }}  
 ~\equiv~
   \lim_{m^2_{01} \to 0} \left(
          X\big|_{\Delta m^2=0} \right)~ 
\label{formaldef}
\eeq
for any quantity $X$.
Thus the \mbox{$\xi\to 0$} limit will always be a smooth 
one with \mbox{$\theta=\pi/4$}, even for \mbox{$\Delta m^2=0$}.


\subsection{Temporal properties of the model:  Basic~features}


Given the definition of our toy model,
it is now relatively straightforward to study the corresponding dynamics.
Indeed, this can be done numerically if not analytically,
and certain features are entirely as expected and relatively easy to understand.
However, many other features are surprising and will play a significant role in 
what follows.
In the remainder of this section, therefore,
we shall discuss general features of the time-evolution of this model.
In particular, we shall focus on the time-dependence of the mass eigenvalues,
mixing angles, and mass eigenstates --- quantities upon
which our future results will rest.

\begin{figure}[!t]
\centering
\hspace*{-0.4cm}\includegraphics[width=0.45\textwidth,keepaspectratio]{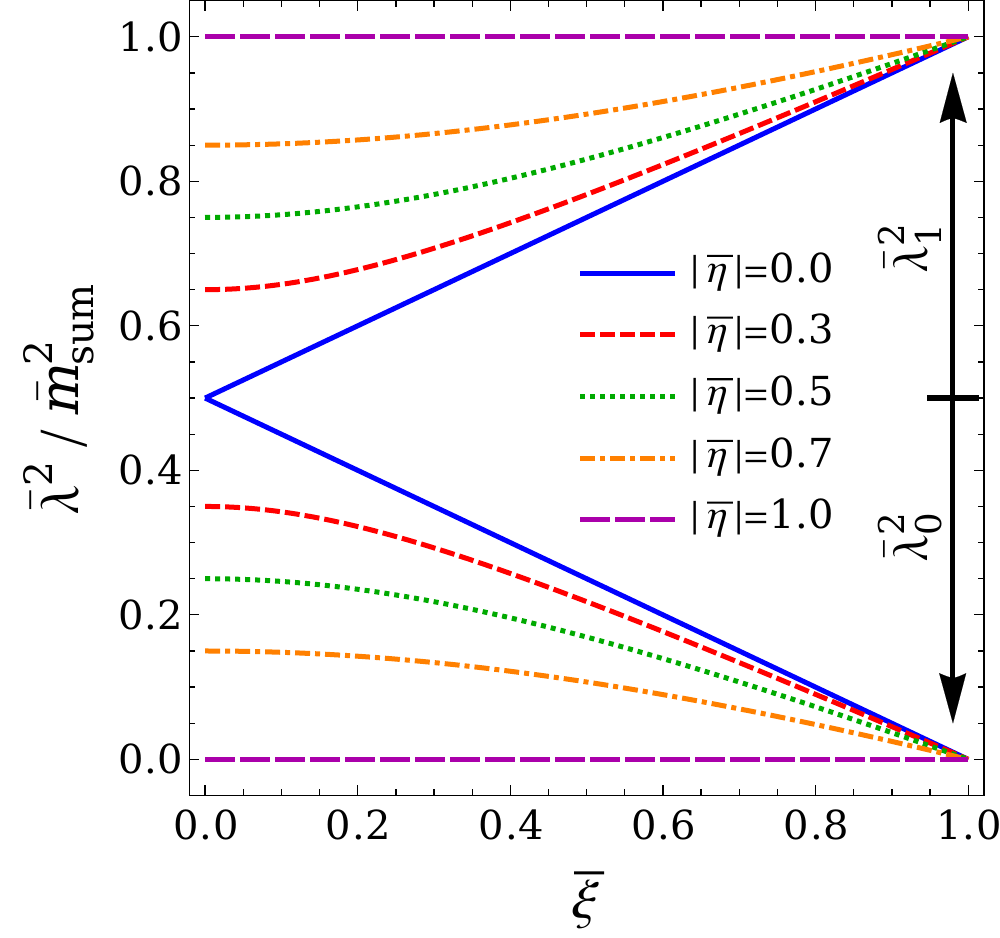}
\caption{The asymptotic (late-time) eigenvalues $\overline\lambda_i$ plotted as functions of
the late-time splitting and mixing parameters $\overline\eta$ and $\overline\xi$.
For all values of $\overline\eta$, we see that   
\mbox{$\overline\lambda^2_0\to 0$} and \mbox{$\overline\lambda_1^2 \to \overline m^2_{\rm sum}$} as \mbox{$\overline\xi\to 1$}.}
\label{fig:lambdares2}
\end{figure}

\begin{figure*}[!t]
\centering
\includegraphics[width=0.329\textwidth,keepaspectratio]{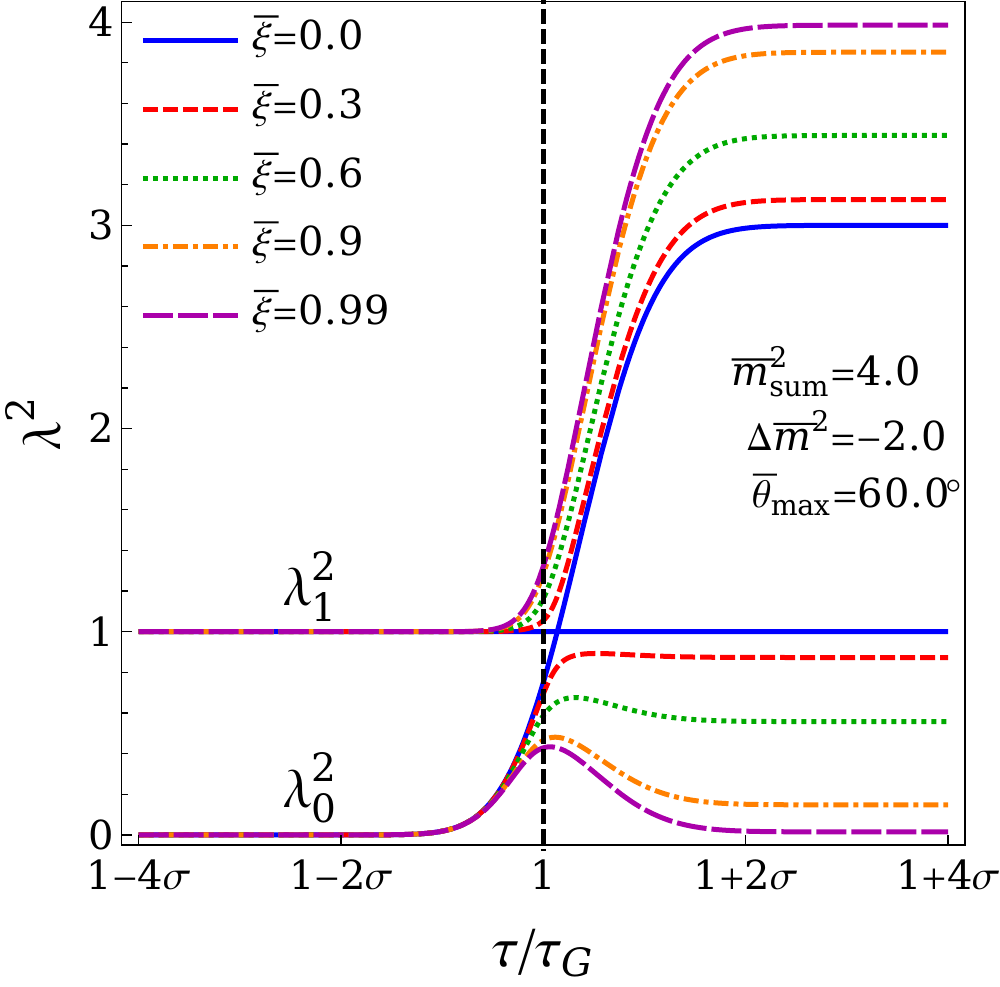}
\includegraphics[width=0.329\textwidth,keepaspectratio]{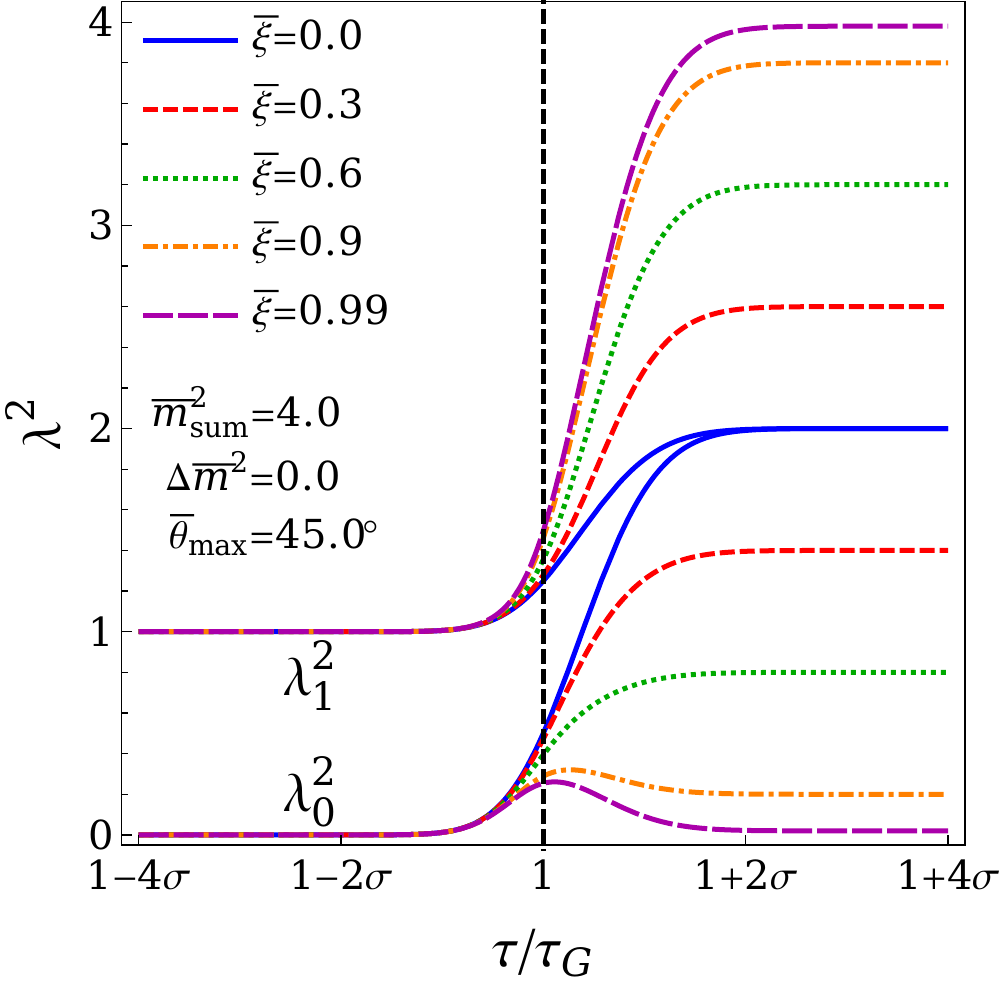}
\includegraphics[width=0.329\textwidth,keepaspectratio]{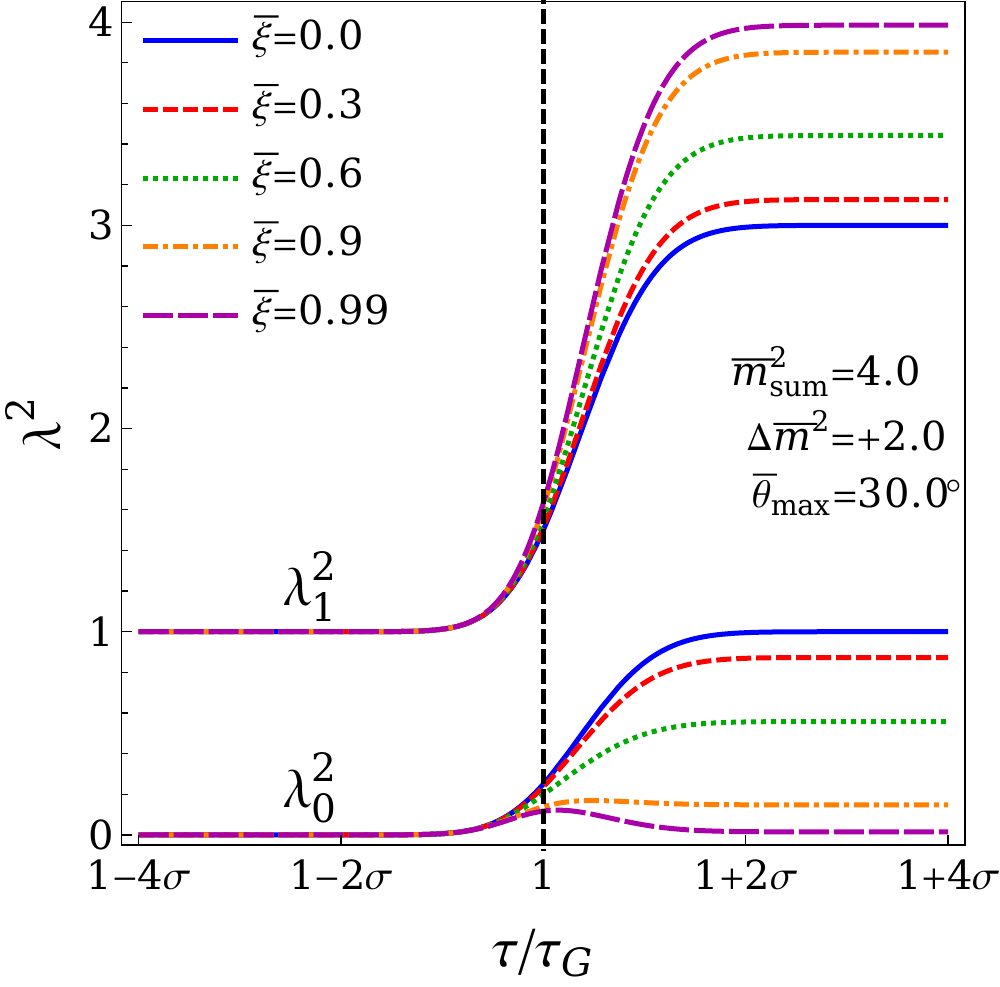}
\caption{The time-evolution of the two mass eigenvalues $\lambda_{0,1}^2$ as the phase
transition unfolds, plotted for different mixing saturations.
Typical behaviors are shown for 
\mbox{$\Delta \overline{m}^2 <0$} (left panel),
\mbox{$\Delta \overline{m}^2 =0$} (middle panel),
and 
\mbox{$\Delta \overline{m}^2>0$} (right panel). 
In all cases, the eigenvalues begin at $\lbrace 0,1\rbrace$, as expected.  For \mbox{$\Delta \overline{m}^2>0$},
the eigenvalues slowly transition to their late-time values without any tendency towards level-crossing.
For \mbox{$\Delta \overline{m}^2<0$}, by contrast, the eigenvalues meet and rebound off each other in the case of zero mixing (\mbox{$\overline \xi=0$})
but such a meeting is thwarted by {\it level repulsion}\/ for all non-zero mixing.
In general, the strength of the level repulsion increases with degree of mixing $\overline\xi$.
Despite these different features at intermediate times, we observe that the eigenvalues ultimately
arrive at the same late-time values regardless of whether $\Delta \mbar^2$ is positive or negative, in accordance
with Eq.~(\protect\ref{lambdabarvalues}).
Finally, for \mbox{$\Delta \overline{m}^2 =0$},
we see that there is no tendency towards level-repulsion,
but for \mbox{$\xibar=0$} the two eigenvalues approach each other asympotically.}
\label{fig:lambdares1}
\end{figure*}

We begin by studying the mass eigenvalues in our model.
At any moment in time, these masses are given by
\beqn
   \lambda_{0,1}^2 &=& \half \, m_{\rm sum}^2\, 
             \left[   1 \mp   \sqrt{   \eta^2 + (1-\eta^2) \xi^2 } \right]~\nonumber\\
         &=& \half \, m_{\rm sum}^2\, (1\mp \eta\, \sec 2\theta)~. 
\label{lambdabarvalues}
\eeqn
The late-time values of these masses are shown in 
Fig.~\ref{fig:lambdares2},
while the evolution of these masses from early to late times
is shown in 
Fig.~\ref{fig:lambdares1}.
Numerous features are immediately apparent. 
First, as predicted from Eq.~(\ref{lambdabarvalues}), 
we see that the eigenvalues at early and late times (\mbox{$\tau\ll \tau_G$} and \mbox{$\tau\gg \tau_G$},
respectively) 
are independent of the sign of $\Delta m^2$ (or $\eta$).
[In this connection we recall from Fig.~\ref{fig:thetamaxfig}
that \mbox{$\sec(2\theta) <0$} when \mbox{$\eta<0$}.]
Thus, 
the early- and late-time values of $\lambda_{0,1}^2$ are identical in the 
left and right panels of Fig.~\ref{fig:lambdares1}.
However, we see that the time-evolution of 
the eigenvalues 
 {\it between}\/ these two endpoints is highly sensitive to the sign of $\Delta \mbar^2$.
For \mbox{$\Delta \mbar^2>0$}, the eigenvalues evolve from initial to final values without 
any tendency towards level-crossing.
For \mbox{$\Delta \mbar^2<0$}, by contrast, the eigenvalues 
initially head towards each other as if to experience a level-crossing.
However, whether this level-crossing actually occurs depends on the value of the mixing.
For \mbox{$\xibar>0$}, 
the non-zero mixing between mass eigenstates
induces a {\it level repulsion}\/ which ultimately prevents a direct level-crossing.
As a result,
the eigenvalues veer away from each other, and ultimately assume the same late-time
values that they had for \mbox{$\Delta \mbar^2>0$}.

The case with \mbox{$\Delta \mbar^2<0$} and \mbox{$\xibar=0$} is more subtle, and deserves special discussion.
As \mbox{$\xibar\to 0$}, we see from the left panel of Fig.~\ref{fig:lambdares1}
that our eigenvalues $\lambda_0$ and $\lambda_1$ actually {\it meet}\/ near $\tau_G$ before bouncing off
each other.
[Note, in this connection, that $\lambda_0$ is always defined as the {\it lighter}\/ eigenvalue, consistent
with Eq.~(\ref{lambdabarvalues}).]
In this sense, no actual level-{\it crossing}\/ occurs, even for \mbox{$\xibar=0$};  instead,
each eigenvalue develops a ``kink'' --- \ie, a discontinuous slope --- at their meeting point near \mbox{$\tau=\tau_G$}. 

Finally, for \mbox{$\Delta \overline{m}^2 =0$},
we see from Fig.~\ref{fig:lambdares1} that there is no tendency towards level-crossing.
However, for \mbox{$\xibar=0$}, the two eigenvalues approach each other asymptotically.

\begin{figure}[t!]
\centering
\hspace*{-0.4cm}\includegraphics[width=0.45\textwidth,keepaspectratio]{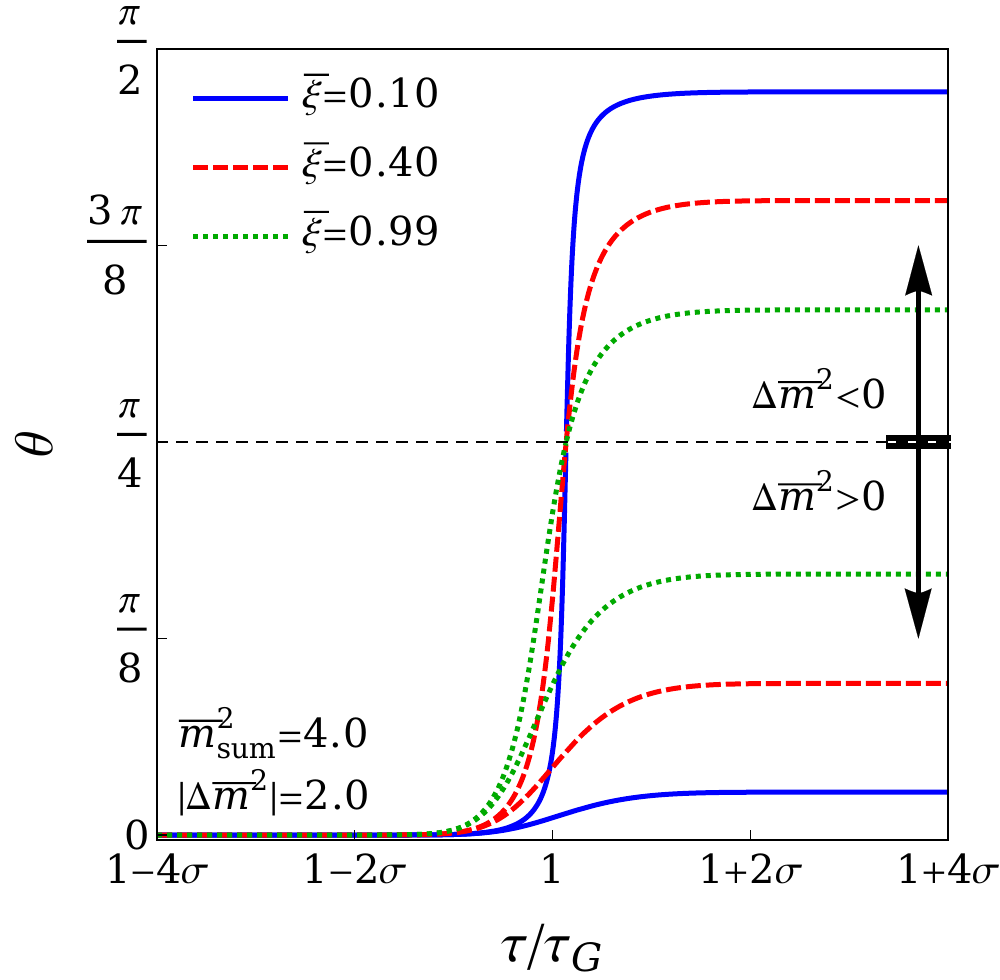}
\caption{The mixing angle $\theta$, plotted as a function of time 
for $\Delta \overline{m}^2=\pm 2$
and different values of $\overline{\xi}$.
For early times \mbox{$\tau\ll \tau_G$}, we generally have 
\mbox{$\Delta m^2 = m_{\rm sum}^2=1$} and \mbox{$\theta=0$}.
The subsequent time dependence then depends critically on the sign of the late-time value $\Delta \overline{m}^2$.
If \mbox{$\Delta \overline{m}^2 <0$}, then $\Delta m^2$ eventually switches from positive to negative 
values;  this occurs at  \mbox{$\tau=\tau^\ast$}, where $\tau^\ast$ is
defined in Eq.~(\ref{tauast}).
At this point, the mixing angle $\theta$ 
passes from the lower ``lobe'' of Fig.~\ref{fig:thetamaxfig} to the upper lobe
through the only allowed transition point between the two lobes at $\Delta m^2=0$ and $\theta=\pi/4$.   
For \mbox{$\Delta \overline{m}^2>0$}, by contrast, the mixing angle 
stays entirely within the lower lobe of Fig.~\ref{fig:thetamaxfig} and thus remains below $\pi/4$.
Finally, for \mbox{$\Delta m^2=0$}, the angle $\theta$ moves to the central point between the two lobes,
consistent with our expectation that \mbox{$\thetabar=\pi/4$}.
However, as discussed in the text, it takes infinite time to do this as $\xibar\to 0$.}
\label{fig:thetavstau}
\end{figure}

We can also study the mixing angle $\theta$ as a function of time.
The result is shown in Fig.~\ref{fig:thetavstau}.
As we see, for \mbox{$\overline{\xi}>0$}
the mixing 
angle begins at zero, as expected, and rises to 
a non-zero late-time value $\thetabar$;  this corresponds to $\Delta m^2$ transitioning
from its initial value \mbox{$\Delta m^2 = m_{\rm sum}^2=1$} 
to its final value $\Delta \overline m^2$.
Once again, however,
the behavior of the angle $\theta$ is highly sensitive to the sign of
the late-time value $\Delta \overline{m}^2$.
For \mbox{$\Delta \overline{m}^2 >0$}, the angle $\theta$ remains below $\pi/4$ (\ie, within the lower right ``lobe''
of Fig.~\ref{fig:thetamaxfig}).
For \mbox{$\Delta \overline{m}^2 <0$}, by contrast,
the angle $\theta$ eventually grows above $\pi/4$, moving from the lower right lobe of 
Fig.~\ref{fig:thetamaxfig} to the upper left.    
The case with \mbox{$\Delta \mbar^2=0$} will be discussed below.

It is natural to define the time $\tau_\theta$ at which 
$\theta$ crosses its midpoint value $\overline\theta/2$;
in general, this is the time for which 
\beq
  h(\tau_\theta) ~=~ {1\over \sqrt{1+ \lambdabar_1^2 -\lambdabar_0^2 }}~.
\label{tautheta}
\eeq
Likewise, for \mbox{$\Delta \mbar^2<0$}, 
we find that the 
transition between lobes at \mbox{$\theta=\pi/4$} 
occurs at the time $\tau^\ast$ for which
\beq
       h(\tau^\ast) ~=~ {1\over \sqrt{1-\Delta \overline{m}^2}}~.
\label{tauast}
\eeq
In general, this ``lobe-crossing'' time $\tau^\ast$ is distinct from $\tau_G$ and $\tau_\theta$.   
However, $\tau^\ast$ and $\tau_\theta$ coincide when $\thetabar=\pi/2$.  
As we see from Fig.~\ref{fig:thetavstau}, this happens only
when \mbox{$\Delta \mbar^2 <0$} and \mbox{$\xibar=0$}.

Once again, this case with \mbox{$\Delta \mbar^2 <0$} and \mbox{$\xibar=0$} deserves special mention.
At early times we have
\mbox{$\Delta m^2 = m_{\rm sum}^2 =1$}, and our mixing angle $\theta$ begins at zero.
This angle then remains at zero all the way until the eigenvalue meeting time shown in the left panel
of Fig.~\ref{fig:lambdares1};  note that this eigenvalue meeting time is indeed 
nothing but \mbox{$\tau^\ast=\tau_\theta$}.
At this time, $\theta$ changes
instantaneously to \mbox{$\theta=\pi/2$}, consistent with the fact that after $\tau^\ast$
we have \mbox{$\Delta m^2<0$}.   Thus we see that $\theta$ behaves as a step function in this limit,
with an effectively instantaneous field rotation.
Note that this ``instantaneous'' behavior in the $\xi\to0$ limit applies only for \mbox{$\Delta \mbar^2<0$}.

The situation with 
$\mbox{$\Delta \mbar^2=0$}$ also deserves special attention.
As usual, we have \mbox{$\theta=0$} at early times.
For all \mbox{$\xibar\not=0$}, this mixing angle $\theta$ then transitions to \mbox{$\thetabar=\pi/4$} at late times.
Moreover, as discussed above,  
it follows formally from Eq.~(\ref{formaldef})  
that \mbox{$\thetabar=\pi/4$} even for $\xibar=0$.
However, 
there is an important subtlety in the latter case.
As \mbox{$\xibar\to 0$} with \mbox{$\Delta \mbar^2=0$}, it turns out that 
\mbox{$h(\tau_\theta)=1$}.
This in turn implies that \mbox{$\tau_\theta\to \infty$}!   
Thus, even though $\thetabar=\pi/4$ in this case, we never actually reach the point at which
the corresponding mixing occurs.  
Rather, our two original states $\phi_0$ and $\phi_1$ remain unmixed at all finite times.
Indeed, in some sense, the transition to $\thetabar=\pi/4$ occurs only when the two corresponding
eigenvalues $\lambda_0$ and $\lambda_1$ actually meet.
Note, in this connection, that even though the field rotation does not occur at any finite time,
the mass eigenvalues $\lambda_{0,1}^2$ nevertheless experience their normal evolution 
in the neighborhood of $\tau_G$, a result which follows directly from the time-dependence 
of the original mass matrix and which is independent of mixing.
In other words, in this special case, the two timescales $\tau_G$ and $\tau_\theta$ are
maximally separated.

Our discussion thus far has focused on the eigenvalue/mixing structure 
of the mass matrix as a function of time.
However, the corresponding field values also generally behave as one might expect,
at least as far as their grossest features are concerned.
For times \mbox{$\tau\ll \tau_\theta$}, our two fields $\phi_{\lambda_{0,1}}$ are effectively uncoupled: 
$\phi_{\lambda_0}$ evolves independently of $\phi_{\lambda_1}$,
and indeed $\phi_{\lambda_1}$ remains vanishing.
For times \mbox{$\tau \sim \tau_\theta$}, by contrast,
the phase transition generates a non-zero, time-dependent mixing 
which couples the two fields together and thereby causes
$\phi_{\lambda_1}$ to accrue a non-zero value as well.
Finally, for times \mbox{$\tau \gg \tau_\theta$}, 
our mixing angle $\overline{\theta}$ is non-zero but essentially constant.
This means that a similar decoupling exists during this period as well,
except that our decoupled fields are now those
linear combinations which are rotated relative our original fields by $\overline\theta$. 
In general, each of these linear combinations will have a non-zero field value, and
will evolve independently according to whether it is individually overdamped or underdamped.

The limiting case with \mbox{$\Delta \mbar^2 <0$} and \mbox{$\xibar=0$} is again worthy of special note.
In this case, $\phi_{\lambda_0}$ retains its original amplitude until the eigenvalue meeting time \mbox{$\tau^\ast=\tau_\theta$}.
The entire amplitude of $\phi_{\lambda_0}$ then transfers instantaneously to $\phi_{\lambda_1}$, where it remains for all later times.
(This follows from the observation that in this limit we 
have an instantaneous rotation of our fields from $\theta=0$ to $\theta=\pi/2$.)
This rapid amplitude transfer is consistent with the instantaneous infinite value of $\dot\theta$ 
at \mbox{$\tau^\ast=\tau_\theta$}. 
As a result of this amplitude transfer, quantities which depend on the field amplitudes (such as the associated
energy densities) transfer instantaneously from $\phi_{\lambda_0}$ to $\phi_{\lambda_1}$
at this time.  In other words, they ``ride'' smoothly across the eigenvalue collision
in the left panel of Fig.~\ref{fig:lambdares1}, transitioning from $\phi_{\lambda_0}$  
to $\phi_{\lambda_1}$.
Of course, this is completely as expected, reflecting nothing more than the fact that the energy density
began in our original field $\phi_0$ and remains there when $\xibar=0$.
All that has suddenly changed at the eigenvalue collision time 
is the identification between this field $\phi_0$ and our mass-eigenstate fields $\phi_{\lambda_{0,1}}$.

One could, in principle, continue along this line of inquiry.
For example,  one could proceed to study the 
phase-space trajectories of our mass-eigenstate fields $\phi_{\lambda_0}$ and 
$\phi_{\lambda_1}$, and map out how these trajectories depend on the different defining parameters of 
our model.
However, an exhaustive study of this toy model is not our purpose in this paper.
Rather, as stated in the Introduction, 
in this paper our interest in this toy model 
stems from the fact that --- despite its simplicity --- it gives rise to certain features which 
have the potential to transcend
our typical expectations when interpreted in a cosmological setting.
It is therefore to these new features that we now turn. 

In keeping with the above observations, 
in the rest of this paper we shall 
mostly concentrate on those regions of parameter space in which our fields are 
already underdamped --- or are in the process of {\it becoming}\/ underdamped  ---
during the mass-generating phase transition.  This is important, since the time-dependent effects of our
mass-generating phase transition tend to be washed out if our fields remain overdamped while 
it occurs.
Thus, in this way, 
we shall be focusing on precisely 
those parameter-space regions of interest:  those which are likely to exhibit
a non-trivial interplay between the width of the mass-generating phase transition,
the transition between overdamped and underdamped regimes,
and the mixing between all of the fields experiencing these effects.

\FloatBarrier

\section{Total late-time energy density}
\label{sec:LateTimeEnergyDensity}


The total energy density $\rho$ is the quantity of central interest in this paper. The role it plays as a cosmological
observable gives it direct importance in any analysis of our toy model --- particularly at late times, after our phase
transition has been completed.
In general, we are interested not only in the total late-time energy density $\overline{\rho}$, 
but also in its distribution between the two individual components $\overline{\rho}_{\lambda}$.  
Moreover, in each case, we are particularly interested 
in knowing the extent to which our mass-generating phase transition
might leave imprints on these late-time energy densities.
These are the issues that we shall study in this section.

Calculating the energy density of the system proceeds
directly from the equations of motion for our two fields and their derivatives.
In the original $\lbrace \phi_0,\phi_1\rbrace$ basis,
the total energy density is given by
\beq
    \rho ~=~ \frac{1}{2}\left( \sum_i \dot{\phi}^2_i + \sum_{ij}\phi_i\mathcal{M}^2_{ij}\phi_j\right)~.
\label{eq:rhoKKbasis}
\eeq
By contrast, in the mass-eigenstate basis 
$\lbrace \phi_{\lambda_0},\phi_{\lambda_1}\rbrace$
introduced in Eq.~(\ref{eq:Udefinition}),
the total energy density is given by
\beq
    \rho ~=~ \frac{1}{2}\sum_{\lambda}\left[\dot{\phi}_{\lambda}^2 + 
             \left(\lambda^2+\dot{\theta}^2\right)\phi_{\lambda}^2\right] 
            ~+~ \dot{\theta}\,\sum_{\lambda\lambda'}\dot{\phi}_{\lambda}\epsilon_{\lambda\lambda'}\phi_{\lambda'}~,~~~ 
\label{eq:rhomassbasis}
\eeq
where $\epsilon_{\lambda\lambda'}$ is the Levi-Civita symbol with \mbox{$\epsilon_{\lambda_0\lambda_1}\equiv +1$}.

The first thing we observe is that in neither case can we express our total energy density 
as a sum of individual contributions.
Indeed, we cannot write $\rho$ in the form \mbox{$\rho =\sum_i \rho_i$} or \mbox{$\rho=\sum_\lambda \rho_\lambda$}.
The reason for this, as most evident from Eq.~(\ref{eq:rhomassbasis}),
is ultimately that our fields experience a {\it time-dependent}\/ mixing, with \mbox{$\dot \theta\not=0$}.
Without mixing --- or even with only a constant mixing --- such a decomposition could have been done
and individual contributions identified.
{\it Thus, in this paper we shall never refer to the individual contributions 
to the total energy density except at late times when the mixing has essentially 
stabilized and \mbox{$\dot\theta=0$}}\/.
Indeed, at late times, we can then identify
\beq
\overline{\rho}_{\lambda} ~=~ \frac{1}{2}\left(\dot{\phi}^2_{\lambda} + \overline{\lambda}^2\phi_{\lambda}^2\right) \ ,
\label{eq:rhocomponentdefinition}
\eeq
where the overbar indicates late-time values in accordance with
Eq.~(\ref{overbardefns}).

As already discussed at the end of Sect.~\ref{sec:AToyModel},
the effects of the mass-generating phase transition 
are washed out if our fields are still overdamped
when it occurs.
Likewise, given the formulation of our toy model,
there is no meaningful way for
the mass-generating phase transition
to occur very much later than the critical overdamped/underdamped transition
for either of our two fields:
the lighter field is massless prior to the mass-generating phase
transition and as such remains overdamped during this entire period,
while the heavier field, either formally overdamped  
or underdamped, has no amplitude of its own until the mass-generating phase
transition.
For this reason,
in this section we shall 
focus on those regions of parameter space in which 
our fields become underdamped
near the time at which the phase transition begins.

\begin{figure}[t!]
\centering
\hspace*{-0.4cm}\includegraphics[width=0.45\textwidth,keepaspectratio]{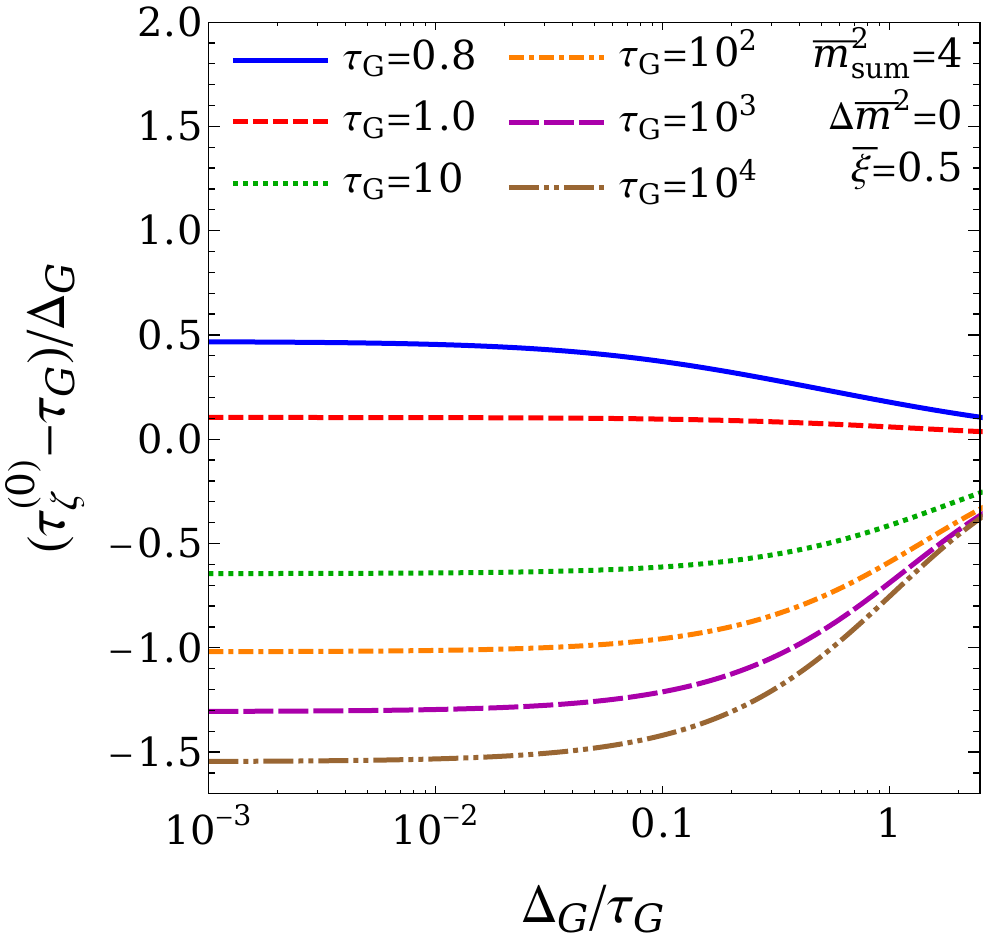}
\caption{The critical damping time $\tau_\zeta^{(0)}$ for the lighter field $\phi_{\lambda_0}$ 
in a matter-dominated universe, expressed
as a number of widths $\Delta_G$ prior to the central phase-transition time $\tau_G$ and
plotted as a function of $\Delta_G/\tau_G$ for different values of $\tau_G$.
For \mbox{$\tau_G\sim 1$}, the critical damping transition
essentially occurs  during the central portion of the  
mass-generating phase transition for all $\Delta_G$;
for large $\tau_G$, by contrast,
the critical damping transition occurs
just prior to the mass-generating phase transition.
These two cases thus survey our main regions of interest in this paper.}
\label{fig:tauzeta}
\end{figure}

The most complete way of surveying physics within this regime is to
vary both $\tau_G$ and $\Delta_G$, with $\tau_G$ limited to the range 
\mbox{$\tau_G\gsim \tau^{(i)}_\zeta$}, where each $\tau^{(i)}_\zeta$ is implicitly defined 
(as in Fig.~\ref{fig:3Hmfig}) as the time at which 
\mbox{$3H=2\lambda_i$}.
This method, which by construction surveys all possibilities, 
thus involves variations in two independent parameters. 
However, another way of surveying many aspects of the physics within this regime 
is to fix $\tau_G$,
and instead to vary the width of the phase transition $\Delta_G$.  
By considering fiducial values of $\tau_G$ which are carefully chosen with respect
to the $\mbar^2_{ij}$ values, 
we can reach situations in which $\tau_\zeta^{(i)}$ are near
$\tau_G$ or just prior to it.
These values of $\tau_\zeta^{(0)}$ are illustrated in Fig.~\ref{fig:tauzeta}, and 
we see that this method also allows us to reach the desired values of $\tau_\zeta^{(0)}$.
It is therefore this latter approach that we shall follow in the next two sections.

Given our choices for $\tau_G$ and $\Delta_G$, we can then calculate 
the late-time total energy density $\overline\rho$ for different values of the $\overline m^2_{ij}$ parameters ---
\ie, for different values of $\overline m_{\rm sum}^2$, $\Delta \overline m^2$, and $(\overline\xi, \overline\theta_{\rm max})$. 
Our goal is to understand the effects on the total late-time energy densities $\rhobar$ that emerge when 
a non-zero mixing, parametrized by $\xibar$, and a non-zero width  $\Delta_G$ 
for our mass-generating phase transition are present simultaneously. 

\begin{figure*}
\centering
\hspace*{-0.4cm}
      \includegraphics[width=0.329\textwidth,keepaspectratio]{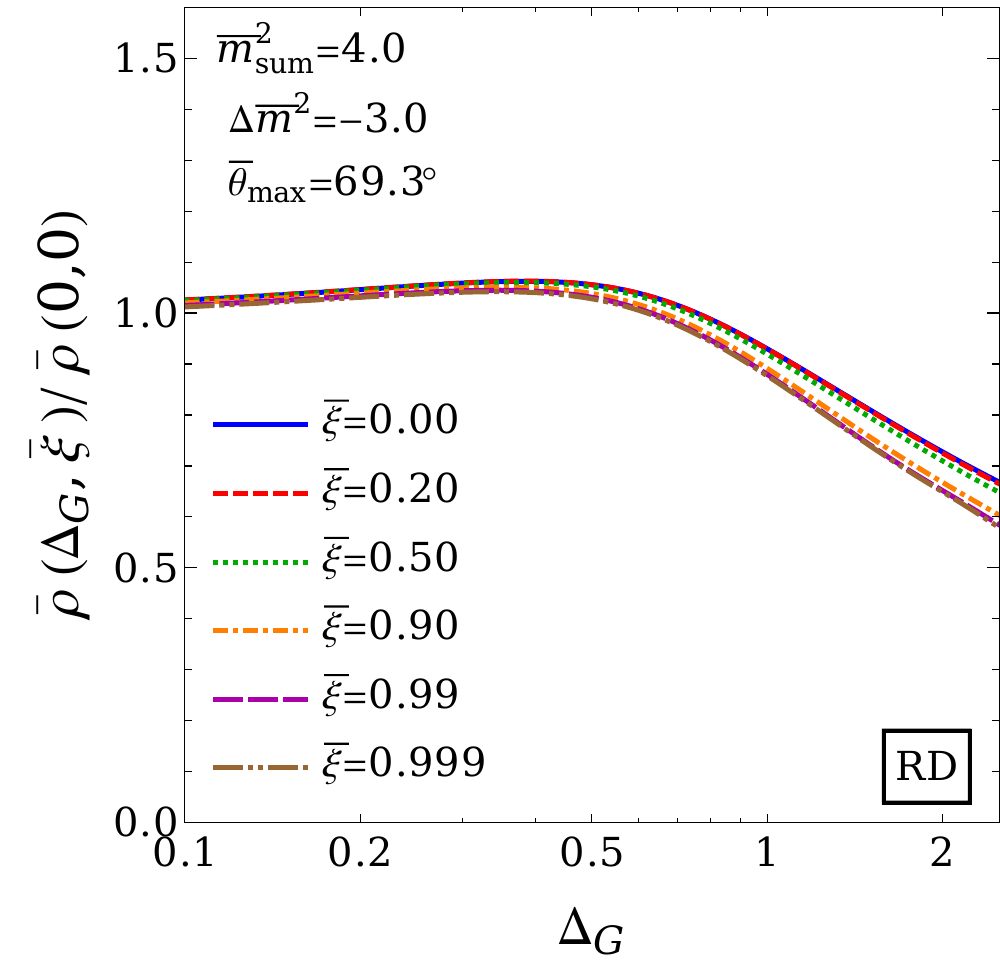}  
      \includegraphics[width=0.329\textwidth,keepaspectratio]{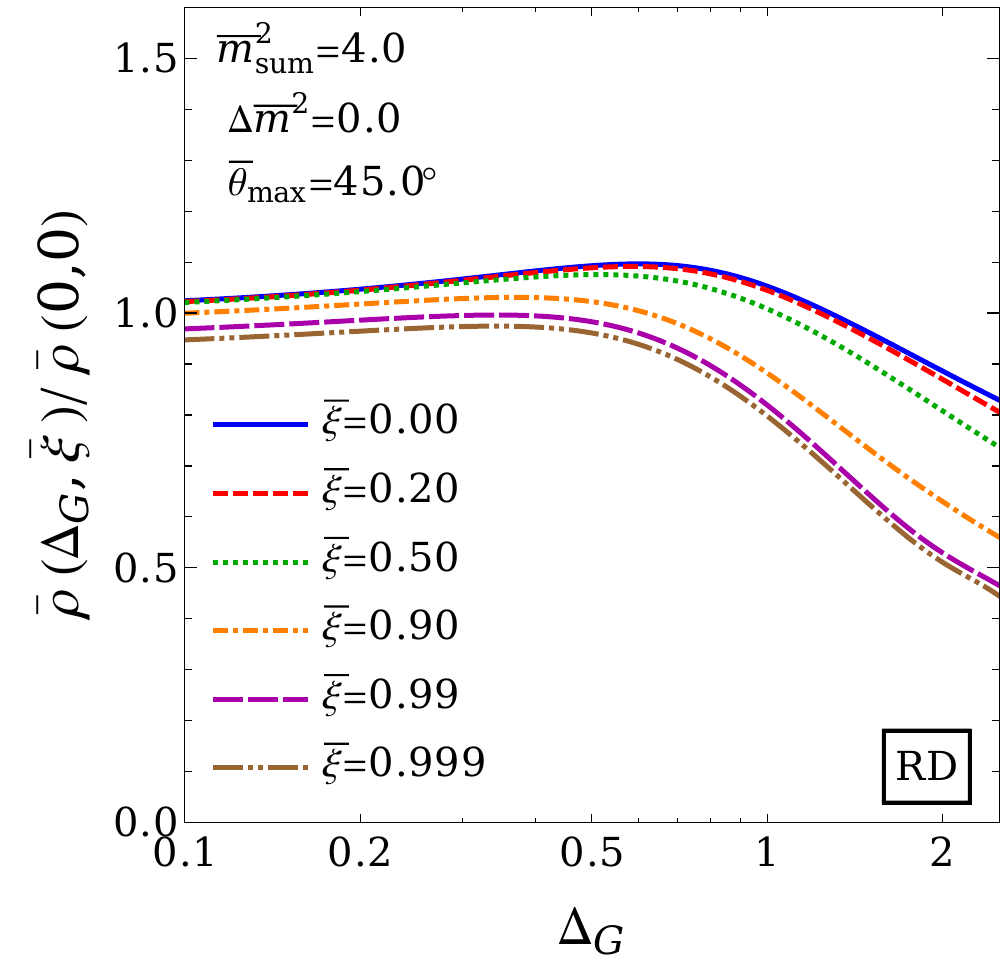} 
      \includegraphics[width=0.329\textwidth,keepaspectratio]{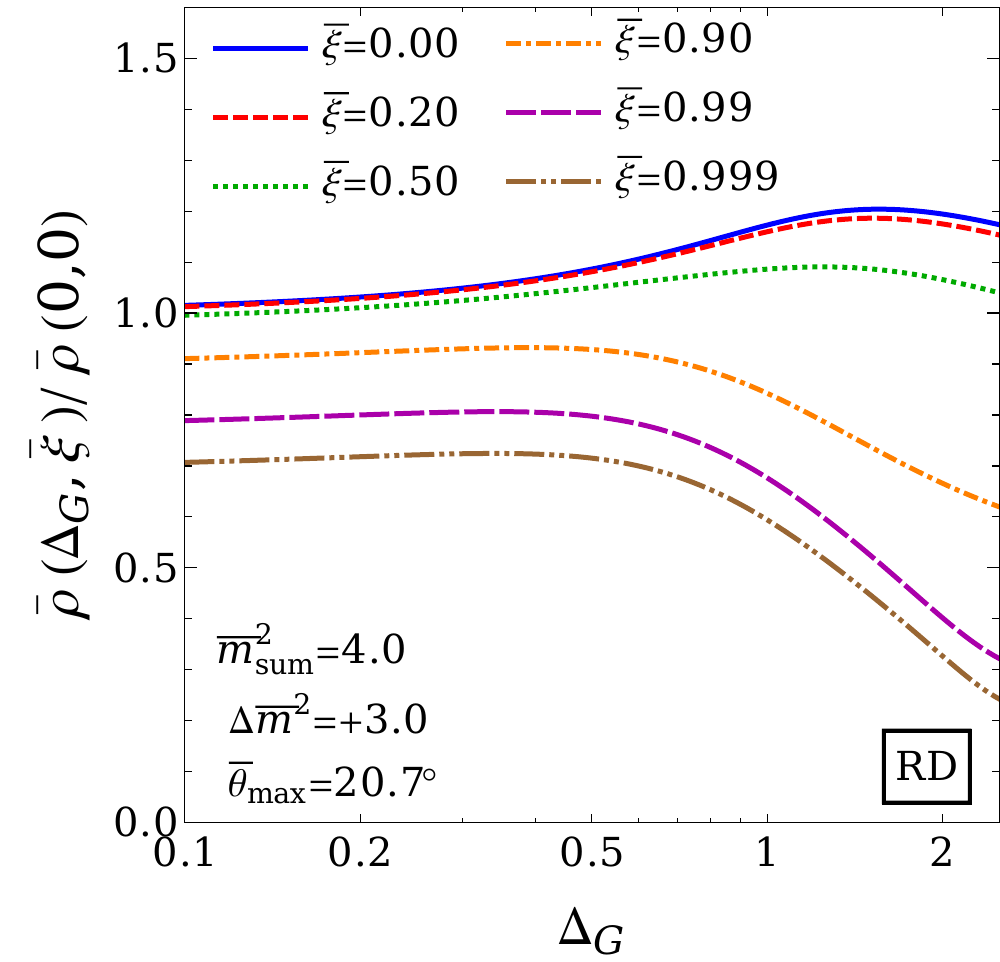}
\caption{Behavior of the total late-time energy density $\overline\rho(\Delta_G,\xibar)$ when mixing effects,
    parametrized by $\xibar$, are 
       combined with a non-zero width $\Delta_G$ for the mass-generating phase transition.
   For these plots we adopt a relatively {\it small}\/ fiducial time \mbox{$\tau_G= 1.0$} and 
   assume a radiation-dominated universe;
   we also hold \mbox{$\overline m_{\rm sum}^2 =4$} and take \mbox{$\Delta\overline m^2 = -3, 0, +3$} 
   for the left, center, and right panels, respectively.
   In each case we plot the late-time energy density $\overline\rho$ as a function of 
   the phase-transition width $\Delta_G$ for different values of the mixing saturation $\xibar$, 
   where in each panel $\overline\rho$ is normalized to its value for \mbox{$\Delta_G=0$} and \mbox{$\xibar=0$}.} 
\label{fig:figname1}
\bigskip
\bigskip
\hspace*{-0.4cm}
      \includegraphics[width=0.329\textwidth,keepaspectratio]{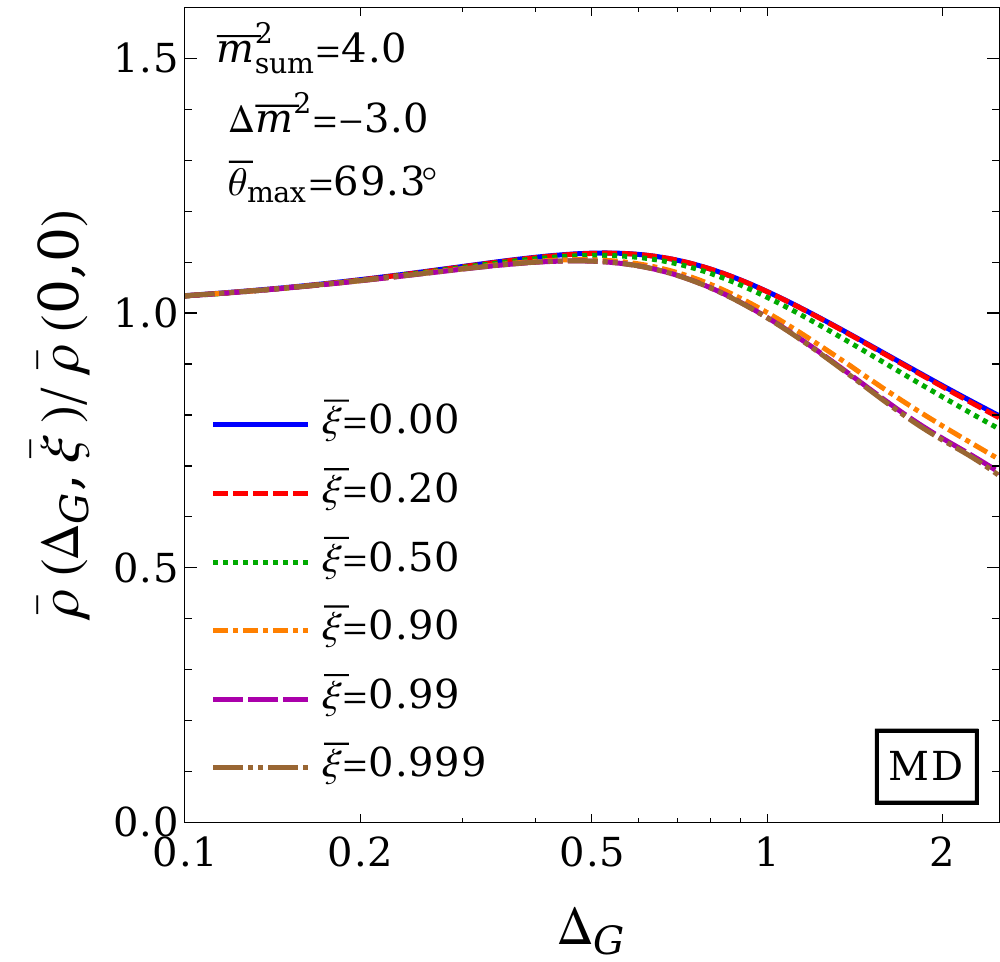}  
      \includegraphics[width=0.329\textwidth,keepaspectratio]{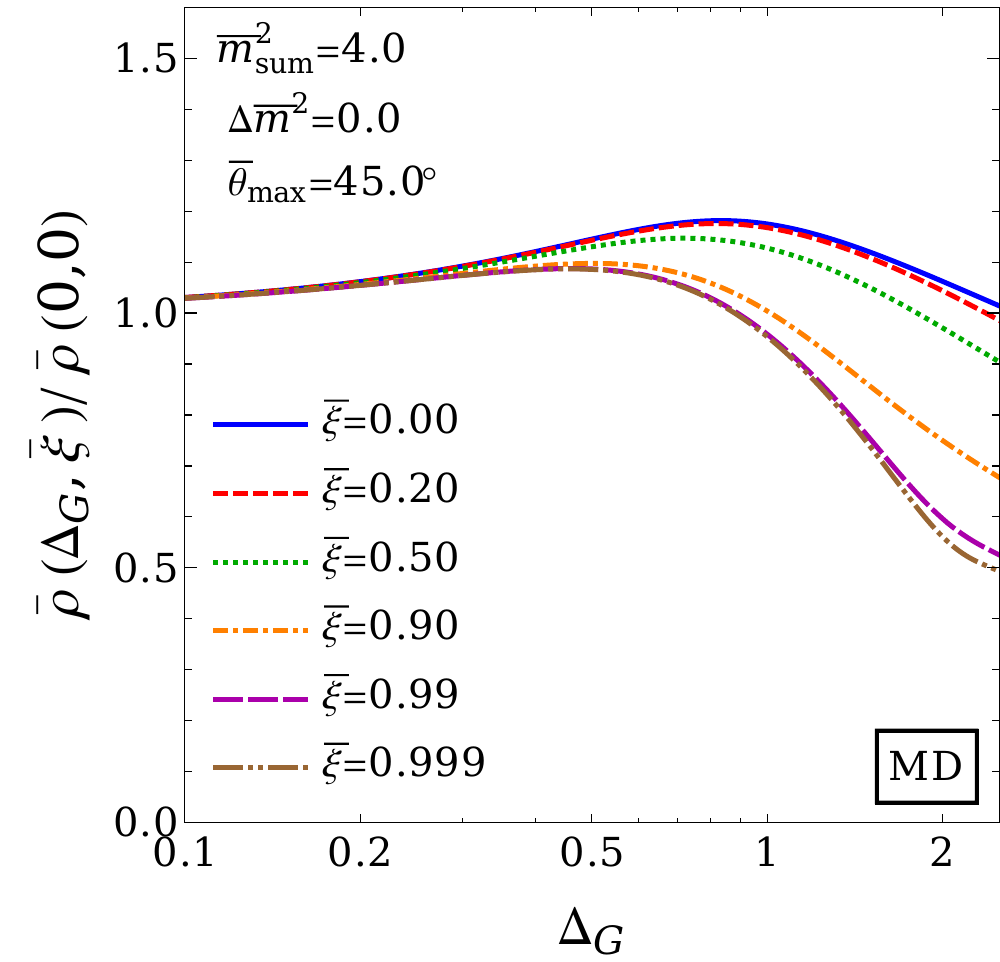} 
      \includegraphics[width=0.329\textwidth,keepaspectratio]{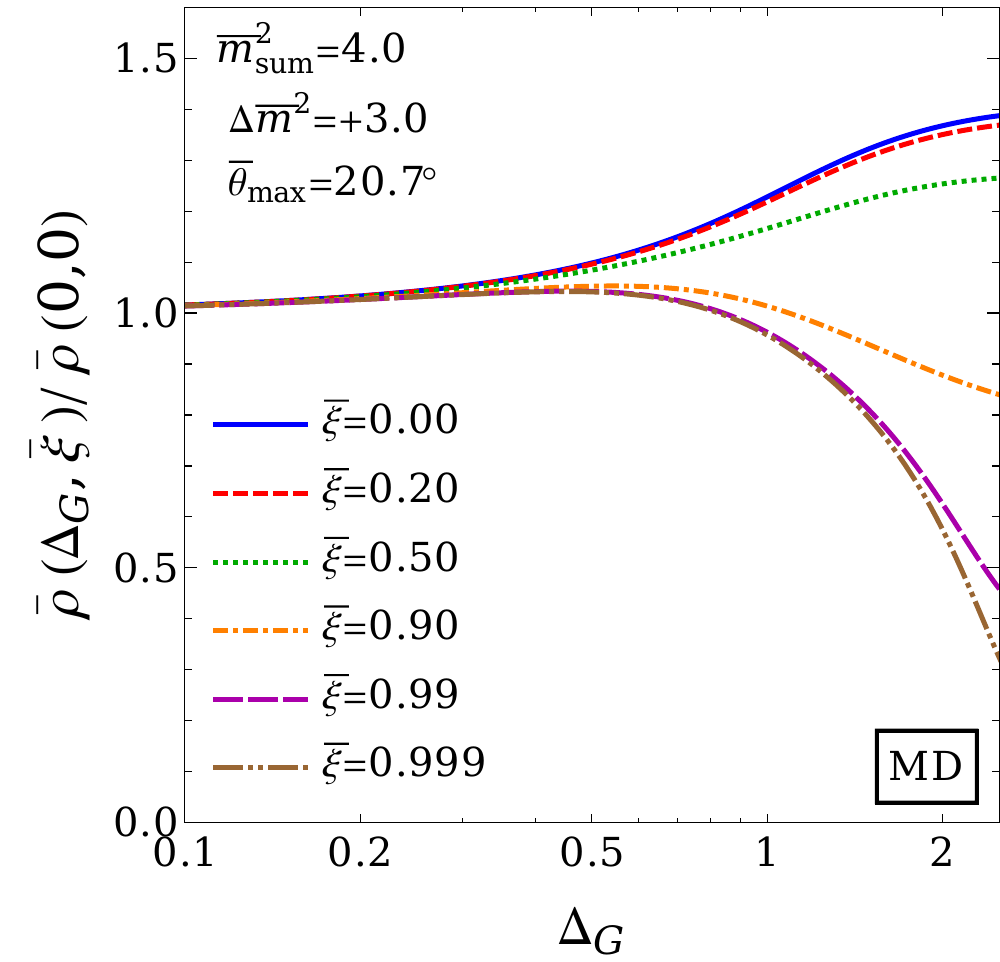}
\caption{Same as Fig.~\ref{fig:figname1}, but for a matter-dominated universe.}
\label{fig:figname2}
\bigskip
\bigskip
\hspace*{-0.4cm}
      \includegraphics[width=0.329\textwidth,keepaspectratio]{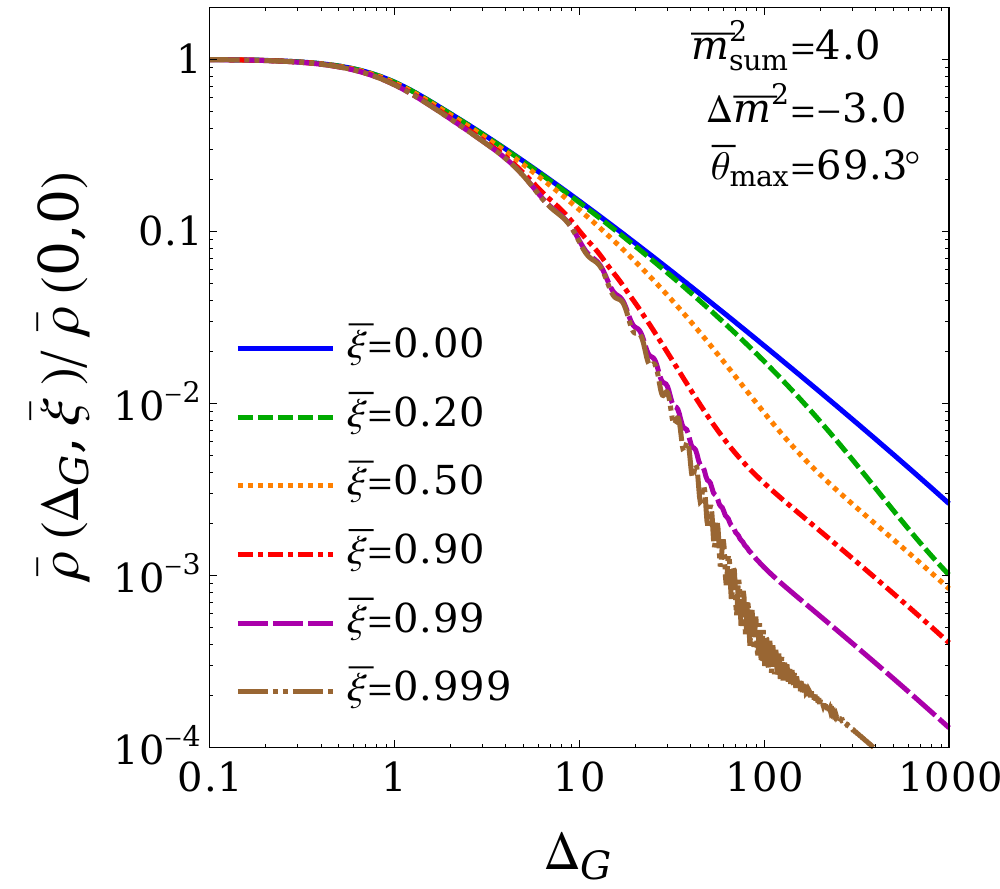}  
      \includegraphics[width=0.329\textwidth,keepaspectratio]{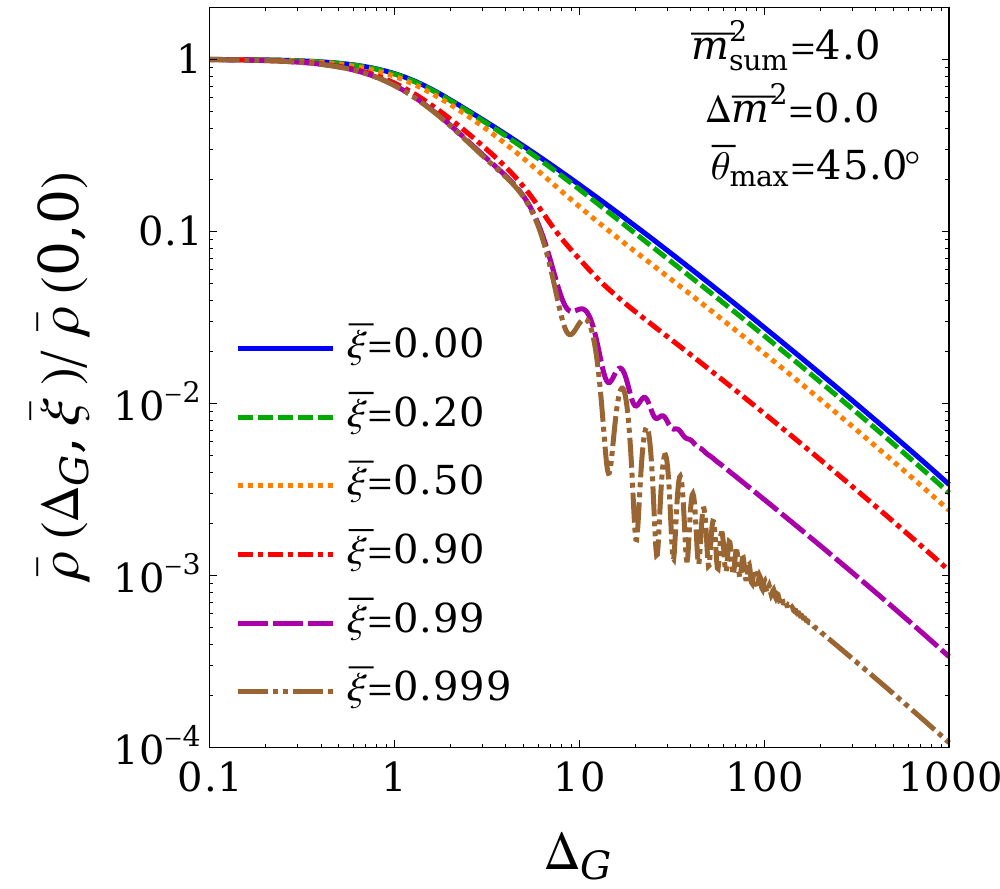} 
      \includegraphics[width=0.329\textwidth,keepaspectratio]{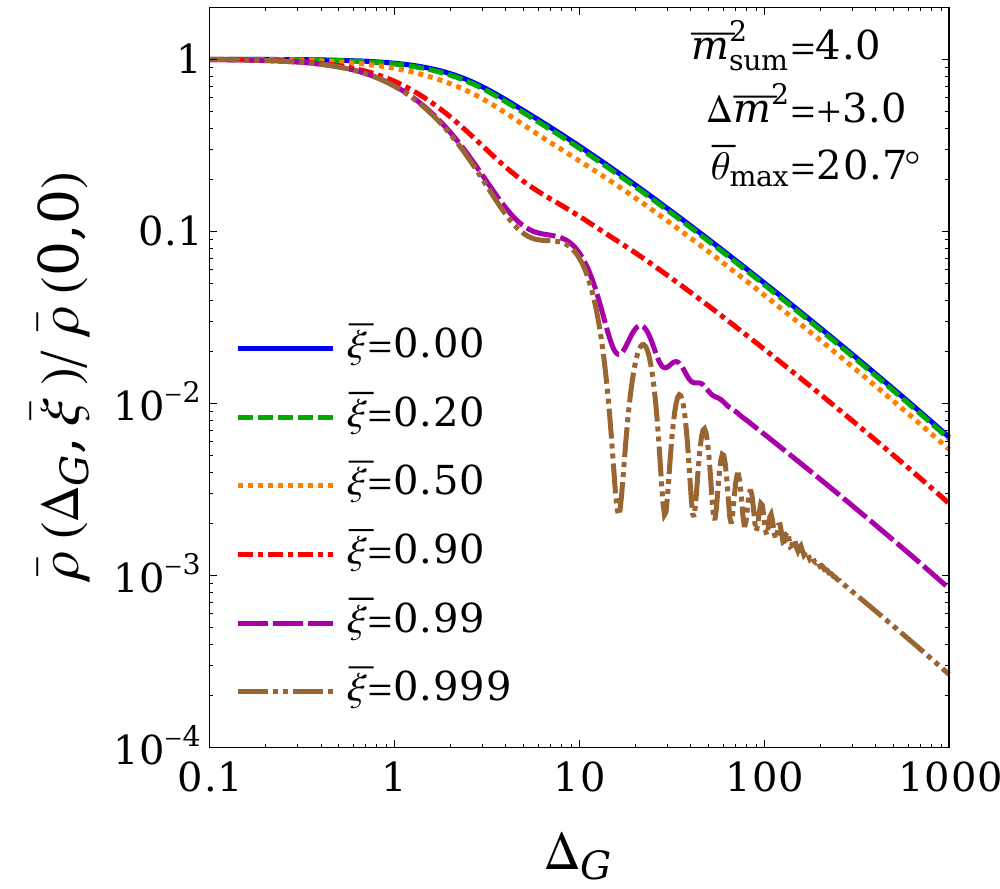}
\caption{Same as Figs.~\ref{fig:figname1} and \ref{fig:figname2}, except that we now adopt a
relatively {\it large}\/ fiducial time \mbox{$\tau_G=10^4$}.   For such large values of $\tau_G$, the results are 
nearly the same for both radiation- and matter-dominated universes.  Note that unlike Figs.~\ref{fig:figname1} and \ref{fig:figname2}, the vertical axes here are {\it logarithmic}\/.}
\label{fig:figname3}
\end{figure*}

Our results are shown in Figs.~\ref{fig:figname1} through \ref{fig:figname3}:
In Figs.~\ref{fig:figname1} and \ref{fig:figname2}
we assume \mbox{$\tau_G=1$}
(for radiation- and matter-dominated universes respectively), while
in Fig.~\ref{fig:figname3} we assume \mbox{$\tau_G\gg 1$} (for which
matter- and radiation-dominated universes yield the same results).
In each of these figures we survey values of the late-time mixing saturation $\xibar$ within
the range \mbox{$0\leq \xibar<1$};   
note that we refrain from considering the actual limiting case with \mbox{$\xibar=1$} 
when discussing late-time quantities such as the 
late-time energy density $\rhobar$  because the lighter field $\phi_{\lambda_0}$ remains massless at late times when \mbox{$\xibar=1$}
(as evident from Fig.~\ref{fig:lambdares2})
and therefore never technically enters the asymptotic underdamped region which characterizes our
definition of ``late'' times.
We also note that although the left, center, and right panels within each figure correspond to different
values of the maximum allowed mixing angle $\thetabar_{\rm max}$, 
we see that it is only in terms of the mixing {\it saturation}\/ $\xibar$ that we can make 
sensible comparisons across the different
panels.   Indeed, although a fixed change in $\xibar$ 
within a panel with large $\thetabar_{\rm max}$ corresponds to a much larger change in the 
absolute mixing angle $\thetabar$
than it does within a panel with small $\thetabar_{\rm max}$,  we see that it
is only the changes in $\xibar$ rather than $\thetabar$ which can be compared meaningfully across panels.
For this reason, even though the value of $\theta_{\rm max}$
is important in order to express our results in terms of absolute mixing angles, 
in the following we shall simply describe our mixings 
as small or large depending on the corresponding values of $\xibar$. 

These figures illustrate the effects of turning on a finite width $\Delta_G$ for the
mass-generating phase transition in conjunction with non-zero mixing between our two fields.
For the small-$\tau_G$ regime plotted in Figs.~\ref{fig:figname1} and \ref{fig:figname2},
we observe that {\it small}\/ mixing actually {\it enhances}\/ the late-time energy density $\rhobar$.
By contrast, we see that {\it large}\/ mixing actually {\it suppresses}\/ the late-time energy density
$\rhobar$.  Indeed, we see from 
Figs.~\ref{fig:figname1} and \ref{fig:figname2}
that these effects (in both directions) are more pronounced  
for matter-dominated universes than radiation-dominated universes
and for situations in which $\Delta \mbar^2$ is positive rather than zero or negative.

For the large-$\tau_G$ regime plotted in Fig.~\ref{fig:figname3}, by contrast,
we see that matter- and radiation-dominated universes give rise to identical results.
All late-time energy densities $\rhobar$, regardless of the mixing saturation $\xibar$, share a common value when 
\mbox{$\Delta_G=0$}.   However, unlike the 
small-$\tau_G$ case, the late-time energy densities experience no enhancement at all, 
even in the absence of mixing (\mbox{$\xibar=0$}).
Indeed, $\rhobar$ experiences only a suppression 
for non-zero $\Delta_G$ --- a
suppression which,
given the logarithmic $\rhobar$ axes within Fig.~\ref{fig:figname3},
 grows much more severe than it was for small $\tau_G$.
Moreover, we see that this suppression of the late-time energy density $\rhobar$ is stronger 
for {\it negative}\/ $\Delta \overline m^2$ (\ie, for negative $\etabar$) than for zero or positive --- 
a feature which is completely reversed relative to the small-$\tau_G$ case!
Finally, we observe the emergence of a non-monotonic ``oscillatory'' behavior 
for $\rhobar$ as a function of $\Delta_G$ when $\tau_G$ is large 
and when the mixing saturation $\xibar$ grows close to $1$.
This feature is most pronounced when 
$\Delta \mbar^2$ is positive, but exists for all $\Delta \mbar^2$.
These oscillations will be discussed in Sect.~\ref{sec:TheParametricResonance}, and indicate
a strong sensitivity of the suppression factor relative to even 
small variations in the phase-transition width $\Delta_G$.

All of these results illustrate the dramatic consequences that
ensue when we consider the timescales associated with both our mass-generating phase transition
and the mixing it generates.  For example, we see from the \mbox{$\Delta \mbar^2>0$} plot in Fig.~\ref{fig:figname2} that mixing
has no effect on the total late-time energy density $\rhobar$ when the mass-generating phase transition 
is rapid (\ie, when \mbox{$\Delta_G=0$}).  
 {\it Thus, it is only the existence of a non-zero phase-transition timescale which allows the mixing
to leave a non-trivial imprint at late times!}\/
Moreover, we see from these figures that the enhancements and suppressions experienced by
$\rhobar$ are typically quite large, stretching from 20\% or 30\% 
in the case of enhancements 
all the way to many orders of magnitude in the case of suppressions!
These effects can thus have dramatic implications for the relative size of the corresponding slice 
of the ``cosmic pie'' --- \ie, for the overall composition 
of the total late-time energy budget of the universe. 

\begin{figure}[t!]
\centering
\hspace*{-0.4cm}\includegraphics[width=0.45\textwidth,keepaspectratio]{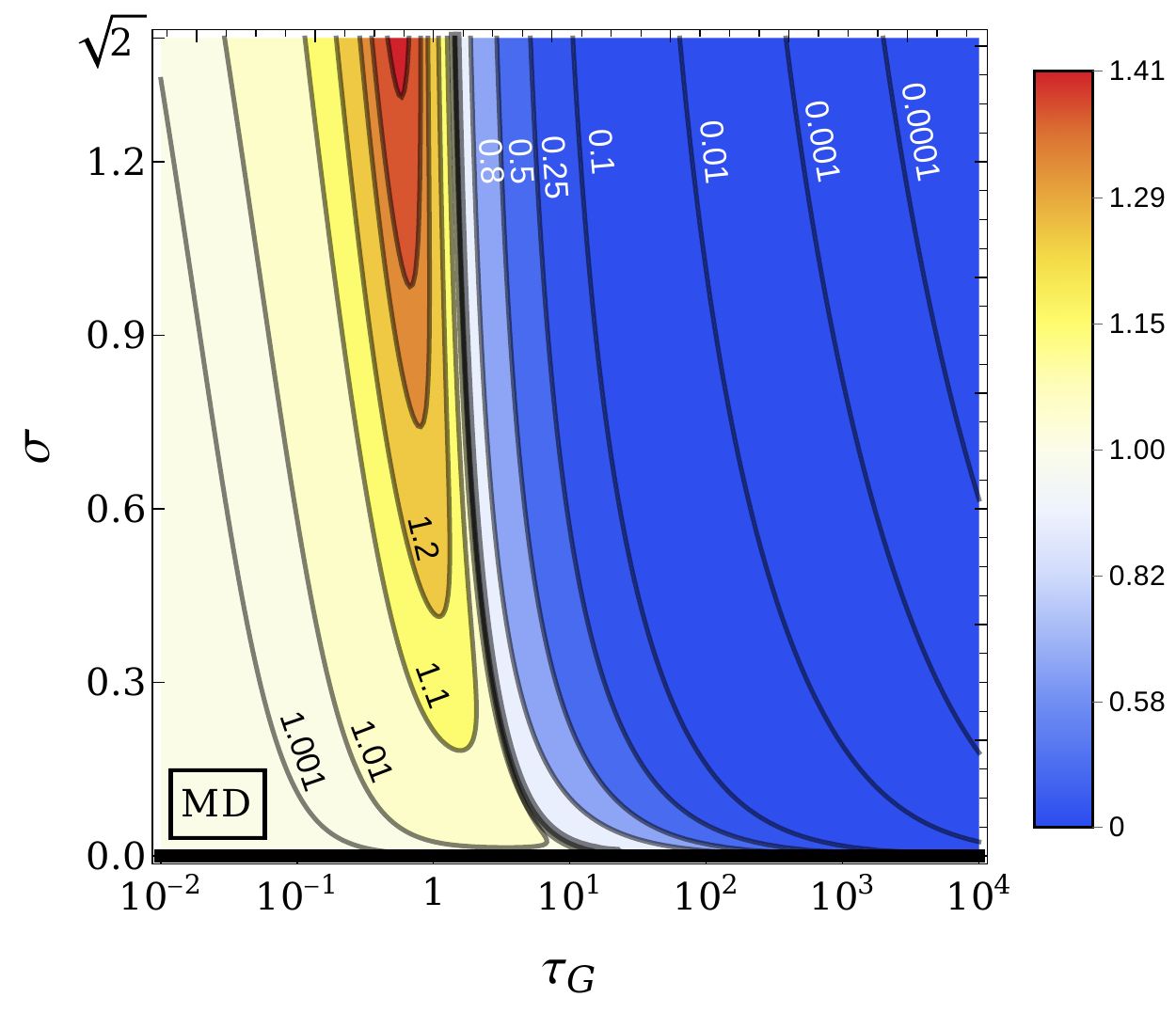}
\caption{Contours of the total late-time energy-density enhancement/suppression factor $\rhobar(\Delta_G)/\rhobar(0)$, plotted in the $(\tau_G,\sigma)$ plane, assuming a matter-dominated universe without mixing (\mbox{$\xibar=0$}) and \mbox{$\mbar_{00}^2=1$}.  The thick black line running vertically and along the \mbox{$\sigma=0$} axis is the contour for which we have neither enhancement nor suppression.  
The corresponding contours for a radiation-dominated universe are similar.   We observe that in the absence of mixing, a finite width for the mass-generating phase transition tends to modify the late-time energy density $\rhobar(\sigma)$ compared to what it would have been for an instantaneous phase transition;   this suppresses $\rhobar(\sigma)$ rather significantly for \mbox{$\tau_G\gsim 5$}, but enhances $\rhobar(\sigma)$ for \mbox{$\tau_G\lsim 1$}.   Remarkably, these results are generally insensitive to $\sigma$ as long as \mbox{$\sigma \gsim 0.1$},
and follow an approximate
scaling behavior \mbox{$\rhobar \sim \Delta_G^{-1}$} in this region.}
\label{fig:singlefield}
\end{figure}

It is also instructive 
to understand those limits of the above results 
in which one or the other of our two variables $\xibar$ and $\Delta_G$ is taken to zero. 
We begin by focusing on  the effects that emerge solely due to the presence of a non-zero width
$\Delta_G$ for the mass-generating phase transition (\ie, the effects that occur when \mbox{$\xibar=0$}).
Results are shown in Fig.~\ref{fig:singlefield},
where we plot contours of the total normalized late-time energy density $\rhobar(\sigma)/\rhobar(0)$
within the $(\tau_G,\sigma)$ plane, where \mbox{$\sigma\equiv \Delta_G/(\sqrt{\pi}\tau_G)$} 
as in Eq.~(\ref{eq:sigDelta}).
Generally, we see from 
Fig.~\ref{fig:singlefield} 
both suppression and enhancements in the late-time
energy density are possible,
depending on whether \mbox{$\tau_G\gsim 5$} or \mbox{$\tau_G\lsim 1$} respectively.   
These results therefore explain the behaviors of the \mbox{$\xibar=0$} curves in Figs.~\ref{fig:figname1}, \ref{fig:figname2},
and \ref{fig:figname3}.
Furthermore, we see from Fig.~\ref{fig:singlefield} that $\rhobar(\Delta_G)$ is largely insensitive to
$\sigma$, and instead  depends almost exclusively on the value of $\tau_G$.  
Indeed, for $\tau_G\gsim 1$, we see from  
Fig.~\ref{fig:singlefield} that $\rhobar(\Delta_G)$ scales approximately as $\tau_G^{-1}$.
This in turn implies that
$\rhobar(\Delta_G)\sim \Delta_G^{-1}$ in this region.
This observation will be discussed further in Sect.~\ref{sec:indiv}.

\begin{figure}[t!]
\centering
\hspace*{-0.4cm}\includegraphics[width=0.45\textwidth,keepaspectratio]{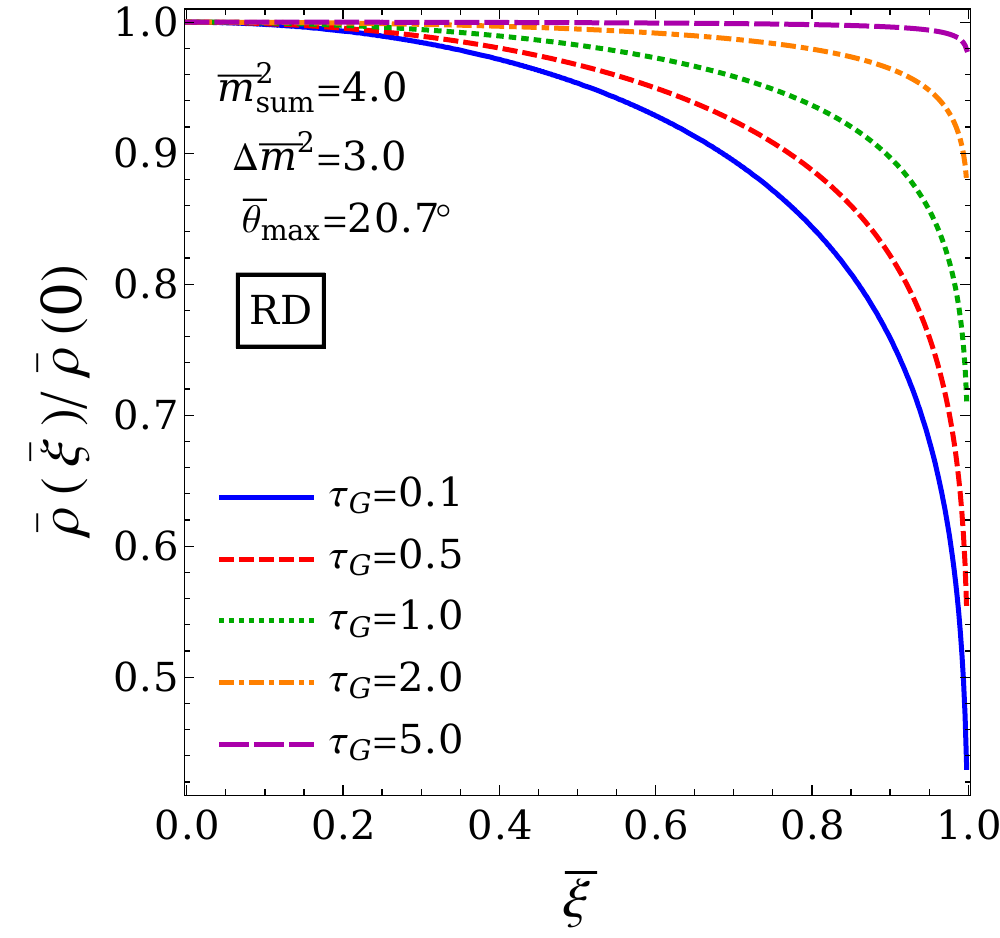}
\caption{Suppression of the late-time energy density $\rhobar$ due to mixing effects alone, with 
\mbox{$\Delta_G=0$}, in a radiation-dominated universe.
We assume an instantaneous phase transition at \mbox{$\tau=\tau_G$},
and plot the corresponding
late-time energy density $\rhobar(\xibar)$ as a function of the late-time mixing saturation $\xibar$, normalized
to its value at \mbox{$\xibar=0$}, for different choices of $\tau_G$.   We see that in all cases
the late-time energy density $\rhobar(\xibar)$ experiences a suppression which grows increasingly severe
as the mixing saturation is increased.
As discussed in the text, 
the magnitude of this effect increases as
\mbox{$\etabar\equiv \Delta \mbar^2/\mbar_{\rm sum}^2\to 1$} (\ie, as \mbox{$\mbar^2_{00}\to 0$}), and is 
entirely absent if the phase transition instead occurs during a matter-dominated epoch or more generally
if \mbox{$\tau_G \gg 1$}.}
\label{fig:DDG}
\end{figure}

Conversely, we can also
study the effects that arise due to mixing alone (\ie, with \mbox{$\Delta_G=0$}).
It turns out that when \mbox{$\Delta_G=0$}, the late-time energy density $\rhobar$ can actually 
be calculated {\it analytically}\/ for arbitrary mixing;
the general result is given in Eq.~(\ref{eq:asymptoticrho}).
In Fig.~\ref{fig:DDG}, we choose values
\mbox{$\Delta \overline{m}^2 = 3$} and \mbox{$\mbar_{\rm sum}^2=4$}, and plot the corresponding late-time 
energy density $\rhobar (\xibar)$
as a function of the late-time mixing saturation $\xibar$, normalized to its unmixed value \mbox{$\rhobar(\xibar=0)$}, 
for a variety of different $\tau_G$.
For \mbox{$\xibar=0$}, this corresponds to a situation in which our two original unmixed fields $\phi_{0,1}$ remain uncoupled,
with all of the energy density remaining in $\phi_0$;
by contrast, as $\xibar$ increases,
increasing amounts of energy density are shared between our two fields during the 
mass-generating phase transition.
Surprisingly, we see from Fig.~\ref{fig:DDG} that increasing $\xibar$ results in an {\it enhanced dissipation}\/
of the total energy density, so that the total late-time energy density $\rhobar(\xibar)$ is suppressed relative
to what it would have been in the absence of mixing.
Indeed, we see from Fig.~\ref{fig:DDG} that this suppression is strongest 
for relatively small $\tau_G$ and relatively large mixing saturations $\xibar$.
Moreover, as evident from the calculation in Appendix~\ref{notation}, 
this effect exists in all cases except for a matter-dominated epoch.
This effect was first observed in Ref.~\protect\cite{Dienes:1999gw} within the context of an infinite tower
of Kaluza-Klein axion modes, where it was exploited in order to permit a higher-dimensional loosening of the usual
four-dimensional overclosure bounds on the Peccei-Quinn scale $f_{\rm PQ}$.
However, we now see that this effect is completely general, and persists even when only two modes are involved.

 {\it We thus conclude that for phase transitions occurring during a radiation-dominated epoch,
mixing alone induces a significant suppression in the late-time energy density.
Indeed, this is true even if the phase transition is treated as instantaneous.
This effect grows in significance as \mbox{$\Delta \overline m^2 \to \overline m_{\rm sum}^2$} 
and  \mbox{$\overline \xi\to 1$}.  This effect is also largest when \mbox{$\tau_G\sim \tau_\zeta^{(i)}$}, and ultimately
vanishes for \mbox{$\tau_G\gg \tau_\zeta^{(i)}$}.}

Note that all of these observations are consistent with the plots shown in Figs.~\ref{fig:figname1},
\ref{fig:figname2}, and \ref{fig:figname3}.
For \mbox{$\Delta_G=0$}, we see that all of the plotted curves 
begin at 
\mbox{$\rhobar(\Delta_G,\xibar)/\rhobar(0,0)=1$}
 {\it except}\/ for those in Fig.~\ref{fig:figname1}, where
the mere fact of
having \mbox{$\xibar\not=0$} induces a suppression
of the late-time energy density,
even for \mbox{$\Delta_G=0$}.

\begin{figure}[]
\centering
\hspace*{-0.4cm}\includegraphics[width=0.45\textwidth,keepaspectratio]{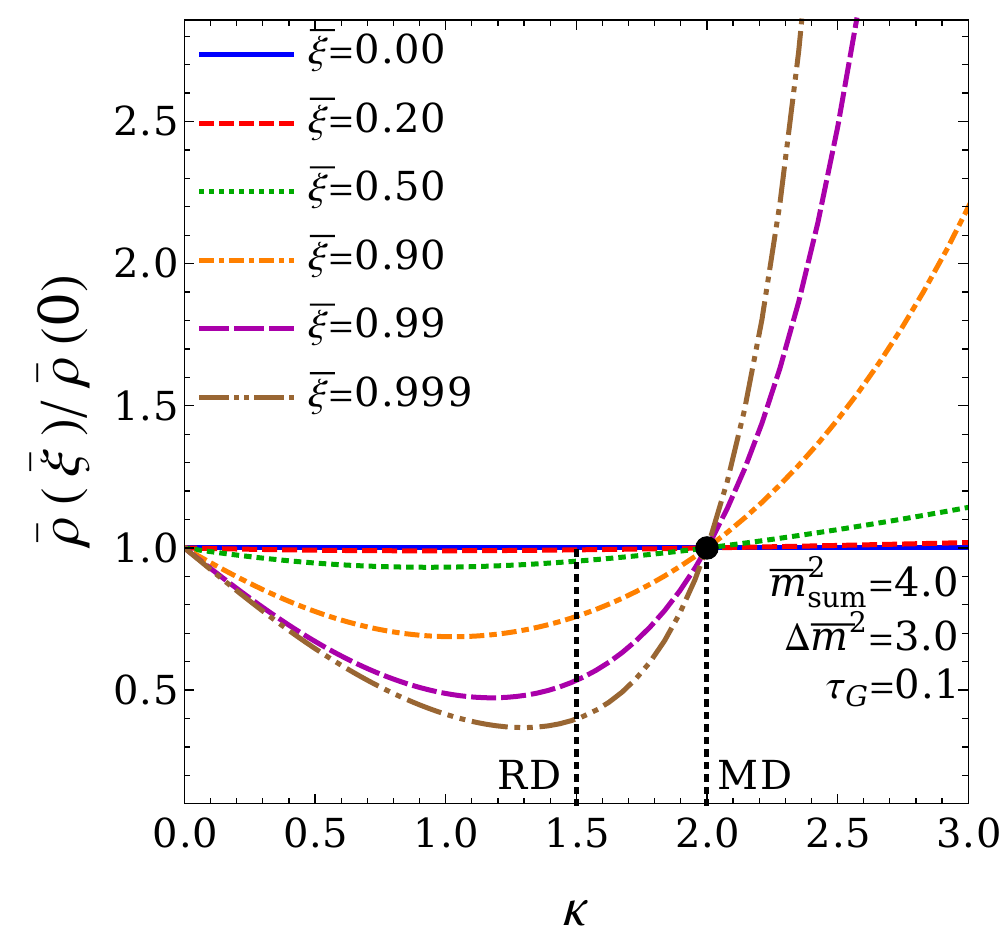}
\caption{Suppression of the late-time energy density $\rhobar$ due to mixing effects alone, with 
\mbox{$\Delta_G=0$}, for arbitrary universes parametrized by $\kappa$.
As in Fig.~\ref{fig:DDG},
we assume an instantaneous phase transition at \mbox{$\tau=\tau_G$},
but we now plot the corresponding
late-time energy density $\rhobar(\xibar)$ as a function of $\kappa$, 
normalized to its value at \mbox{$\xibar=0$}, for different choices of $\xibar$.
We see that mixing causes a {\it suppression}\/ of the late-time energy density within \mbox{$\kappa<2$} universes 
but an {\it enhancement}\/ within \mbox{$\kappa>2$} universes --- indeed, 
these effects disappear only
in the special case of a matter-dominated universe.
These effects can therefore produce significant modifications of the final late-time energy density,
even for instantaneous mass-generating phase transitions.}
\label{fig:DDG2}
\end{figure}

Finally,
along the same lines,
it is interesting to contemplate 
what happens for more general universes beyond those that are radiation-dominated.
The corresponding results for universes with general 
values of $\kappa$ are shown in  Fig.~\ref{fig:DDG2}.
We see that mixing alone indeed produces a suppression of the late-time energy density
for universes with \mbox{$\kappa <2$}, but this suppression actually becomes an {\it enhancement}\/
for universes with \mbox{$\kappa >2$}!
Moreover, as expected, both effects become stronger as the mixing saturation $\xibar$ increases ---
strong enough to change the late-time energy density by factors of two or three or even more.
Indeed, it is only for a matter-dominated universe (corresponding to \mbox{$\kappa=2$}) that this effect
disappears.
 
\FloatBarrier

\section{Individual late-time energy densities\label{sec:indiv}}


In the previous section, we focused on only one quantity:  the total late-time energy density $\rhobar$.
However, another important feature is the actual {\it distribution}\/ of this total energy density $\rhobar$ 
amongst the two fields of our system.  
We shall now proceed to study this issue.

\begin{figure*}
\centering
\hspace*{-0.4cm}
\includegraphics[width=0.328\textwidth,keepaspectratio]{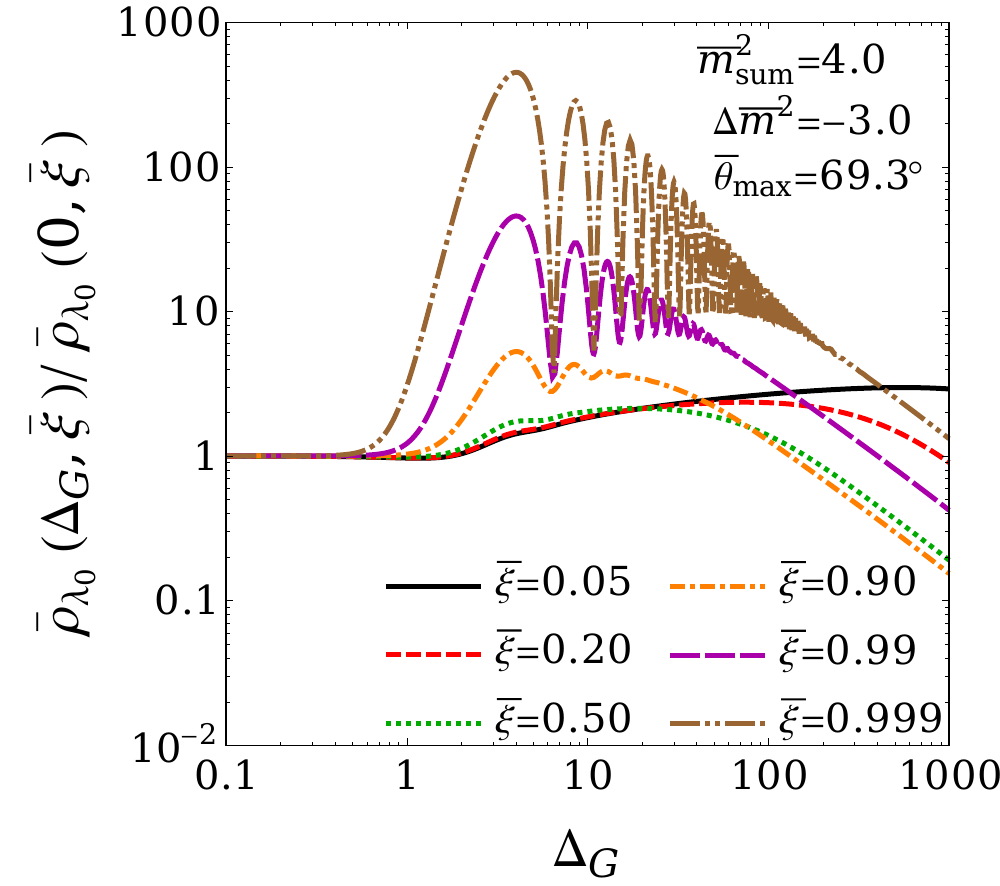} 
\includegraphics[width=0.328\textwidth,keepaspectratio]{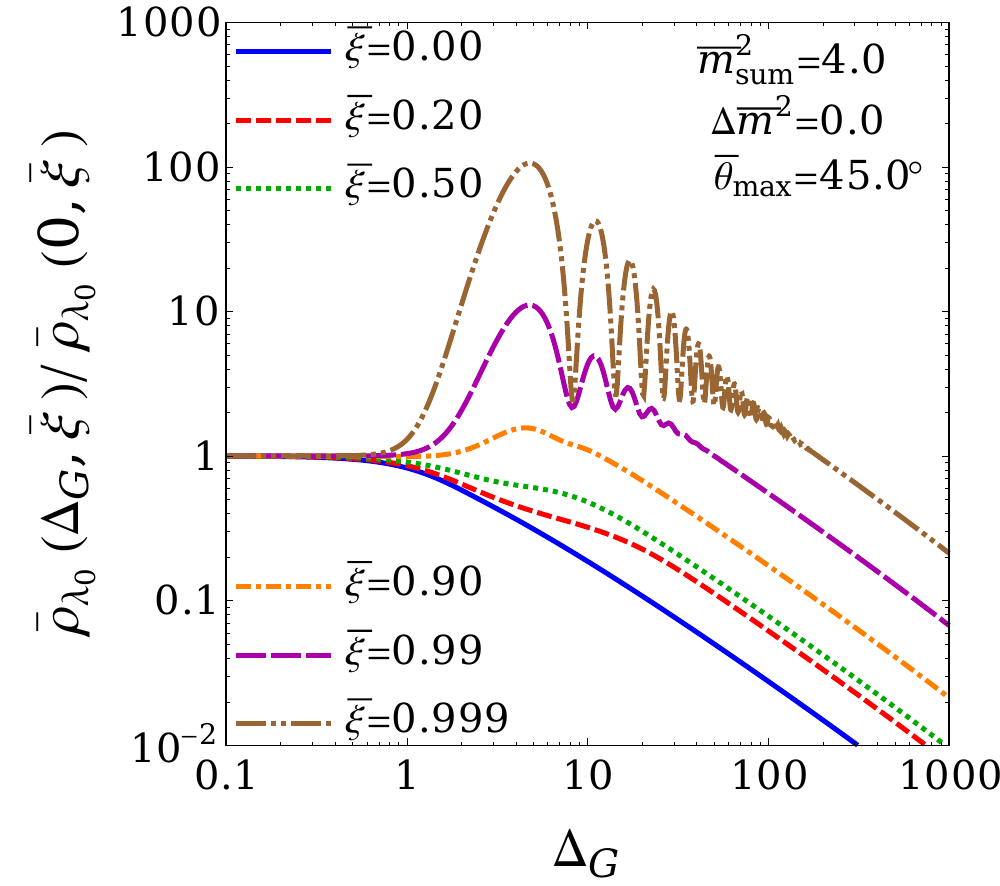} 
\includegraphics[width=0.328\textwidth,keepaspectratio]{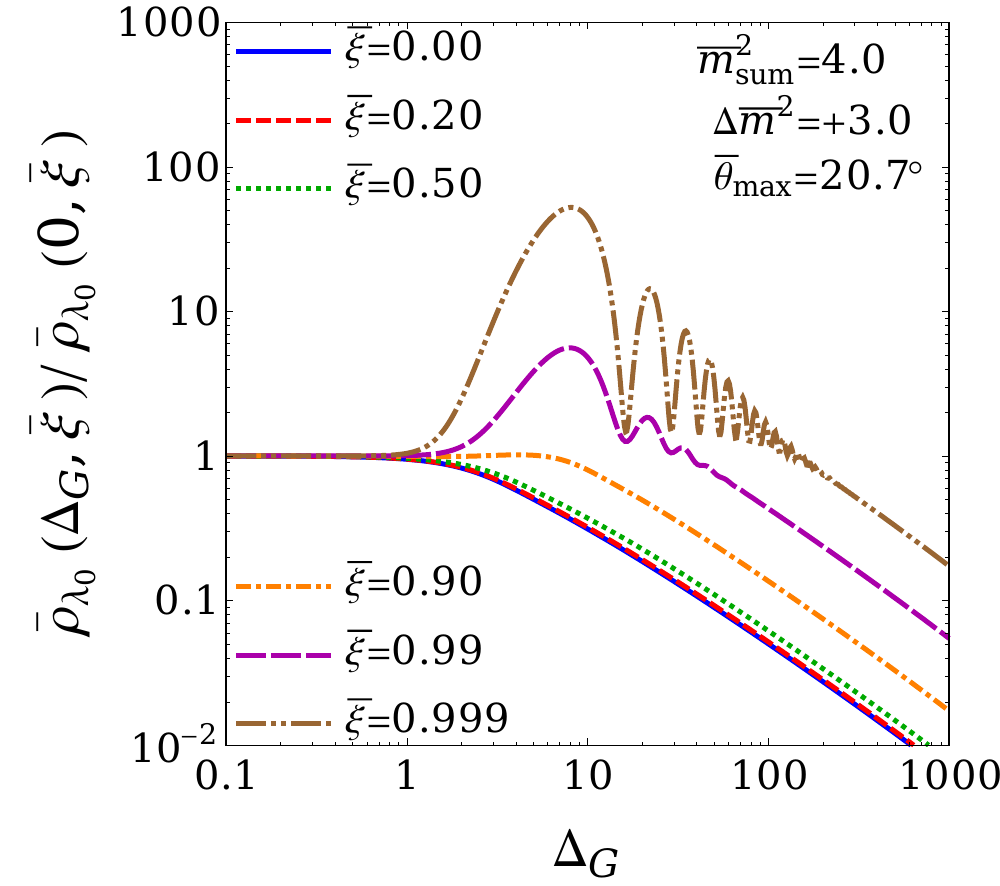}
\centering
\vskip 0.2 truein
\includegraphics[width=0.328\textwidth,keepaspectratio]{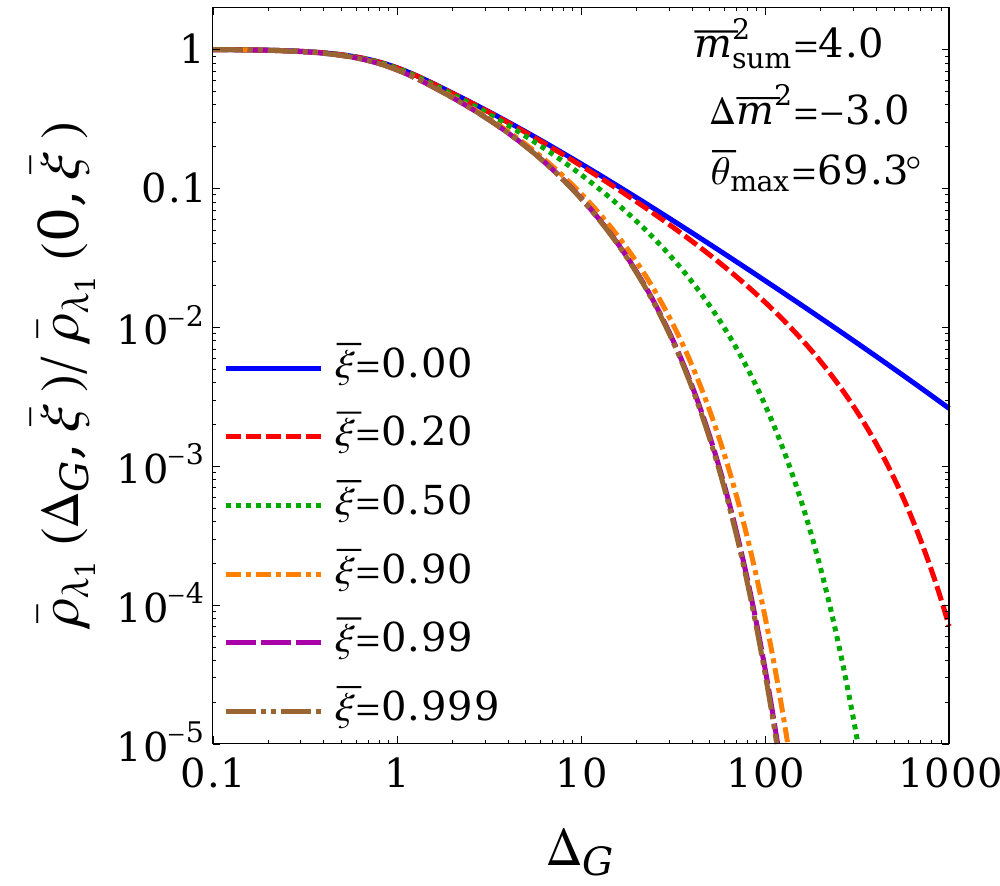} 
\includegraphics[width=0.328\textwidth,keepaspectratio]{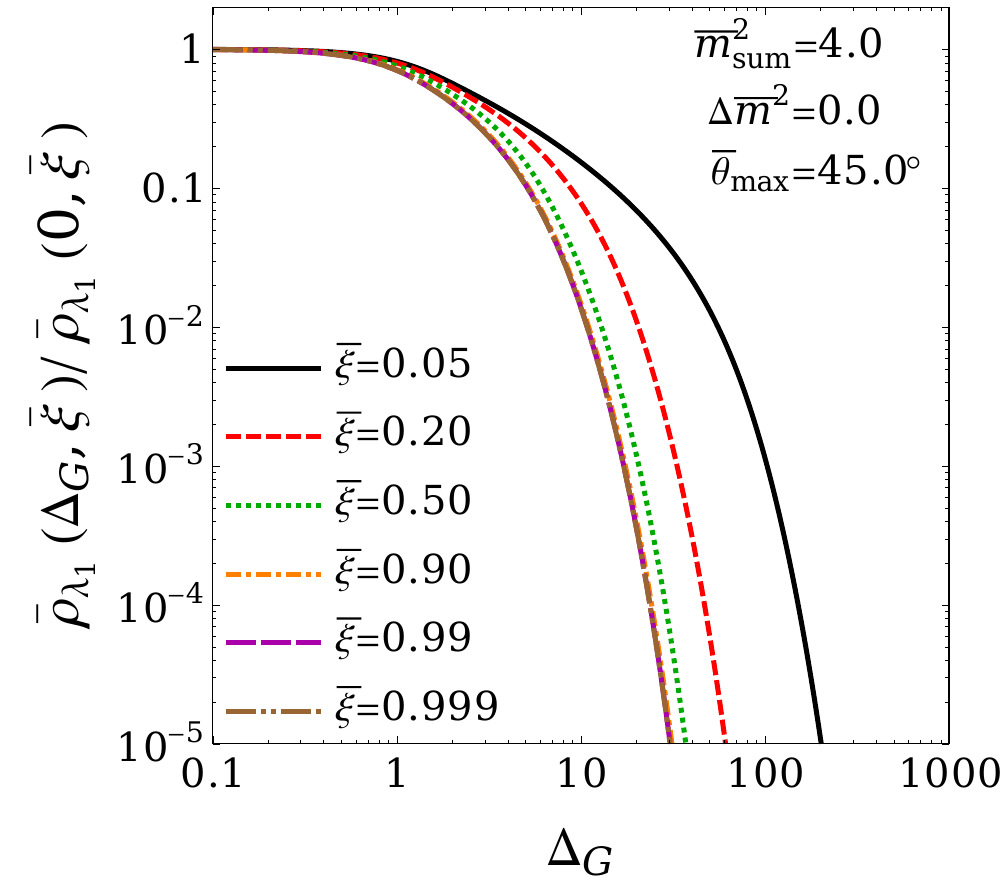} 
\includegraphics[width=0.328\textwidth,keepaspectratio]{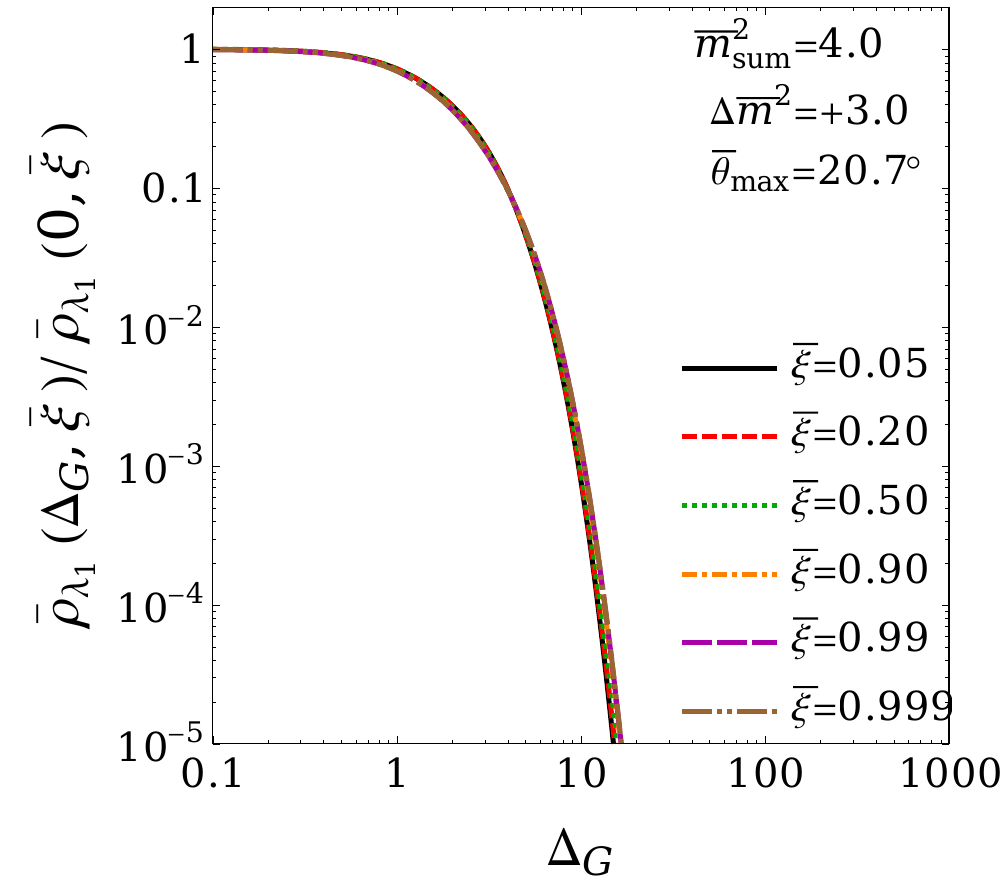}
\caption{Behavior of the {\it individual}\/ late-time energy densities 
     $\overline\rho_{\lambda_{0,1}}(\Delta_G,\xibar)$ when mixing effects,
    parametrized by $\xibar$, are 
       combined with a non-zero width $\Delta_G$ for the mass-generating phase transition.
   For these plots we adopt the large fiducial time \mbox{$\tau_G= 10^4$} 
    utilized for Fig.~\ref{fig:figname3}, 
     which leads to identical results
   for radiation- and matter-dominated universes.
   The top row shows results for $\rhobar_{\lambda_0}$ (associated with the lighter state in all cases),
   while the bottom row shows results for $\rhobar_{\lambda_1}$ (associated with the heavier state);
   we have also held \mbox{$\overline m_{\rm sum}^2 =4$} and take \mbox{$\Delta\overline m^2 = -3, 0, +3$} 
   for the left, center, and right columns, respectively.
   In each case we plot the late-time energy density $\overline\rho_{\lambda}$ as a function of 
   the phase-transition width $\Delta_G$ for different values of the mixing saturation $\xibar$, 
   normalized to its value for \mbox{$\Delta_G=0$}.   Note that we refrain from plotting curves 
   for \mbox{$\xibar=0$} (and instead plot curves for \mbox{$\xibar=0.05$}) in cases where the 
     corresponding energy densities vanish.}
\label{fig:rhocomponents}
\end{figure*}

In Fig.~\ref{fig:rhocomponents}, we show the behavior of the individual late-time energy densities associated
with $\phi_{\lambda_0}$ (the lighter field)
and $\phi_{\lambda_1}$ (the heavier field) for a mass-generating phase transition 
occurring at \mbox{$\tau_G=10^4$}.
It is therefore the sum of these two energy densities which produces the results in Fig.~\ref{fig:figname3};
recall that these results are the same for matter- and radiation-dominated universes.
It turns out that there are 
many features illustrated within Fig.~\ref{fig:rhocomponents} 
which will be important for our future results.
We shall therefore step through these features, one by one.

We begin by concentrating on the special case without mixing (\ie, with \mbox{$\xibar=0$}).
This will allow us to study the effects of the non-zero phase-transition width $\Delta_G$
for a single component alone without the complications due to mixing.
For \mbox{$\Delta \mbar^2\geq  0$}, all of the energy density arising due to the mass-generating phase transition
accrues to the lighter field $\phi_{\lambda_0}$.    For \mbox{$\Delta \mbar^2 < 0$}, by contrast,  
all of the late-time energy density is associated with 
the heavier field $\phi_{\lambda_1}$, as discussed at the end of Sect.~\ref{sec:AToyModel}.

In either case, we observe from
Fig.~\ref{fig:rhocomponents}
that the corresponding late-time energy density 
$\rhobar_\lambda$ remains essentially constant (\ie, independent of $\Delta_G$) 
for \mbox{$\Delta_G\lsim 2\pi/\lambda$}.   
This makes sense, as
the effects of introducing a non-zero width for our phase transition
will be essentially invisible if the 
mass generation occurs more rapidly than the natural timescale of
oscillations of our (underdamped) field $\phi_\lambda$.  
In other words, 
for \mbox{$\Delta_G\lsim 2\pi/\lambda$},   
the process of pumping energy density into our system
occurs with what may be considered to be maximum efficiency, since the phase transition
appears to be effectively instantaneous with respect to the natural oscillation timescale. 
However, for \mbox{$\Delta_G\gsim 2\pi/\lambda$},
the field oscillations tend to compete against the process of mass generation.
As a result, the mass-generating phase 
transition is less efficient in pumping energy density into our system,
thereby inducing a suppression in the late-time energy $\rhobar_\lambda$ which ultimately
scales as an inverse power of $\Delta_G$:
\beq
        \rhobar_{\lambda}(\Delta_G,0) ~\sim~ 1/\Delta_G~ ~~~~~~ {\rm for}~~ \Delta_G \gg 2\pi/\lambda~.
\label{eqone}
\eeq
Thus, once $\Delta_G$ exceeds the natural oscillation period of our mass eigenstate, 
increasing the width of the phase transition has the effect of introducing 
a power-law suppression of the corresponding late-time energy density.

These results hold for the unmixed scenarios with \mbox{$\xibar=0$}.
However, for the lighter fields $\phi_{\lambda_0}$ (corresponding
to the top row of Fig.~\ref{fig:rhocomponents}),
we see that the above asymptotic behavior continues to hold {\it regardless of the value of the mixing}\/:
\beq
        \rhobar_{\lambda_0}(\Delta_G,\xibar) ~\sim~ 1/\Delta_G~ ~~~~~~ {\rm for}~~ \Delta_G \gg 2\pi/\lambda_0~.
\label{eqtwo}
\eeq
Thus, we see that increasing the width of the phase transition continues to be associated
with a power-law suppression of the late-time energy density of the lighter field --- 
even in the presence of non-zero mixing.

Turning on a mixing between our two fields
also has a number of other important effects on their individual late-time energy densities. 
For the lighter fields $\phi_{\lambda_0}$ (as considered along
the top row of Fig.~\ref{fig:rhocomponents}),
the most prominent effect
is of course the 
set of very strong oscillations which are induced
for very large mixing saturations \mbox{$\xibar \lsim 1$}
and ``intermediate'' widths \mbox{$1\lsim \Delta_G\lsim 100$}.
We shall defer our discussion of these oscillations until the next section,
but we see from Fig.~\ref{fig:rhocomponents} that these oscillations are relatively large,
occasionally enhancing the corresponding late-time energy densities above
what they would have been in the absence of mixing {\it by several orders of magnitude}\/!
Indeed, this enhancement 
of the corresponding late-time energy densities above
their \mbox{$\xibar=0$} values
persists even for values of $\Delta_G$ which lie beyond the
actual oscillations themselves.

As a result, 
we see that
in the presence of a non-zero width for the phase transition,
mixing has a general tendency to {\it enhance}\/ the late-time energy density of the lighter field --- all 
without disturbing the power-law suppression discussed above.
Indeed,
for the lighter fields $\phi_{\lambda_0}$, the effects of the non-zero transition
width $\Delta_G$ and the non-zero mixing saturation $\xibar$ tend to pull in opposite directions:
the width tends to suppress the corresponding late-time energy density while the mixing tends to enhance it.
As evident from the plots along the top row of Fig.~\ref{fig:rhocomponents},
the nature of the net result (either an overall
enhancement or overall suppression) therefore depends non-trivially
on the precise values of $\Delta_G$ and $\xibar$ involved.
 
By contrast, for the {\it heavier}\/ fields $\phi_{\lambda_1}$ (for which the corresponding energy densities
are plotted along 
the bottom row of Fig.~\ref{fig:rhocomponents}),
the effects of non-zero mixing are quite the opposite:   
the enhancement  discussed above is gone, and instead there is now an {\it additional suppression}\/ 
which helps to drive the corresponding energy densities to even smaller values, as
functions of $\Delta_G$, than we had for the lighter fields!

\begin{figure}[t]
\centering
\hspace*{-0.4cm}
\includegraphics[width=0.45\textwidth,keepaspectratio]{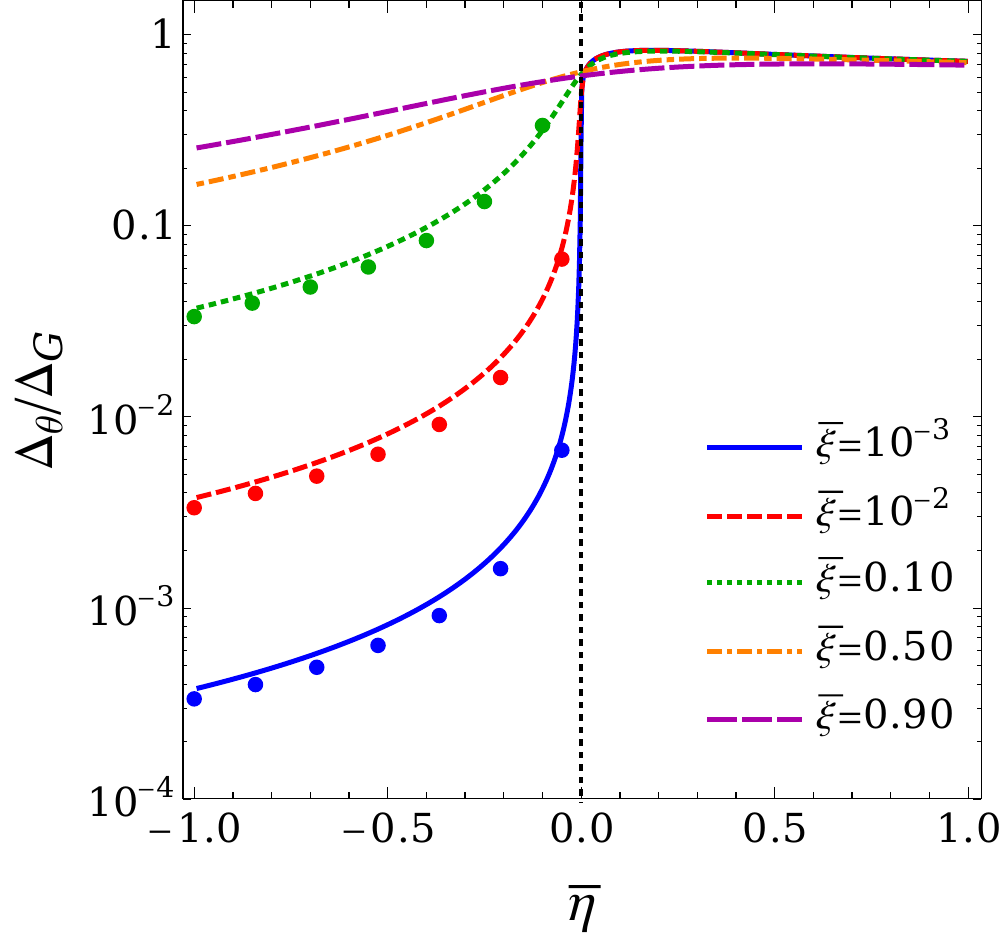}
\caption{The ratio of the two timescales $\Delta_G$ and $\Delta_\theta$, 
  plotted as a function of $\etabar$ for different mixing saturations $\xibar$.    
  In all cases, we see that \mbox{$\Delta_\theta < \Delta_G$}, with \mbox{$\Delta_\theta\ll \Delta_G$}
  for \mbox{$\etabar<0$} and \mbox{$\xibar\ll 1$}.
  The dots represent the approximate analytical result in Eq.~(\ref{approxanalytical}).
  The distinction between $\Delta_\theta$ and $\Delta_G$ has important implications for
  the energy density associated with the heavier field $\phi_{\lambda_1}$, since
  $\Delta_G$ governs the rate at which the {\it mass}\/ of this field is modified during the phase transition
   while 
  $\Delta_\theta$ governs the rate at which the {\it amplitude}\/ of this field is generated 
   as a result of its mixing
   with the lighter field.}
\label{fig:DToverDG}
\end{figure}

It is important to understand the origins of this additional suppression.
Unlike the lighter field $\phi_{\lambda_0}$,
which obtains its energy density 
directly from the phase transition
through the generation of a non-zero mass,
the heavier field $\phi_{\lambda_1}$ actually gains 
its energy density when it gains an overall field amplitude
through mixing with the lighter field.
The magnitude of the resulting energy density 
is then governed by two factors: the magnitude of
the field amplitude generated, and the masses that are also generated by the 
phase transition.

Thus, while the natural timescale governing the rate at which the energy
density is originally pumped into the lighter field   
is given directly by $\Delta_G$ --- and thus
by the function $h(\tau)$ plotted
in Fig.~\ref{fig:hsamples} ---
the natural timescale governing the rate 
at which the energy
density is pumped into the heavier field   
is governed not only
by $\Delta_G$ but also by
the rate of change of the {\it mixing angle}\/ $\theta$
plotted in Fig.~\ref{fig:thetavstau}.
In analogy with Eq.~(\ref{eq:hwidthderivation}),
we may even define a corresponding width $\Delta_\theta$ for the 
mixing angle $\theta$ via the slope of $\theta(\tau)$ at its midpoint 
$\overline{\theta}/2$:
\beq
     \Delta_{\theta} ~\equiv~ \left.(\overline{\theta}/\dot{\theta})\right|_{\tau_\theta} ~;
\label{eq:Deltathetadef}
\eeq
here $\tau_\theta$, as defined in Eq.~(\ref{tautheta}), is defined analogously to $\tau_G$ as the time
at which \mbox{$\theta=\overline{\theta}/2$}.
A rough analytical approximation for 
$\Delta_\theta$ when \mbox{$\etabar<0$} is given by
\beq
  \Delta_{\theta} ~\approx~ \frac{\xibar}{3|\etabar|}\, \Delta_G ~~~~~ {\rm for}~~\etabar<0~.
\label{approxanalytical}
\eeq
The magnitudes of $\Delta_G$ and $\Delta_\theta$ are compared in Fig.~\ref{fig:DToverDG},
from which we see that $\Delta_\theta$ never exceeds $\Delta_G$.
Indeed, we see 
from Fig.~\ref{fig:DToverDG}
that $\Delta_\theta$ can be much smaller than $\Delta_G$ 
for \mbox{$\etabar<0$} and \mbox{$\xibar\ll 1$}. 
We thus see that during our mass-generating phase transition,
the mixing angle $\theta$ always changes at least as rapidly  
as do the eigenvalues $\lambda_i$, and can in fact under certain circumstances
change even more rapidly, 
the latter despite  the fact that 
both sets of changes arise due to the same phase transition.

This observation is directly relevant for the manner in which
the heavier field accrues its energy density from the mass-generating phase transition.
As such, this feature is thereby directly relevant for 
the resulting behavior of the late-time 
energy density $\rhobar_{\lambda_1}$.
There are three distinct cases to consider.

\begin{itemize}

\item \mbox{$\Delta_G \ll 2\pi/\lambda_1$}:   In this case, the 
oscillations of the heavier field are so slow during the phase transition
that all aspects of the phase transition appear effectively instantaneous.
This is true for the generation of the {\it amplitude}\/ of the heavier 
field as well as the change in its {\it mass}\/.
Energy density is thereby delivered to the heavier field
with maximum efficiency, in a manner which is completely
insensitive to the non-zero timescales $\Delta_\theta$ and $\Delta_G$
associated with the phase transition and which is thereby protected
against all of their associated dissipating effects.
This behavior is evident for sufficiently small $\Delta_G$ within
the plots along the lower row of 
Fig.~\ref{fig:rhocomponents}.

\item \mbox{$\Delta_{\theta} \lesssim 2\pi/\lambda_1\lesssim \Delta_G$}:  In this case,
the oscillations of the heavier field are sufficiently slow that 
the generation of its field amplitude appears to be effectively instantaneous;
these oscillations are nevertheless 
sufficiently rapid that full oscillations can occur during the change in field mass.
The net result is that the heavy-field energy density $\rhobar_{\lambda_1}$ experiences
the same suppression as does the light-field energy density $\rhobar_{\lambda_0}$,
with both depending non-trivially on $\Delta_G$ and ultimately scaling inversely with 
$\Delta_G$.
Note that this case exists only for situations in which $\Delta_\theta$ is significantly
smaller than $\Delta_G$, which according to Fig.~\ref{fig:DToverDG}
tend to emerge only for \mbox{$\Delta \mbar^2<0$} and relatively small mixing.

\item \mbox{$2\pi/\lambda_1 \lesssim \Delta_{\theta}$}: In this case, the field 
oscillations of the heavier field are sufficiently rapid that full oscillations  
are occurring not only during the change in its mass {\it but also during the generation of its amplitude}\/.
This latter feature leads to an additional source of suppression for the late-time energy density
beyond what the lighter field experiences, and causes $\rhobar_{\lambda_1}$ to exhibit an even more
dramatic suppression as a function of $\Delta_G$ than that exhibited by $\rhobar_{\lambda_0}$.
This behavior is clearly evident for sufficiently large $\Delta_G$ in the plots along the lower row of 
Fig.~\ref{fig:rhocomponents}.
\end{itemize}

The above discussion describes the
behaviors of the individual components $\rhobar_{\lambda}$
as functions of $\Delta_G$ and $\xibar$.
However, our final task is to determine how much each of these 
individual components contributes to the total energy density $\rhobar$.
Of course, to do so requires that we understand not just the intrinsic behaviors of these individual
components as functions of $\Delta_G$ and $\xibar$, but also the relative sizes of these
individual components.
In other words, we need to understand the {\it relative normalization}\/ of the curves for $\rhobar_{\lambda_0}$
in the top row of Fig.~\ref{fig:rhocomponents} 
relative to those for $\rhobar_{\lambda_1}$ in the bottom row.
However, this relative normalization may easily be determined in the \mbox{$\Delta_G = 0$} limit,
for which analytical results are given in Appendix~\ref{notation}.
Indeed, use of the virial approximation 
\mbox{$\lambda_i \tau_G\gg 1$}, which is valid for the plots in Fig.~\ref{fig:rhocomponents}, 
yields the ratio
\beq
        \frac{\rhobar_{\lambda_0}(0,\xibar)}{\rhobar(0,\xibar)}  = 
         \frac{ \lambdabar_0^2 \cos^2\thetabar}{ \lambdabar_0^2 \cos^2\thetabar + \lambdabar_1^2 \sin^2 \thetabar}
           = \cos^2\theta \left( \frac{ 1-\eta \sec 2\theta}{1-\eta}\right)
\label{ratio}
\eeq
as the fraction of the total late-time energy density remaining in the lighter field.
As expected, this quantity 
evolves monotonically
from unity at \mbox{$\theta=0$} to
zero at \mbox{$\theta=\theta_{\rm max}$} [where $\theta_{\rm max}$ is defined in Eq.~(\ref{thetamaxdeff})].
It may thus be greater or less than 50\%, 
depending on the eigenvalues and mixing angles in question.
Note that in the limit of small mixing (which corresponds to \mbox{$\theta\approx 0$} or \mbox{$\theta\approx \pi/2$}),
the result in Eq.~(\ref{ratio}) implies that whether the bulk of the energy density winds up at late times
associated with the lighter or heavier field depends simply on the sign of $\eta$ rather than its magnitude.
(Recall, in this connection, that \mbox{$\eta\to -\eta$} implicitly changes \mbox{$\theta\to \pi/2-\theta$}.)

\begin{figure*}
\centering
\hspace*{-0.4cm}\includegraphics[width=0.329\textwidth,keepaspectratio]{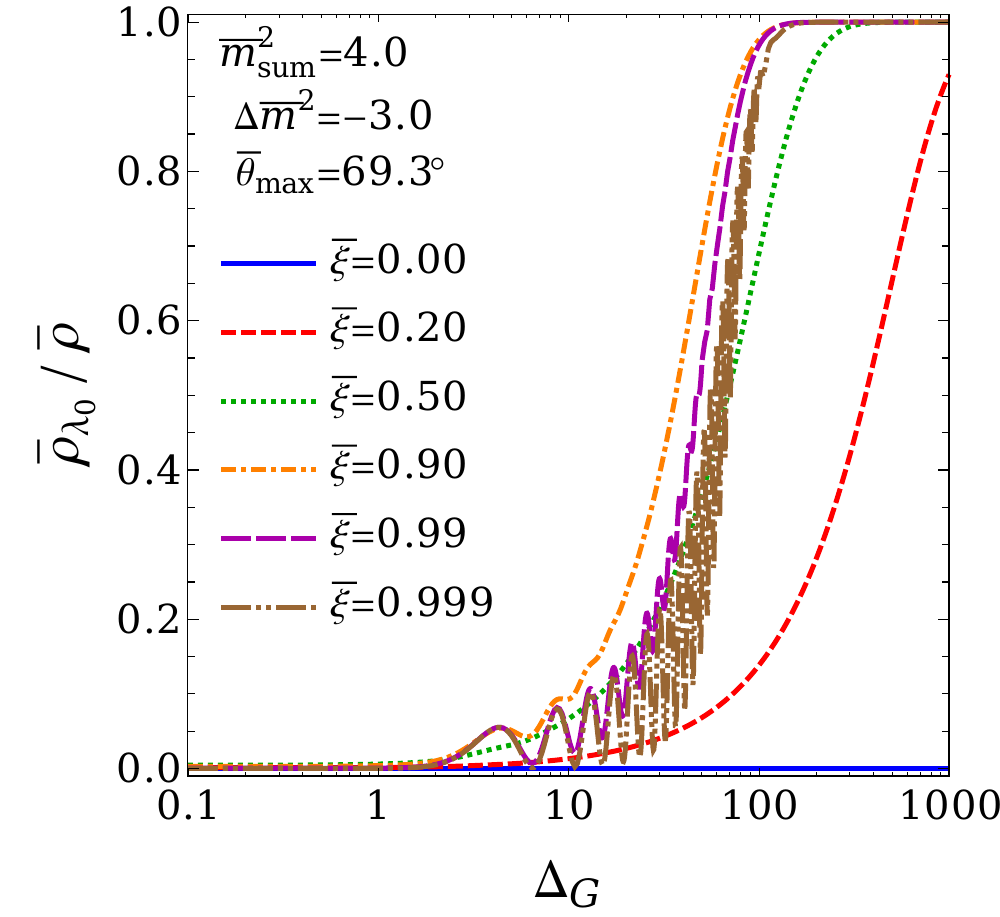} \ \includegraphics[width=0.329\textwidth,keepaspectratio]{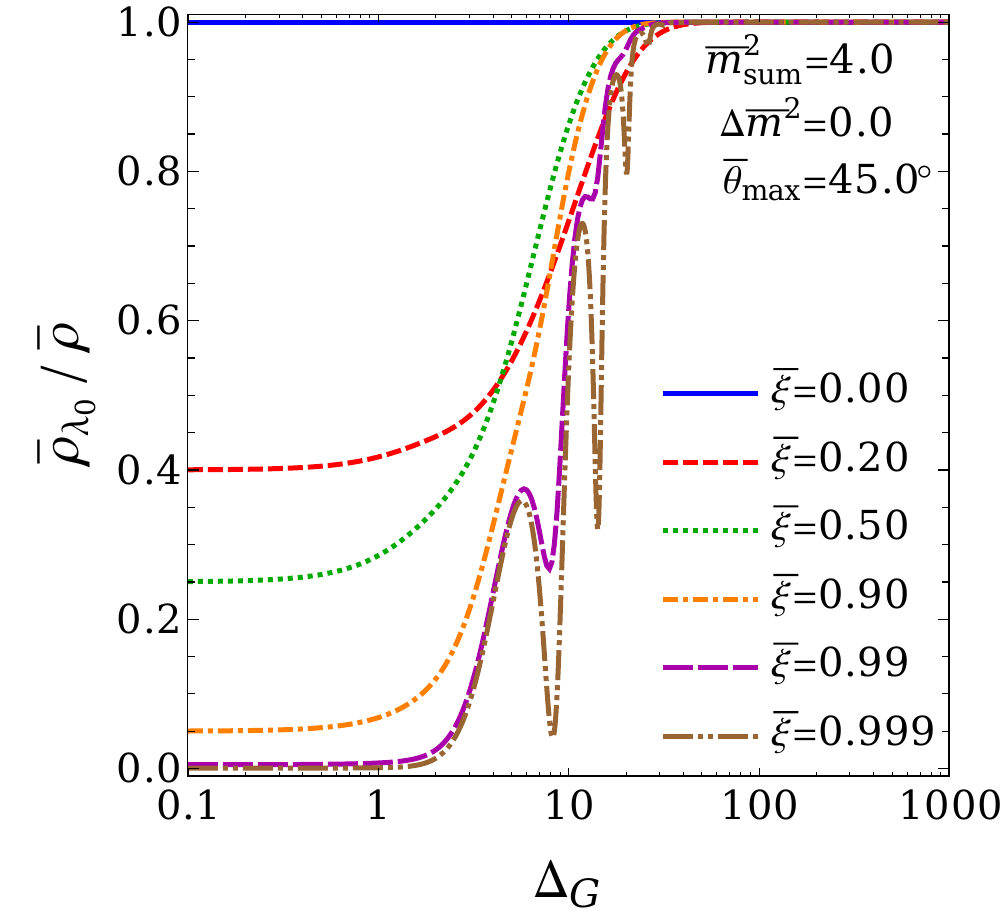}  \ \includegraphics[width=0.329\textwidth,keepaspectratio]{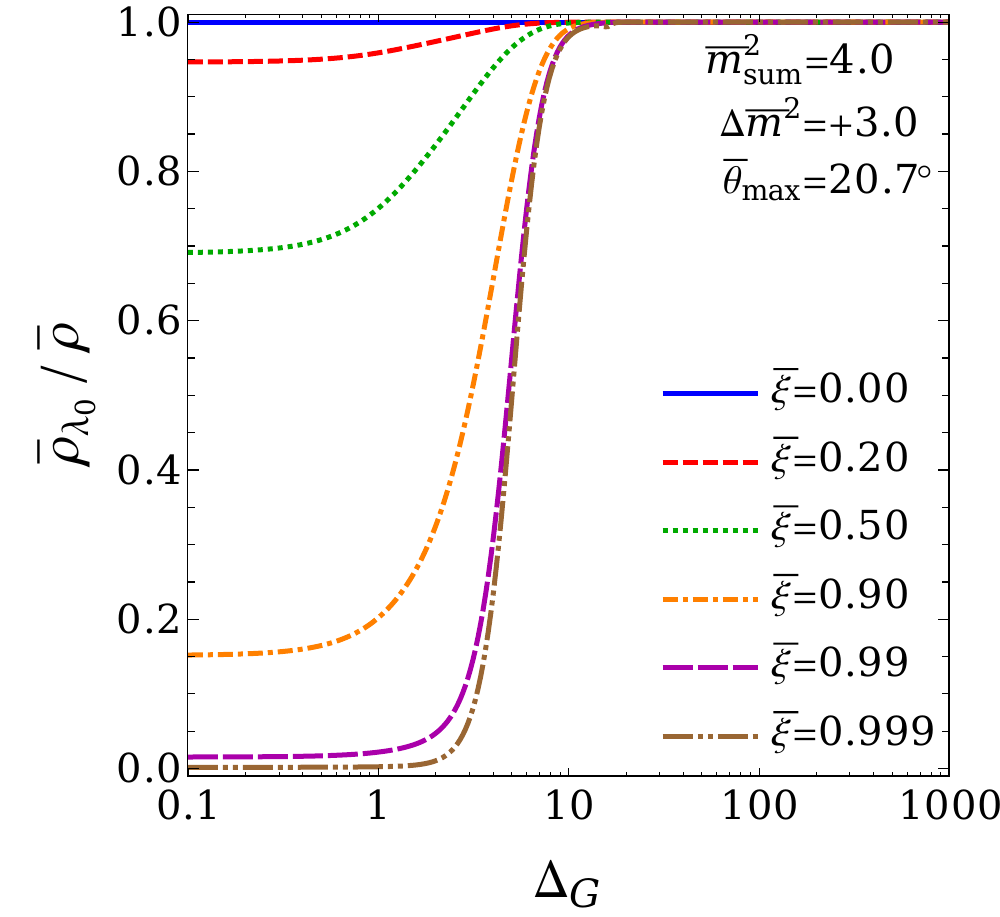}
\caption{The fraction of the total late-time energy density $\rhobar(\Delta_G,\xi)$ 
which is associated with the  
lighter field $\phi_{\lambda_0}$, plotted as functions of $\Delta_G$ for different values of $\xi$, with \mbox{$\tau_G=10^4$}. 
Regardless of the small-$\Delta_G$ values of this fraction [as given analytically in Eq.~(\ref{ratio})],
we see that increasing the width $\Delta_G$ of the phase transition generally 
has the net effect of transferring more and more
of the total energy density into the lighter field.
Thus, simply by adjusting the width of the phase transition, we see 
that we can often entirely reverse the distribution of the late-time energy density.}
\label{fig:rhocomponentsratio}
\end{figure*}

The result in Eq.~(\ref{ratio}) holds only for \mbox{$\Delta_G=0$}.
However, as $\Delta_G$ grows larger, this ratio will change.
For the total energy densities plotted in Fig.~\ref{fig:figname3}, 
the corresponding fractional results are shown in Fig.~\ref{fig:rhocomponentsratio}
as a function of $\Delta_G$.
It is easy to understand the 
behavior shown in Fig.~\ref{fig:rhocomponentsratio}.
Regardless of the fraction of the total late-time energy density associated with the lighter field $\phi_{\lambda_0}$
when \mbox{$\Delta_G=0$}, we have already seen that the energy densities associated with the heavier fields are typically
more suppressed as a function of $\Delta_G$ than are the energy densities associated with the lighter fields.
Thus, {\it regardless of the late-time energy-density configuration when the phase transition is instantaneous,
we see that increasing the width of the phase transition has the effect of throwing an increasingly large share of the total 
late-time
energy density into the lighter field}.   Indeed, in some cases, {\it we see that we can entirely reverse
the distribution of the total energy density from the heavier field to the lighter field, 
simply by adjusting the timescale over which the phase transition occurs}\/!  

\FloatBarrier

\section{Parametric resonance\label{sec:TheParametricResonance}}


We now turn to what is perhaps the most prominent feature within the plots shown in Figs.~\ref{fig:figname3},
\ref{fig:rhocomponents},
and \ref{fig:rhocomponentsratio}:
the appearance of non-monotonicities in the late-time energy densities $\rho_{\lambda}$ associated with the lighter
fields $\phi_{\lambda_0}$ as functions of the phase-transition width $\Delta_G$.
As indicated in Sect.~\ref{sec:LateTimeEnergyDensity},
these oscillations grow particularly large when 
$\tau_G$ is large
and when the mixing saturation $\xibar$ grows close to $1$.
These non-monotonicities indicate that within this region of parameter space 
there exists an extremely strong
sensitivity of the late-time energy density $\rhobar_{\lambda_0}$ to even 
small variations in the phase-transition width $\Delta_G$.
Moreover, as evident from Fig.~\ref{fig:rhocomponents}, these non-monotonicities
can often enhance $\rhobar_{\lambda_0}$  
by several orders of magnitude!

These non-monotonicities are yet another consequence of
the interplay between the width of our phase transition and the mixing it induces.
It is not difficult to understand the origin of these oscillations:   ultimately they are {\it parametric resonances}\/
triggered by our mass-generating phase transition.
Recall that a parametric resonance generically occurs when the mass of an oscillator 
itself exhibits an oscillatory behavior whose frequency is approximately twice the natural frequency
of the oscillator.   (An example of this is a child ``pumping'' a swing by alternatively standing and squatting
on the swing seat at the correct moments during the period of the swing.)
In such systems, the amplitude of oscillation grows exponentially
and without bound unless the oscillator also has a frictional damping term.
A similar behavior can also emerge if the friction term (rather than the mass term) experiences the
oscillatory behavior. 

At first glance, it may seem that our system of coupled 
scalar fields has no means of experiencing a parametric resonance.  
The equations of motion of our model are given in 
Eq.~(\ref{eq:newerdoublefieldequationsofmotion}),
and although both the Hubble damping term $H(\tau)$ 
and the mass matrix $\calM^2_{ij}$ in our model are time-dependent, 
the Hubble term is monotonically falling.
Likewise, the $h(\tau)$-function --- which governs the time-development of our
mass matrix, as in Eq.~(\ref{eq:doubleTDfactoring}) ---
is also monotonic.

However, this is where the mixing plays a critical role.
Because of the non-zero mixing inherent in this sytem, 
the aspects of the mass matrix which are important
are not  its individual components $m^2_{ij}$ but 
rather its {\it eigenvalues}\/ $\lambda_i$.
Indeed, even though the mass-matrix components $m^2_{ij}(\tau)$ are all monotonic as functions of time $\tau$,
the eigenvalues $\lambda_i$ need not be.
These eigenvalues were plotted as functions of time in Fig.~\ref{fig:lambdares1},
and we see that for the smaller eigenvalue $\lambda_0$ 
the mixing induces a level-repulsion that often results in a 
non-monotonic ``pulse''.
This pulse is more prominent for \mbox{$\Delta \mbar^2<0$} than
for \mbox{$\Delta m^2>0$}, but it is ultimately a generic feature of the behavior of the lower eigenvalue
for sufficiently saturated mixing.
Indeed, this pulse grows increasingly dramatic as \mbox{$\xibar\to 1$}, 
and has a width governed by $\Delta_G$.
In general, it is straightforward to show that a pulse will always appear for
the lightest eigenvalue if the extra contribution to the squared-mass matrix
that comes from the mass-generating phase transition has negative determinant --- \ie,
\mbox{$\det \mbar_{ij}^2 <0$}.

\begin{figure}[t]
\centering
\hspace*{-0.4cm}\includegraphics[width=0.5\textwidth,keepaspectratio]{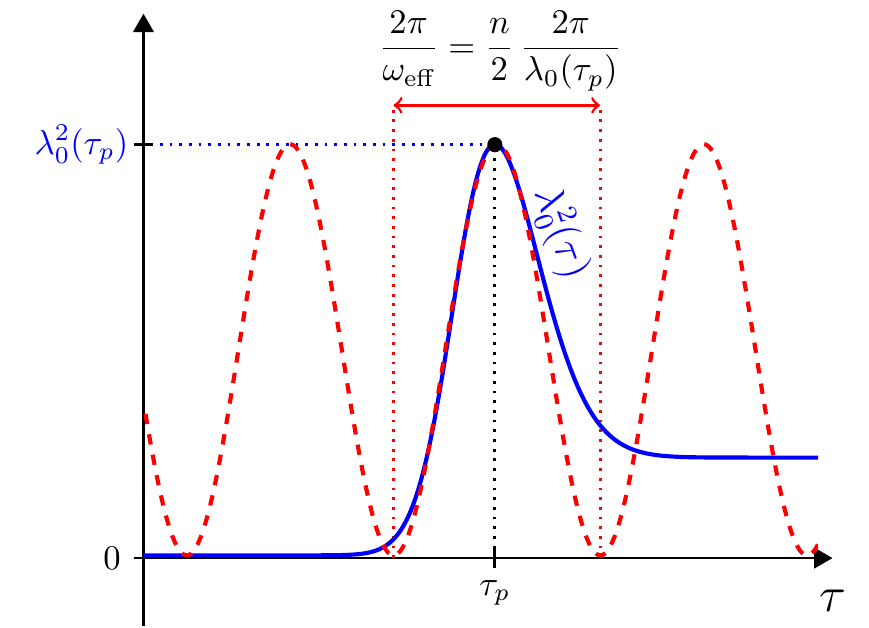}
\caption{A schematic illustration of the ``pulse'' (solid blue curve) experienced by the smaller 
eigenvalue $\lambda_0(\tau)$ near \mbox{$\tau\approx \tau_{{p}}$} 
and a full sinusoidal function (dashed red curve) to which it may 
be approximated in the vicinity of the pulse.
As discussed in the text, this approximation allows us to extract the effective frequency $\omega_{\rm eff}$ associated 
with the pulse and thereby determine the mathematical condition under which an $n^{\rm th}$-order
parametric resonance occurs.} 
\label{fig:resonancefig}
\end{figure}

Strictly speaking, a single pulse does not oscillatory behavior make.
However, during the relevant time interval near $\tau_G$,
this pulse may be regarded as one oscillation within a full sinusoidal pattern.
This situation is illustrated schematically in Fig.~\ref{fig:resonancefig}.
Of course, variations in $\Delta_G$ directly affect the width of this pulse,
and thereby change the effective pulse frequency.
As a result, there will exist a single  value of $\Delta_G$ for which this frequency
is exactly twice the natural frequency of our oscillator,
and for which our primary parametric resonance emerges.
Even greater values of $\Delta_G$ 
then correspond to the higher harmonics of
this resonance.

This, then, is the origin of the parametric oscillations apparent in 
the plots shown in Figs.~\ref{fig:figname3},
\ref{fig:rhocomponents},
and \ref{fig:rhocomponentsratio}.
As such, this entire phenomenon is a prime example of the interplay between the mixing $\xibar$ 
and the width of the mass-generating phase 
transition:   the non-zero mixing 
produces the level repulsion that leads to the pulse,
while the non-zero width of the phase transition endows this pulse
with the specific width/frequency needed for it to potentially trigger an actual resonance.
Moreover, with this explanation, 
we now understand why these resonances occur only for certain values of $\Delta_G$,
and only for the lighter field.
This also explains why these resonances grow stronger as \mbox{$\xibar \to 1$}.
Furthermore, although this pulse exists for both \mbox{$\Delta \mbar^2<0$} and \mbox{$\Delta \mbar^2>0$}
(as evident in Fig.~\ref{fig:lambdares1}),
we have already learned in the previous section
that for the intermediate values of $\Delta_G$ relevant for the parametric
resonance, the energy density of the lighter field 
contributes a greater fraction of the total energy density for 
\mbox{$\Delta \mbar^2  >0$} than it does for 
\mbox{$\Delta \mbar^2  <0$}. 
This behavior is illustrated, for example, in Fig.~\ref{fig:rhocomponentsratio}.
It is for this reason that 
the {\it total}\/ energy density $\rhobar$  
exhibits a stronger parametric resonance for \mbox{$\Delta \mbar^2 >0$} than for \mbox{$\Delta \mbar^2 <0$} (as evident in 
Fig.~\ref{fig:figname3}),
even though the {\it individual}\/ component 
$\rhobar_{\lambda_0}$
exhibits a stronger
parametric resonance 
for \mbox{$\Delta \mbar^2 <0$} 
than 
for \mbox{$\Delta \mbar^2 >0$} 
(as evident in Fig.~\ref{fig:rhocomponents}).

In general, it is easy to determine the values of $\Delta_G$ for which such parametric
resonances occur.
As indicated above, we imagine that $\lambda_0^2(\tau)$ experiences a ``pulse'' centered at \mbox{$\tau=\tau_{{p}}$},
as sketched in Fig.~\ref{fig:resonancefig}.  
Within the region of the pulse,  
we then approximate the behavior of the pulse as part of a sinusoidal function with an effective
 frequency $\omega_{\rm eff}$: 
\beq
  \lambda_0^2(\tau) ~\approx~  \half \lambda_0^2(\tau_{{p}}) \, 
            \biggl\lbrace 1+ \cos [\omega_{\rm eff} (\tau-\tau_{{p}})]\biggr\rbrace~
\label{cosineform}
\eeq
for \mbox{$\tau\approx \tau_{{p}}$}.
From Eq.~(\ref{cosineform}) we then find that
\beq
        \omega_{\rm eff}^2 ~=~  -4 \, \left( \frac{\ddot \lambda_0}{\lambda_0}\right)  \Biggl|_{\tau=\tau_{{p}}}~.
\label{weffvalue}
\eeq

A pulse with this frequency will lead to an $n^{\rm th}$-order parametric 
resonance only if $\omega_{\rm eff}$ is $(2/n)$ times the natural oscillation frequency of our system.
The case with \mbox{$n=1$} produces the primary parametric resonance, while the cases with \mbox{$n\in \mathbb{Z}>1$} produce
its higher-order harmonics.
However, near \mbox{$\tau\approx \tau_{{p}}$},
the natural oscillation frequency associated with the lightest field is nothing but $\lambda_0(\tau_{{p}})$, since
it is the mass of the field which drives its oscillations.
We thus must demand 
\beq
             \omega_{\rm eff} ~=~ \frac{2}{n} \cdot \lambda_0(\tau_{{p}})~.
\eeq
In other words, a parametric resonance will occur only when two timescales are balanced: 
the timescale $1/\omega_{\text{eff}}$ of the pulse, and the timescale
$1/\lambda_0(\tau_{{p}})$
of the field oscillations in the vicinity of the pulse.
Combining this with Eq.~(\ref{weffvalue}),
we then obtain a condition for an $n^{\rm th}$-order parametric resonance: 
\beq
              \left(\frac {\ddot\lambda_0} {\lambda_0^3} \right)\Biggl|_{\tau=\tau_{{p}}} 
       =~ - \frac{1}{n^2} ~,~~~~~~~ n\in\mathbb{Z}^+~.
\label{resonancecondition}
\eeq
Variations in $\Delta_G$ will modify the value of the left side of this equation.
Thus, there exist a discrete set of values $\Delta_G^{(n)}$ for which this equation
can be satisfied, one for each value of $n$.
These are then the phase-transition widths for which parametric resonances
exist.
 
\begin{figure}[t]
\centering
\hspace*{-0.4cm}\includegraphics[width=0.45\textwidth,keepaspectratio]{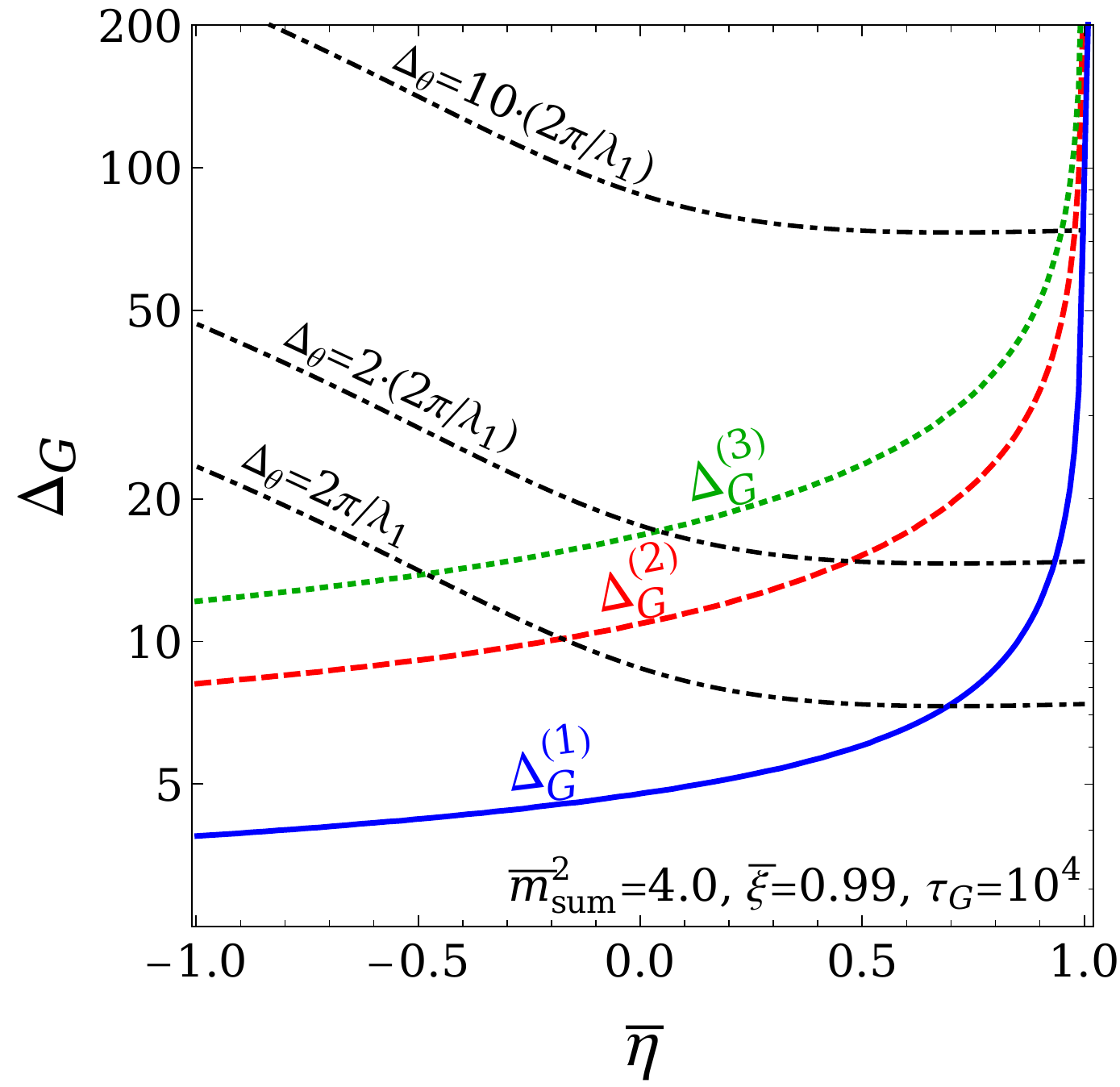}
\caption{The phase-transition widths $\Delta_G^{(n)}$ (\mbox{$n=1,2,3$})  at which the first three parametric resonances occur, plotted as functions of $\etabar$.
Contours showing $2\pi/\lambda_1$ in relation to $\Delta_\theta$ are also superimposed.
The region with large $\etabar$ is thus one in which our parametric resonances occur 
while \mbox{$\Delta_\theta \gg 2\pi/\lambda_1$} --- \ie, a region in which the total late-time energy density $\rhobar$ is
dominated by the contribution from the lighter field experiencing the parametric resonance.}   
\label{fig:sigmares}
\end{figure}

The solutions for $\Delta_G^{(n)}$ can be obtained numerically, and 
our results are shown in Fig.~\ref{fig:sigmares}.  We see
that the values of $\Delta_G^{(n)}$ generally increase with $n$,
as expected.
Likewise, these critical widths also increase as functions of $\etabar$ and
ultimately diverge as \mbox{$\etabar\to 1$}.
Moreover, superimposed on this plot are contours showing 
the values of $2\pi/\lambda_1$ in relation to $\Delta_\theta$. 
As noted in Sect.~\ref{sec:indiv}, 
the smaller $2\pi/\lambda_1$ is in relation to $\Delta_\theta$, 
the more suppressed is the energy-density contribution of the heavier field relative
to that of the lighter field
and hence 
the more the lighter field (which feels the parametric resonance) dominates 
the total energy density.
This then confirms our previous expectation that the parametric resonance, through stronger
within $\rhobar_{\lambda_0}$ when \mbox{$\Delta \mbar^2 <0$},
is nevertheless more pronounced within $\rhobar$ 
when \mbox{$\Delta \mbar^2 >0$}.

Thus, we conclude that parametric resonances not only occur, but can also dominate
the total energy density of our system.   Such parametric resonances can distort
the resulting late-time energy densities by several orders of magnitude, and thus
may play an important role in early-universe cosmology.
 
\FloatBarrier

\section{Re-overdamping\label{sec:ReOverDamping}}


We now turn to another novel feature which results from the confluence between mixing and
a finite phase-transition width:  a phenomenon which we refer to as ``re-overdamping''. 

As discussed at the beginning of Sect.~\ref{sec:AToyModel},
any scalar field $\phi$ with mass $\lambda(t)$ will experience overdamped behavior 
(and thus function as vacuum energy)
at times $t$ 
for which \mbox{$3H(t)/2 \gsim \lambda(t)$}.
By contrast, such a field will experience
underdamped
behavior 
(and thus function as matter)
when \mbox{$3H(t)/2 \lsim \lambda(t)$}.
Of course, \mbox{$H(t)\sim 1/t$} in any post-inflationary epoch.
Thus, in situations for which $\lambda(t)$ is either constant or monotonically increasing,
as illustrated in Fig.~\ref{fig:3Hmfig},
there exists a single, relatively short era of times $t$ for which 
\mbox{$3H(t)/2 \approx \lambda(t)$}.   Indeed, this is the period, as discussed in Sect.~\ref{sec:AToyModel}, 
during which our scalar field $\phi$ is transitioning from overdamped to underdamped behavior.
After this transition, our field then remains underdamped for all future times.

\begin{figure}[b]
\begin{center}
\hspace*{-0.4cm}\includegraphics[width=0.45\textwidth,keepaspectratio]{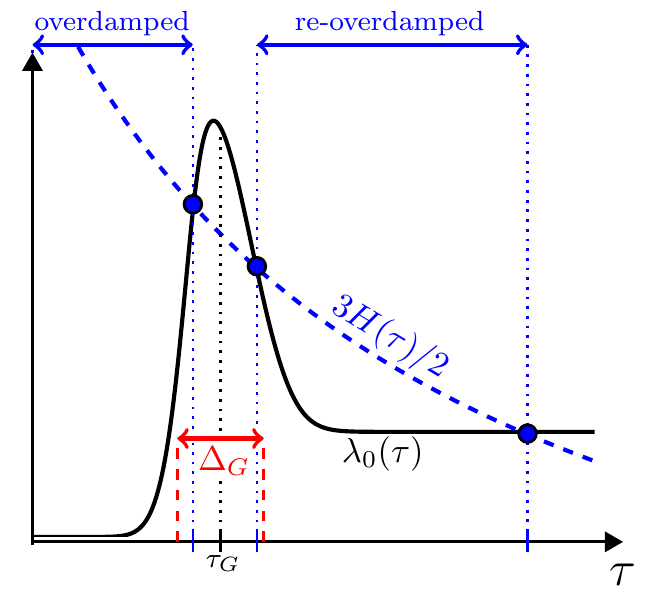}
\end{center}
\caption{A schematic illustrating the behavior of the lower-mass eigenvalue $\lambda_0(t)$ 
in the presence of a ``pulse'', plotted as a function of time with the Hubble parameter $3H(t)/2$ superimposed. 
The corresponding field $\phi_{\lambda_0}$ therefore experiences not just the usual transition from overdamped to underdamped behavior, as in 
Figs.~\ref{fig:rhosinglefield} and \ref{fig:3Hmfig},
but also a novel {\it re-overdamping} transition.   As shown, this re-overdamped phase can persist for considerably longer than the width $\Delta_G$ of the pulse that produced it, and thus can have a significant impact on the resulting
cosmological history.}
\label{fig:backforthlambda}
\end{figure}

This is the standard situation, 
and assumes that 
$\lambda(t)$ is either constant or monotonically increasing.
However, as we discussed in Sect.~\ref{sec:TheParametricResonance},
this need not be true when multiple fields mix in the presence of a 
mass-generating phase transition of non-zero width.  
Indeed, as we have seen 
in Sect.~\ref{sec:TheParametricResonance}
for our two-component toy model, 
the mass of our lighter field can experience a ``pulse'',
as illustrated in Fig.~\ref{fig:resonancefig};
this is the phenomenon underlying the parametric resonances discussed
in Sect.~\ref{sec:TheParametricResonance}.
In the presence of a pulse, however, the $3H(t)/2$ curve may cross the $\lambda(t)$ curve not just
once but {\it three}\/ times, resulting in a field which begins overdamped, then becomes underdamped,
but then experiences a {\it re-overdamped}\/ phase  before eventually settling back into a final underdamped state.  
This behavior is illustrated in Fig.~\ref{fig:backforthlambda}.
Moreover, as sketched in Fig.~\ref{fig:backforthlambda}, 
the period during which the field exists in a re-overdamped phase can be
of extremely long duration --- even relative to the width of the pulse that produced it.
Indeed, as we shall discuss below, this duration can become arbitrarily long, depending on the
values of the underlying input parameters of our model.
As a result, the re-overdamped phase can persist throughout a sizable portion of
cosmological history.

\begin{figure*}[t!]
\centering
\hspace*{-0.1cm}\includegraphics[width=0.31\textwidth,keepaspectratio]{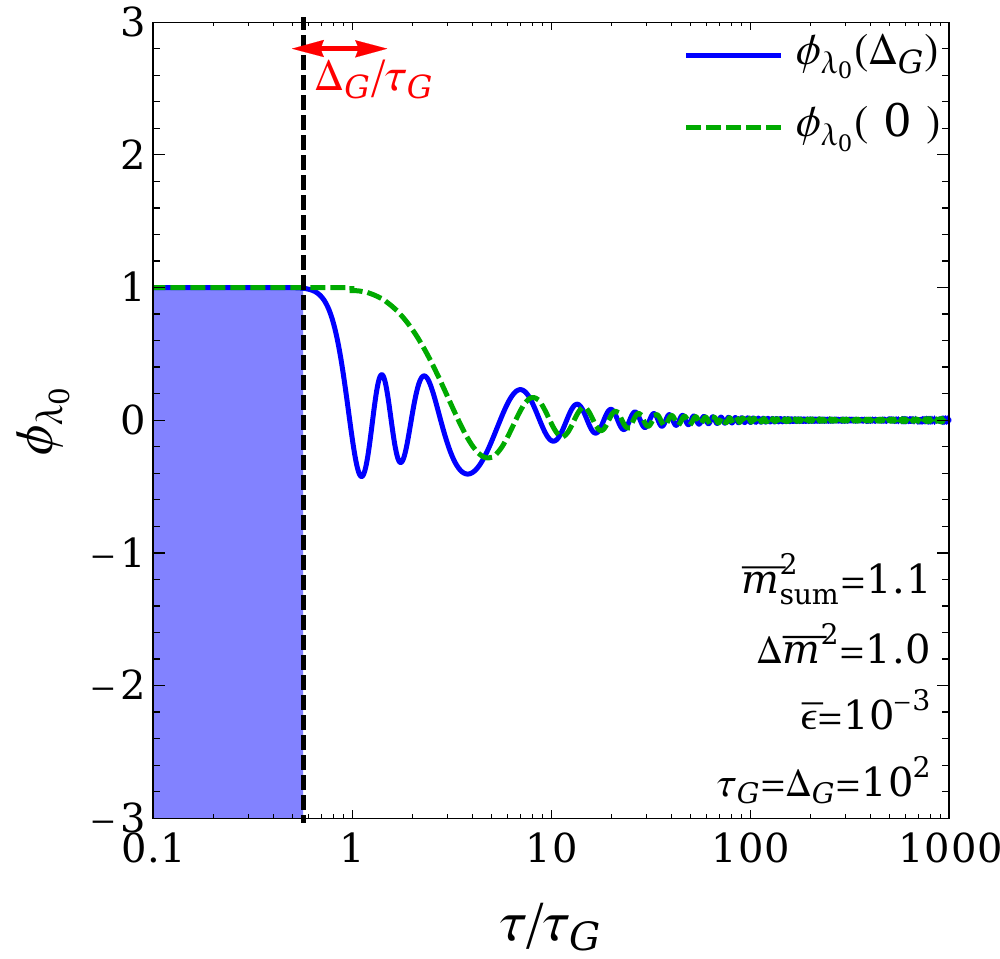} \  \ \ \includegraphics[width=0.31\textwidth,keepaspectratio]{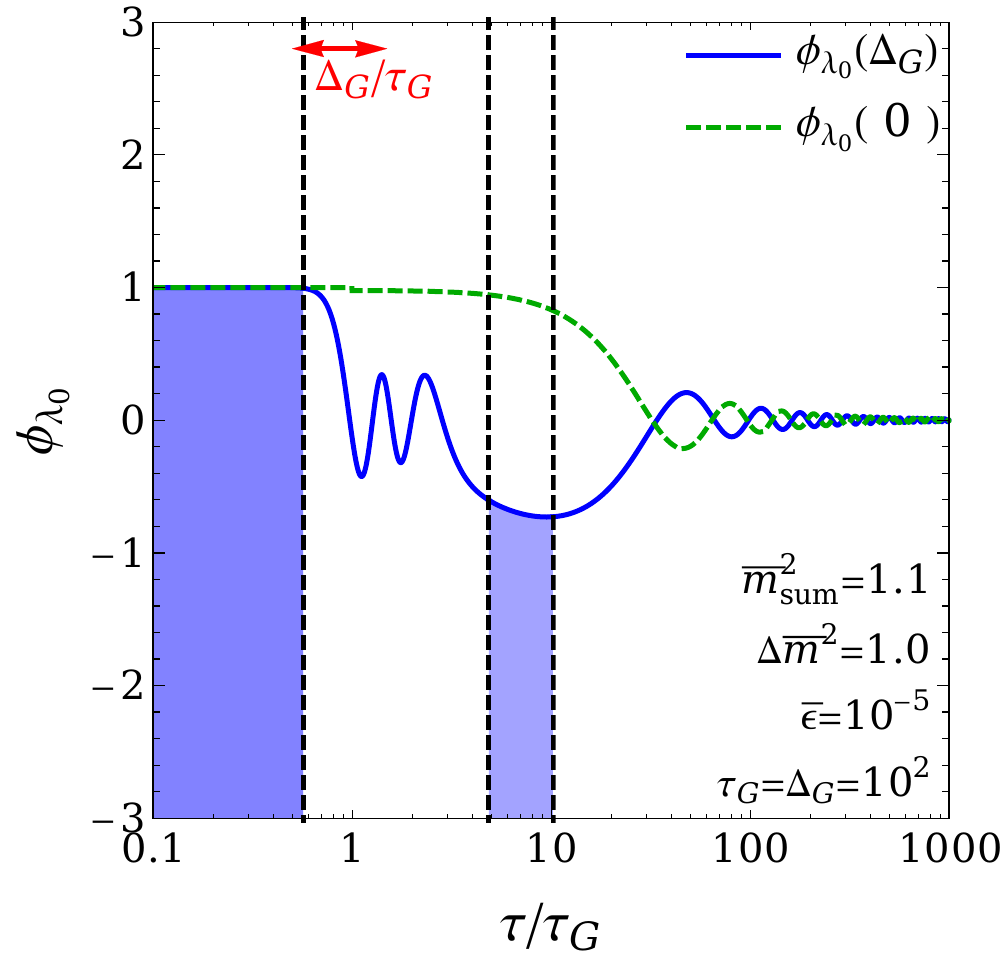} \ \ \ \includegraphics[width=0.31\textwidth,keepaspectratio]{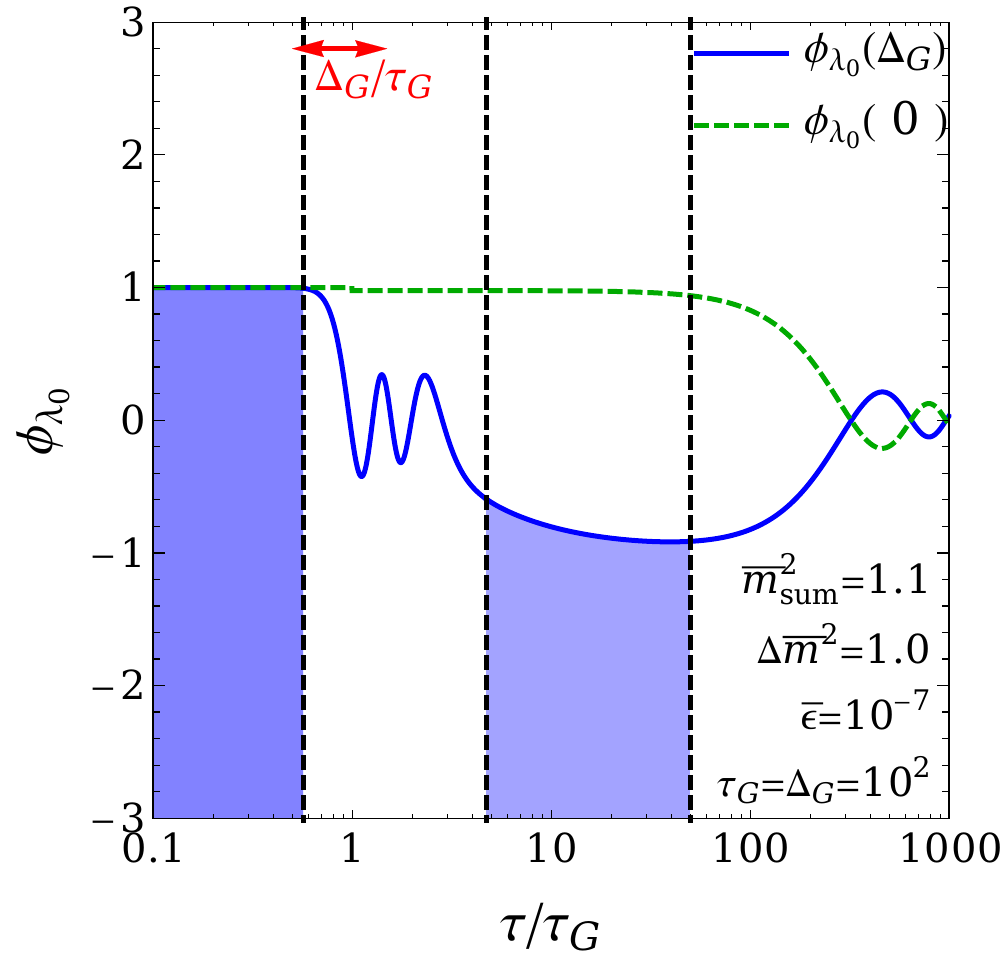}
\\ \mbox{} \\
\centering
\hspace*{-0.4cm}\includegraphics[width=0.329\textwidth,keepaspectratio]{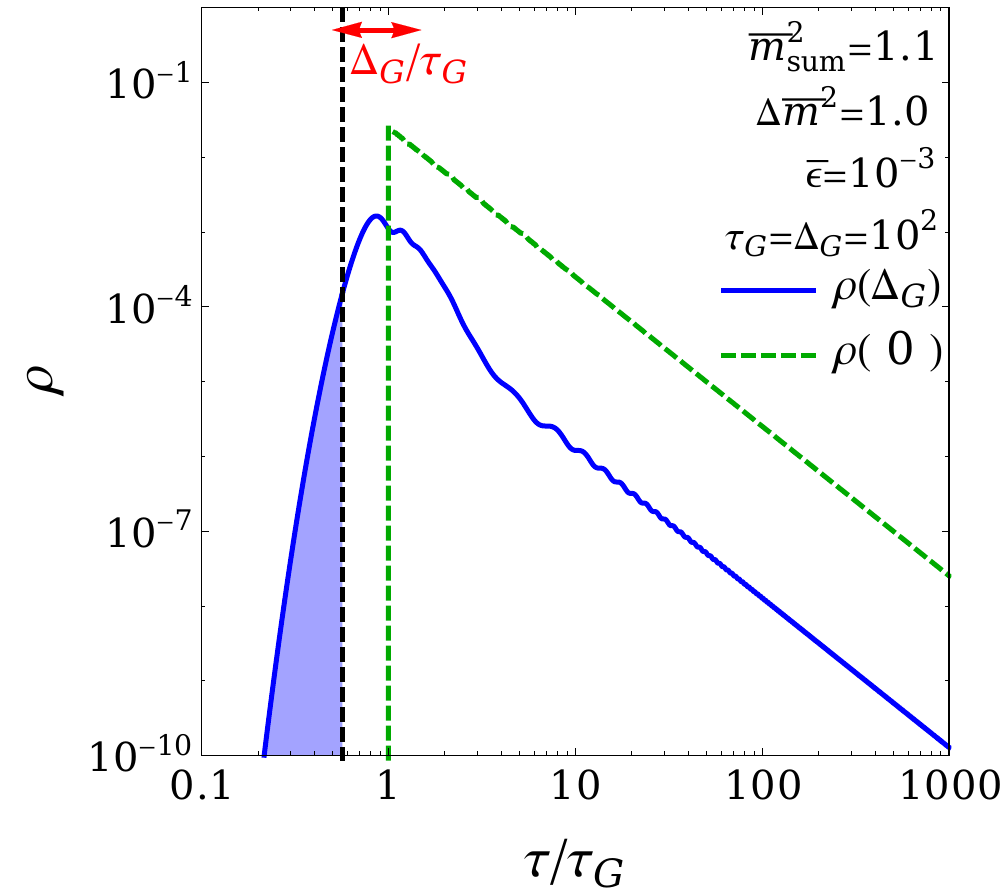} \includegraphics[width=0.329\textwidth,keepaspectratio]{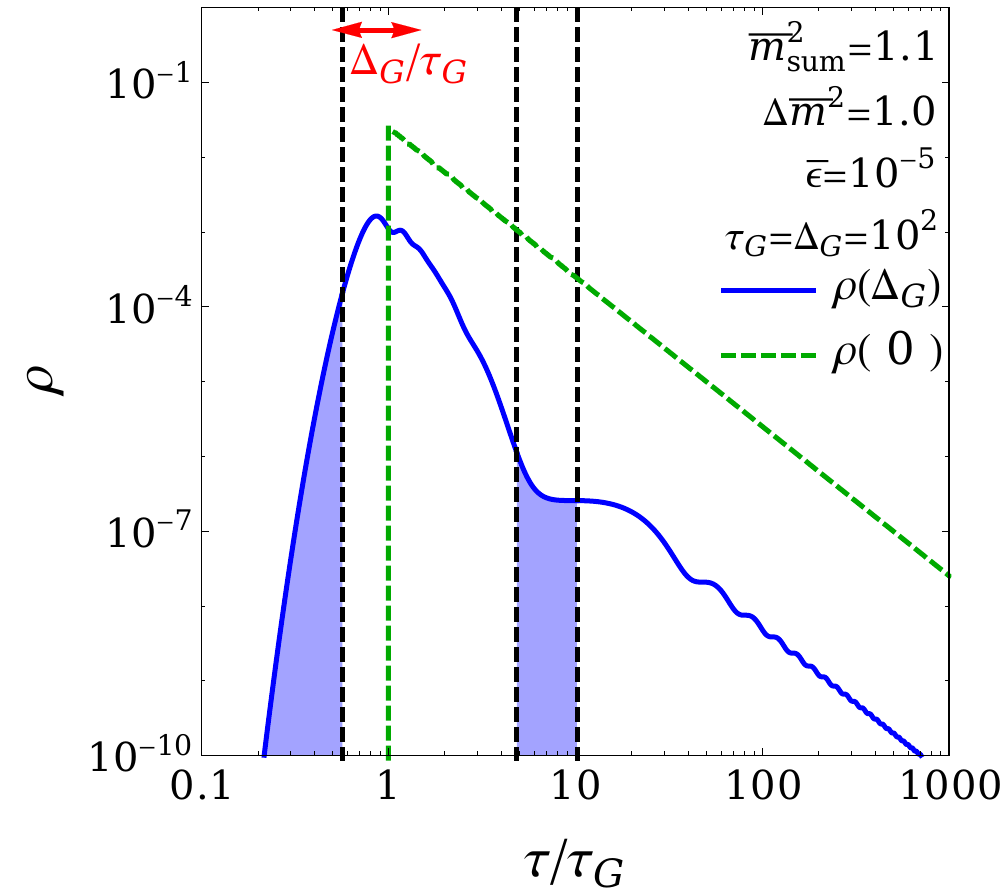} \includegraphics[width=0.329\textwidth,keepaspectratio]{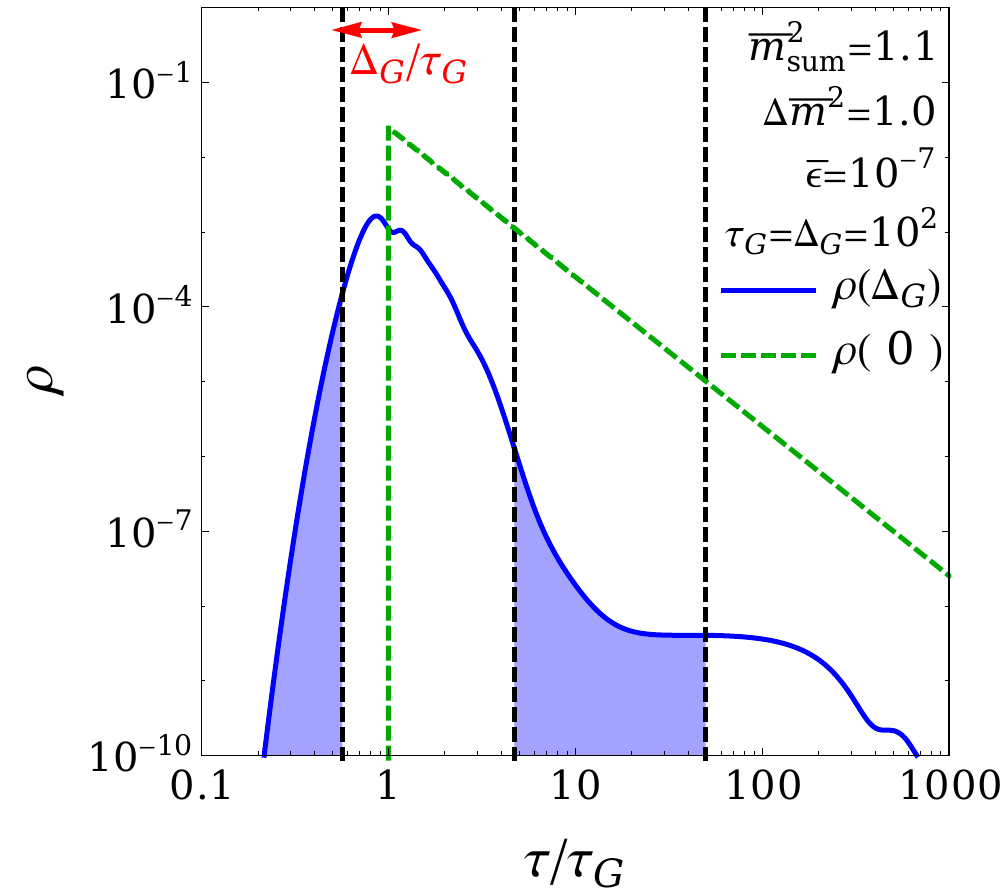}
\caption{The lighter field $\phi_{\lambda_0}$ (top row) and corresponding total energy density $\rho$ (bottom row), 
plotted as functions of time for situations (left to right) in which 
\mbox{$\xi\to 1$} (or \mbox{$\epsilon \equiv 1-\xi \to 0$}).
Note that the fields and corresponding energy densities are normalized so that \mbox{$\phi_{\lambda_0}=1$} for \mbox{$\tau\ll \tau_G$}.  
Time intervals during which the system is overdamped are shaded (blue), while the width $\Delta_G$ of the original mass-generating phase transition is also indicated (red).  As the mixing saturation increases towards its maximum value
(a point of enhanced symmetry for the system),
we see that a re-overdamped phase emerges with increasing duration.
For comparison purposes, superimposed on each plot are the field values and/or total energy densities that our system would have exhibited in
the \mbox{$\Delta_G\to 0$} limit (green);  recall that in this limit it is the energy density of the heavier field which dominates the total energy density.   
We thus see that the re-overdamped phase can easily persist throughout a large period of cosmological history, 
with
a phenomenology that is quite different from those of 
the traditional overdamped or underdamped phases.}
\label{fig:backforthfieldsrhos}
\end{figure*}

It is important to stress that unlike the original overdamped phase,
this {\it re}\/-overdamped phase of the theory
should not be associated with vacuum energy.
Indeed, at the moment of transition into the re-overdamped phase,
oscillatory behavior ceases but our field value re-enters the overdamped
phase carrying a non-zero velocity $\dot\phi$.
Thus, the re-overdamped phase is one in which our field value has a non-zero but steadily decreasing
first derivative. 
In this sense, then, the re-overdamped phase represents 
a truly different behavior for our field values --- one which is neither vacuum energy (constant field values)
nor matter (oscillatory field values), but something completely different which transcends our traditional expectations
for scalar fields in an expanding universe.

The re-overdamped phase tends to emerge in systems in which the mixing is close to its maximum
allowed value, with \mbox{$\xi\to 1$} (or \mbox{$\epsilon\equiv 1-\xi\to 0$}).
The resulting behavior for field values and energy densities in this limit 
are illustrated in Fig.~\ref{fig:backforthfieldsrhos}.
As expected, we see that
the re-overdamped phase has a phenomenology which is quite distinct
from those of the traditional overdamped or underdamped phases.

On the one hand,
this re-overdamped phase can be considered to be a mere ``transient''.   
After all, it cannot last forever:   as long as the field $\phi$ has non-zero mass at
late times, this phase must always give way to a final underdamped phase. 
In other words, the re-overdamped phase
can never represent the eventual end-state for our system as \mbox{$t\to\infty$}.
Moreover, even within this phase,
the non-zero initial field velocity $\dot\phi$ 
which characterizes its unique properties
always eventually dissipates to \mbox{$\dot\phi\approx 0$} as
long as this phase has sufficient duration.
In such cases, this phase eventually comes to resemble vacuum energy.
On the other hand, however, this attitude can be viewed as unnecessarily stringent.
Even though the re-overdamped phase must always 
eventually give way to one which is underdamped,
we have seen that it can,
depending on the specific model under study, 
be made to persist for arbitrarily long duration. 
Likewise, this phase can begin with  
arbitrarily large initial velocity $\dot\phi$.
Thus, in purely practical terms, this phase 
can be considered alongside ``vacuum energy'' and ``matter''
as an equally valid new behavioral phase that a scalar field can 
experience during a significant portion of cosmological evolution.

As a final comment, we remark that this re-overdamping phenomenon also leads to a resonance-like behavior
wherein quantities such as the late-time energy density experience a non-monotonic sensitivity to quantities 
such as $\Delta_G$.   This is because the behavior of our field $\phi$ during the re-overdamped epoch
depends very sensitively on the particular field velocity $\dot\phi$ 
that happened to exist at the precise moment the re-overdamping transition occurs.      
This gives rise to a periodic sensitivity of the final energy density to any quantity which affects the 
positioning of the re-overdamping transition moment.
However, it turns out that the mathematical conditions underlying this periodicity are essentially the same as
those underlying the parametric resonance discussed in Sect.~\ref{sec:TheParametricResonance};   
this makes sense as both concern the fitting of (half-)integral
numbers of field oscillations within the same eigenvalue pulse. 
Moreover, it turns out that the parametric resonance necessarily exists for all situations in which re-overdamping exists
(even though the converse is not true).
As a result, in such situations 
both features are inherent in the solutions to our equations of motion
and 
the effects of this re-overdamping-induced oscillation are difficult to disentangle
from those of the parametric resonance.

\FloatBarrier

\section{Phenomenological example: A~second axion \label{sec:Axion}}


As a phenomenological example of some of the main results of this paper,
let us consider the effects that might come from 
adding a second axion to the standard QCD axion theory.
As we shall see, the introduction of a second possible axion
can severely distort the physics normally associated
with the ordinary QCD axion, even
if the second axion is itself associated with a relatively heavy Peccei-Quinn scale 
or is only mildly coupled to the first.

Our theoretical set up is as follows.
We assume that $\phi_0$ and $\phi_1$ are both axion-like particles --- \ie,
the pseudo-Nambu-Goldstone bosons associated with the breaking of 
global symmetries $U(1)_0$ and $U(1)_1$ at energy scales $f_0$ and $f_1$, respectively.  
We assume that both fields are coupled to the QCD gauge fields  
via the chiral anomaly, with ${\cal O}(1)$ model-dependent coefficients $\xi_0$ and $\xi_1$.  
Moreover, 
in keeping with the assumptions of our toy model, 
we assume that at early times $\phi_0$ is massless 
while $\phi_1$ has a non-zero mass $M$ which possibly arises due to some earlier dynamics.
Eventually, non-perturbative QCD instanton effects 
give rise to a temperature-dependent effective potential $V(\phi_1,\phi_2)$ which takes
the form
\beq
   V(T) ~=~ \mu_\Lambda^4 \, h^2_{\rm QCD}(T)\, \left[ 1 - \cos \left(    
       {\xi_0 \phi_0\over f_0} 
       +{\xi_1 \phi_1\over f_1} 
       +\overline{\Theta}\right)\right]
\label{cosinepotential}
\eeq
where \mbox{$\mu_\Lambda^4\equiv g_s^2 \Lambda_{\rm QCD}^4 /( 32\pi^2)$},
where $h_{\rm QCD}(T)$ describes the temperature-dependence of this effective potential
as it turns on (as illustrated in Fig.~\ref{fig:axioncompare}), 
and where $\overline{\Theta}$ is the QCD theta-angle.
Thus we see that the  instanton-induced QCD confining 
phase transition plays the role of our mass-generating phase transition:  it occurs at time \mbox{$t_G= t_{\rm QCD}$} with
width \mbox{$\Delta_G=\Delta_{\rm QCD}=\sqrt{\pi} \sigma t_{\rm QCD}$}, and gives rise to 
an additional contribution
to the squared-mass matrix of our axion-like fields which at late times takes the form
\beq
    \mbar_{ij}^2 ~\equiv~ {\partial^2 V \over \partial \phi_i \partial \phi_j}\biggl|_{\langle \phi \rangle} ~=~ {\mu_\Lambda^4 \over f_i f_j}~
\eeq 
where we have set \mbox{$\xi_0=\xi_1=1$} for simplicity.

Given this setup, we thus have two 
mass eigenstates, $\phi_{\lambda_0}$ and $\phi_{\lambda_1}$, with 
late-time masses $\lambdabar_0$ and $\lambdabar_1$ respectively.
As we survey the range from \mbox{$M=0$} to \mbox{$M=\infty$},
we find that these late-time masses increase monotonically within the ranges
\beq
     0 ~\leq ~\lambdabar_0 ~\leq~ \mu_\Lambda^2/f_0~ 
\label{lam1range}
\eeq
and
\beq
               \mu_\Lambda^2 \sqrt{{1\over f_0^2} + {1\over f_1^2}} ~\leq ~\lambdabar_1 ~ \leq~ \infty~.
\label{lam2range}
\eeq
For reasons to be discussed shortly,
we shall identify the lighter mass eigenstate as our QCD axion, which means that we shall identify
$\lambdabar_0$ as the axion mass $m_a$.
We shall also define \mbox{$x\equiv \lambda_0 f_0/\mu_\Lambda^2$};  thus \mbox{$0\leq x\leq 1$}.
Note that in this section we are 
keeping $M$ explicit in our expressions;  thus, all quantities 
such as $\lambdabar_i$ have their true physical mass dimensions.  
Our goal is then to understand how the
late-time energy density associated with this field 
is influenced by the presence of the additional axion.

Note that there are two physically distinct ways in which to realize
a limit in which the second axion decouples from the first:  
either we can take \mbox{$M\to \infty$} and leave $f_1/f_0$ arbitrary, 
or     we can take \mbox{$f_1/f_0\to\infty$} and leave $M$ arbitrary.   
In the first case, we find that \mbox{$\xbar= 1$} regardless of the value of $f_1$.
In the second case, by contrast, we find that \mbox{$\xbar=1$} for all \mbox{$M\geq \mu_\Lambda^2/f_0$},
but \mbox{$\xbar= Mf_0/\mu_\Lambda^2< 1$} for all \mbox{$M < \mu_\Lambda^2/f_0$}.
Of course both decoupling limits smoothly merge together in the region for which
both $M$ and $f_1/f_0$ are taken to infinity.

\begin{figure*}[t]
\includegraphics[width=0.45\textwidth]{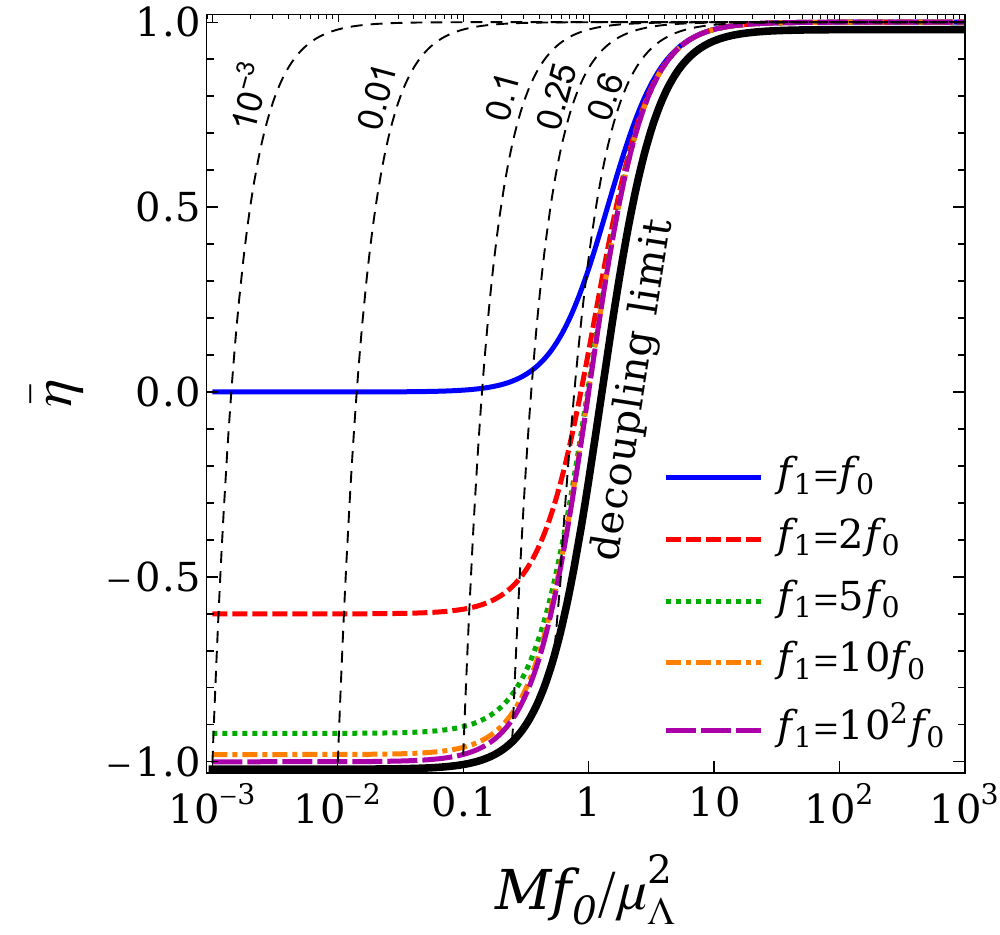}
\hskip 0.3 truein
\includegraphics[width=0.45\textwidth]{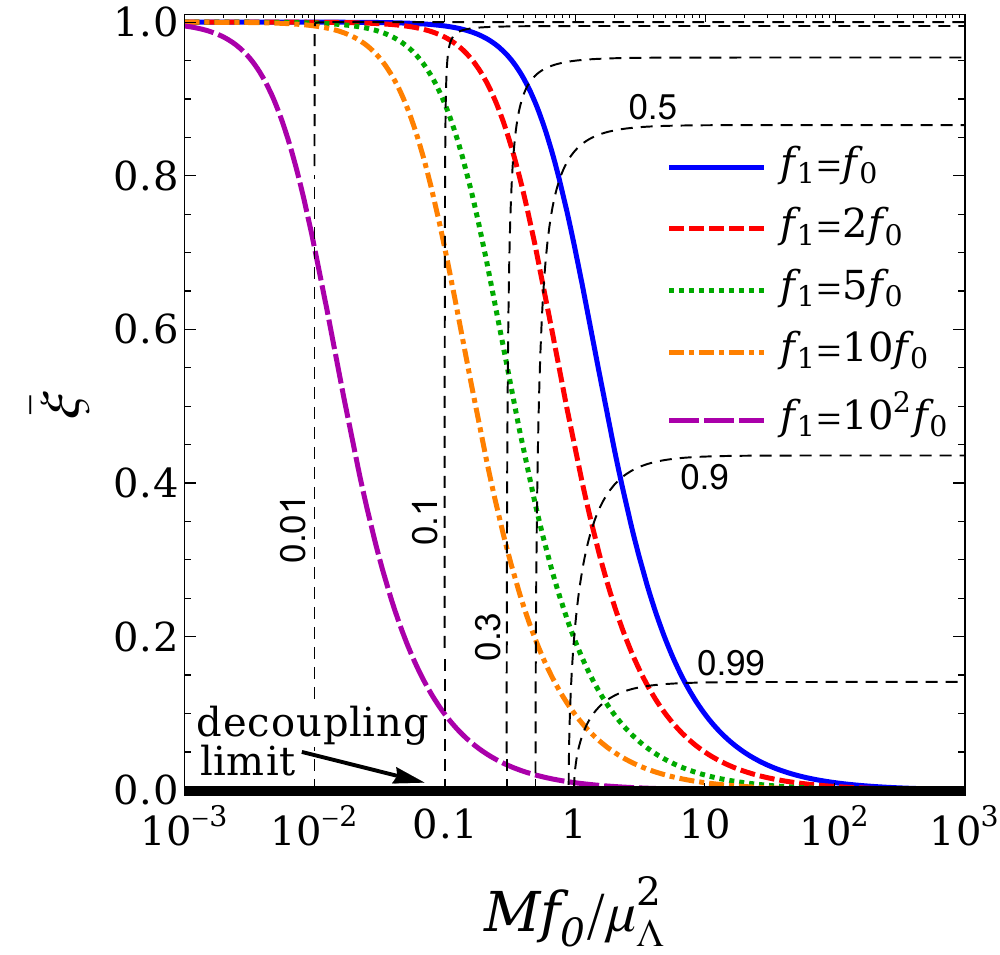}
\caption{The quantities $\etabar$ (left panel) and $\xibar$ (right panel), plotted as functions of $M$ for different values of $f_1/f_0$.
Superimposed on each plot are contours of $\overline{x}$ (dashed black lines).  The decoupling regions in each case correspond to taking \mbox{$f_1/f_0\to\infty$} for any $M$  as well as taking \mbox{$M\to\infty$} for any $f_1/f_0$;  these smoothly connect to form the curves labeled ``decoupling limit'' 
on each plot.  
Note that we have \mbox{$\xibar=0$} along the entire decoupling-limit line, as expected since the decoupling limit is one in which there
is no mixing between the two axions.  
By contrast, the decoupling-limit line surveys all possibilities of $\etabar$.
We also observe from either panel that  \mbox{$\overline{x}=1$} along the decoupling line 
for all \mbox{$M\geq \mu_\Lambda^2/f_0$}, while
\mbox{$\overline{x}<1$} along the decoupling line for all \mbox{$M< \mu_\Lambda^2/f_0$};  this is easier to see in the right panel but is true as a general statement
about the relationship between $M$ and $\overline{x}$ in the decoupling limit.  Thus,
an infinite $M$ implies that 
\mbox{$\overline{x}=1$} but \mbox{$\overline{x}=1$} implies
only that \mbox{$M\geq \mu_\Lambda^2/f_0$}.}
\label{fig:etavslambda0f0}
\end{figure*}

In order to quickly survey the expected phenomenologies of this two-axion system as functions
of \mbox{$\lbrace M,f_0,f_1\rbrace$} --- and also to understand the properties of these different decoupling limits ---
let us calculate the corresponding values of $\etabar$ and $\xibar$.
In terms of 
 \mbox{$\lbrace M,f_0,f_1\rbrace$}, 
these are given by 
\beq
 \etabar ~=~ \frac
 {  1 + (f_1/f_0)^2 \left\lbrack 2 (M f_0/\mu_\Lambda^2)^2 -1\right\rbrack }
 {  1 + (f_1/f_0)^2 \left\lbrack 2 (M f_0/\mu_\Lambda^2)^2 +1\right\rbrack } 
\eeq
and
\beq
 \xibar ~=~ \frac{1}{\sqrt{  1+ 2 (f_1/f_0)^2 (M f_0/\mu_\Lambda^2)^2 }}~.
\eeq
These quantities are shown in Fig.~\ref{fig:etavslambda0f0}.
Although
both decoupling limits result in \mbox{$\xibar\to 0$} (the absence of mixing), as expected,
we immediately see that the decoupling limit with \mbox{$f_1\gg f_0$} and arbitrary $M$ generally gives rise 
to 
all possible values of $\etabar$ in the range \mbox{$-1\leq \etabar\leq 1$}.  
By contrast, the decoupling limit with \mbox{$M\to\infty$} gives rise to \mbox{$\etabar=1$} regardless
of the value of $f_1/f_0$.

This is an important distinction because we have already seen in 
Sect.~\ref{sec:indiv}
that \mbox{$\etabar>0$} is the regime in which the majority of the energy density  
is found in the lighter field at late times, whereas \mbox{$\etabar<0$} is the regime
in which the majority of the energy density is transferred to the heavier field at
late times.
This behavior was in fact discussed explicitly below Eq.~(\ref{ratio}) in the limit of
an instantaneous phase transition.
Thus, it is only the decoupling limit with \mbox{$M\to\infty$} and arbitrary $f_1/f_0$
for which the energy density actually remains in the lighter field where it started,
as we would expect for a proper decoupling limit.

We shall therefore adopt the limit with \mbox{$M\to\infty$} and arbitrary $f_1/f_0$
as our benchmark against which to measure the effects that come from 
the presence of the second axion.
Note that in the \mbox{$M\to\infty$} limit, the physics of our system actually becomes 
 {\it independent}\/ of $f_1/f_0$;   we have already seen this behavior for
the specific quantities
$\etabar$ and $\xibar$ in Fig.~\ref{fig:etavslambda0f0},
but this feature indeed holds more generally for all properties of the system.

Although we have been considering 
the lighter mass eigenstate $\phi_{\lambda_0}$ as our QCD axion,
it is important to realize that 
the entire notion of a ``QCD axion'' no longer strictly applies.
For example, there is no mass eigenstate in our model whose mass is inversely
proportional to $f_0$ or $f_1$, as might be taken to characterize a QCD axion.
Likewise, there is no mass eigenstate in this model which solves the strong CP problem;
in particular, we see from Eq.~(\ref{cosinepotential}) that it is only the linear combination
\mbox{$\phi_0/f_0+\phi_1/f_1$} which dynamically relaxes against the strong CP angle $\overline{\Theta}$. 
However, in the \mbox{$M\to\infty$} decoupling limit, our model reduces to the standard single-axion
model:  the lighter mass eigenstate remains massless prior to the instanton-induced
phase transition, gathers a mass $\mu_\Lambda^2/f_0$ after this transition, and
solves the strong CP problem.
For this reason we shall continue to refer to the lighter mass eigenstate as our ``QCD axion'',
even in the presence of mixing.

Given this, we now calculate the late-time energy density associated 
with the lighter (QCD) axion as a function of the lighter (QCD) axion mass \mbox{$\xbar\equiv \lambdabar_0 f_0/\mu_\Lambda^2$},
normalized to our decoupled benchmark limit with \mbox{$M=\infty$} (for which \mbox{$\xbar=1$}).
(As a technical point, note that we describe this benchmark through the condition \mbox{$M=\infty$} rather
than the condition \mbox{$\xbar =1$}, since the former implies the latter regardless of $f_1/f_0$ whereas
the latter fails to imply the former when \mbox{$f_1/f_0\to\infty$}.) 
In all cases we take \mbox{$t_G=t_{\rm QCD}$} 
and \mbox{$\Delta_G=\Delta_{\rm QCD}$} to have the specific fixed values that correspond to the instanton-induced 
QCD phase transition, as plotted in Fig.~\ref{fig:axioncompare}.

\begin{figure*}[ht]
\includegraphics[width=0.32\textwidth]{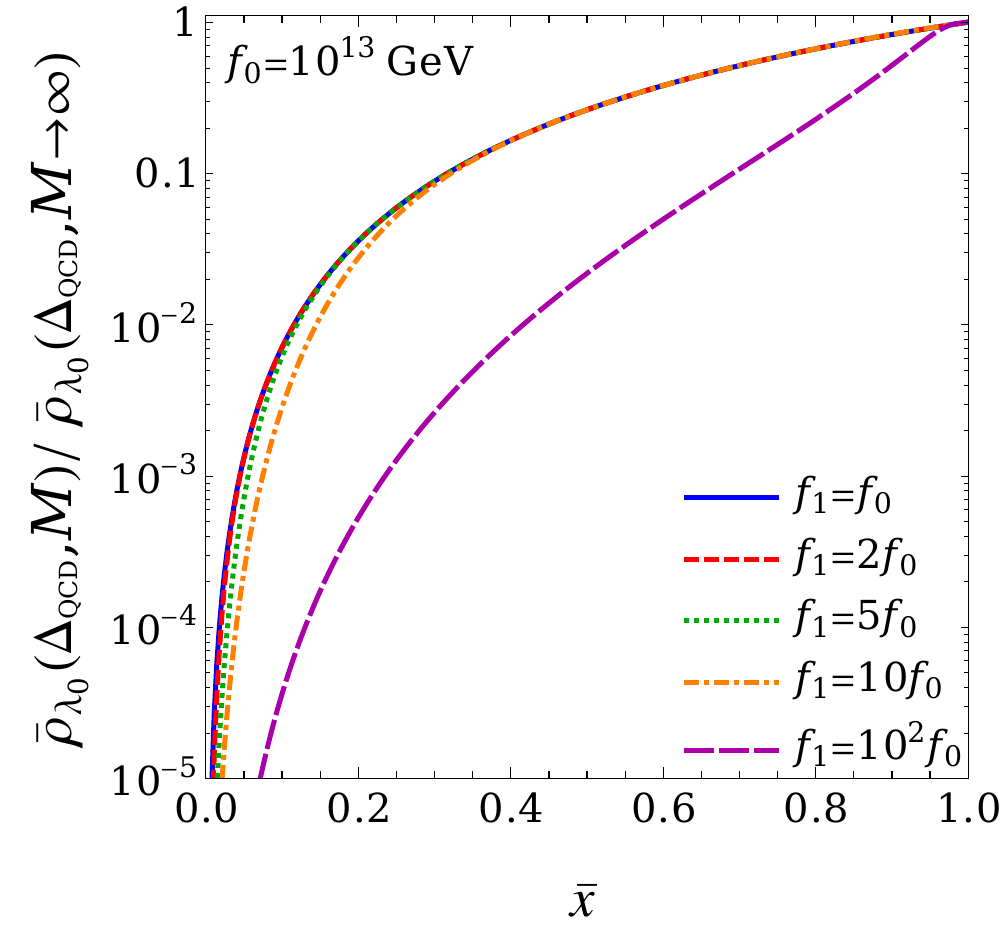}
\includegraphics[width=0.32\textwidth]{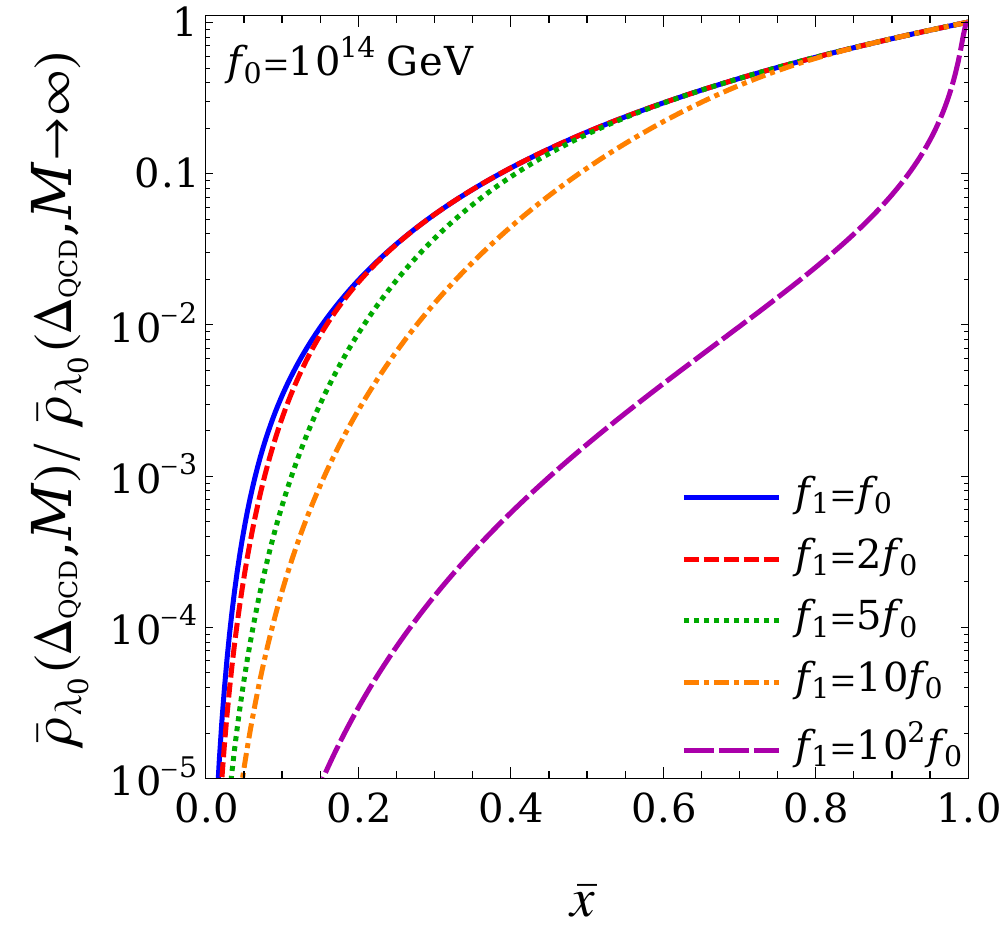}
\includegraphics[width=0.32\textwidth]{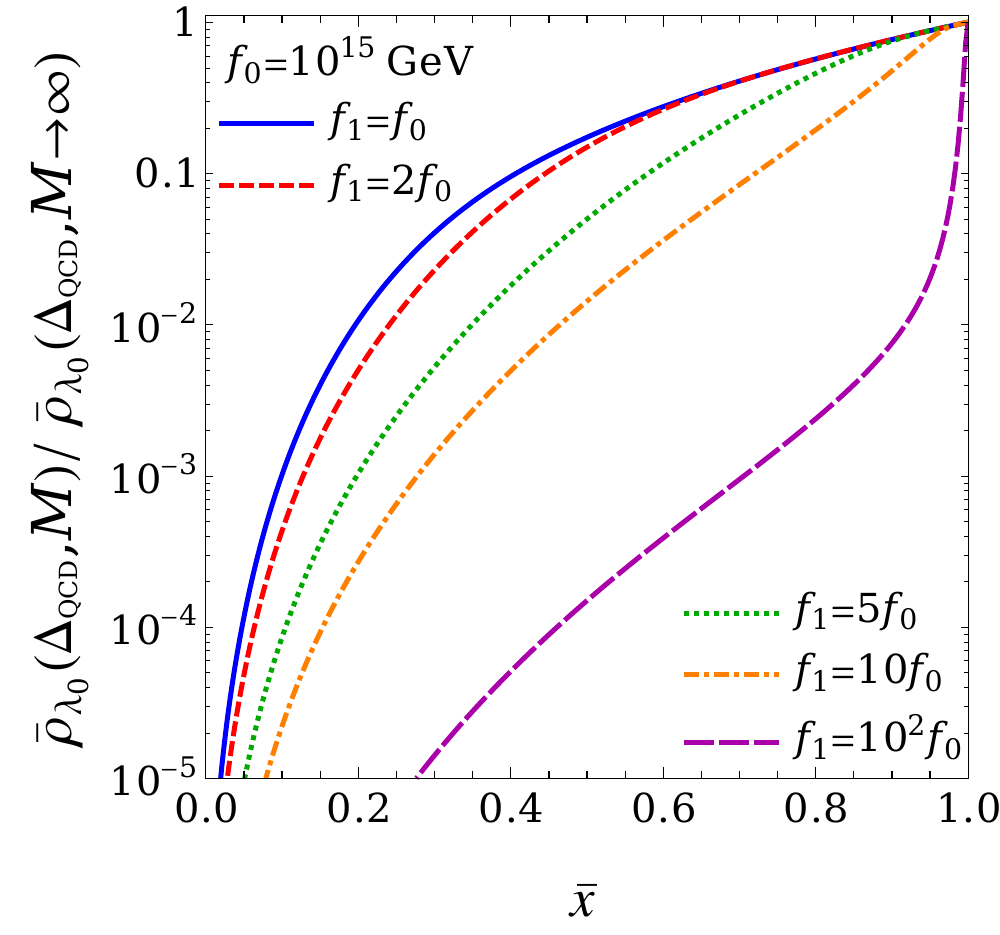}
\caption{The late-time energy density $\rhobar_{\lambda_0}(\Delta_{\rm QCD},M)$ associated with the QCD axion field $\phi_{\lambda_0}$,
plotted as a function of the late-time axion mass \mbox{$\overline{x}\equiv \lambdabar_0 f_0/\mu_\Lambda^2$}
and normalized to what it would have been in the decoupling limit in which the second axion plays no role. 
These plots thus illustrate the effects of the non-zero mixing between our two axion fields in the presence of a non-zero
QCD phase transition width $\Delta_{\rm QCD}$.
Note that although each normalization factor $\rhobar_{\lambda_0}(\Delta_{\rm QCD},M\to\infty)$ 
is in principle different for each different $f_1/f_0$ curve,
physics in the decoupling limit \mbox{$M\to\infty$} is insensitive to the value of $f_1/f_0$.
Thus all curves in each panel share the same normalization factor.}
\label{fig:firstset}
\vskip 0.3 truein
\includegraphics[width=0.32\textwidth]{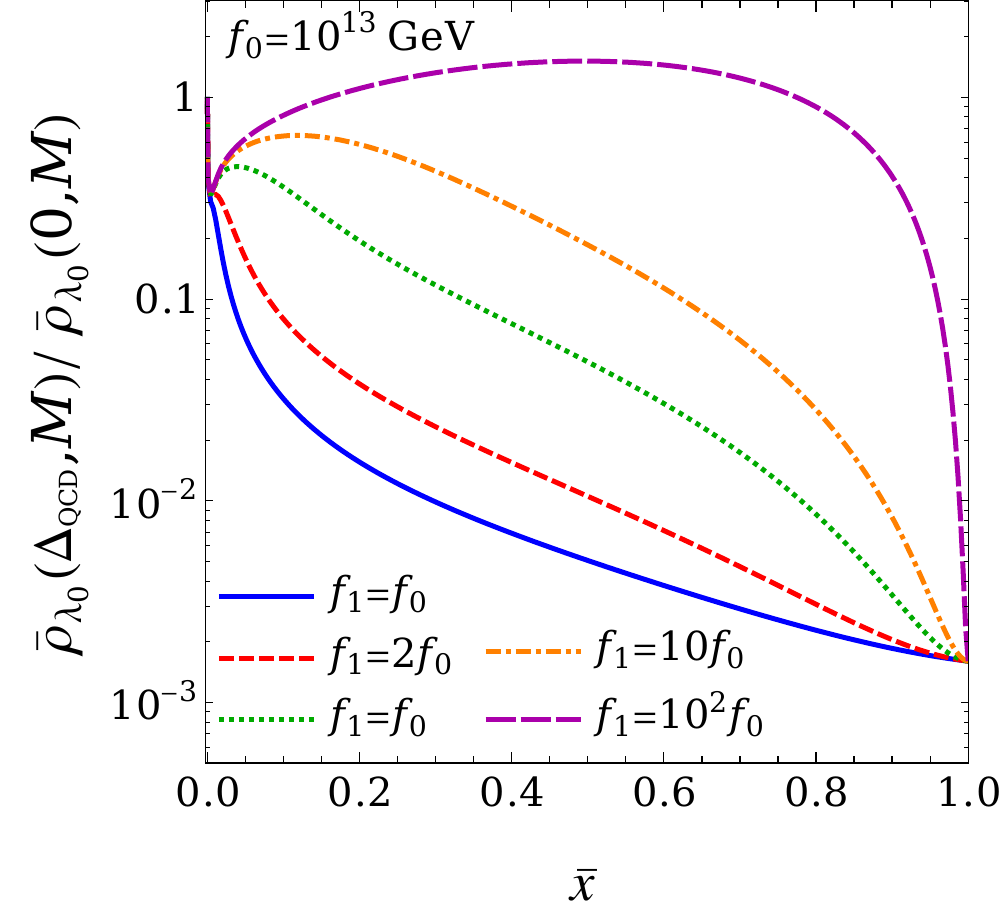}
\includegraphics[width=0.32\textwidth]{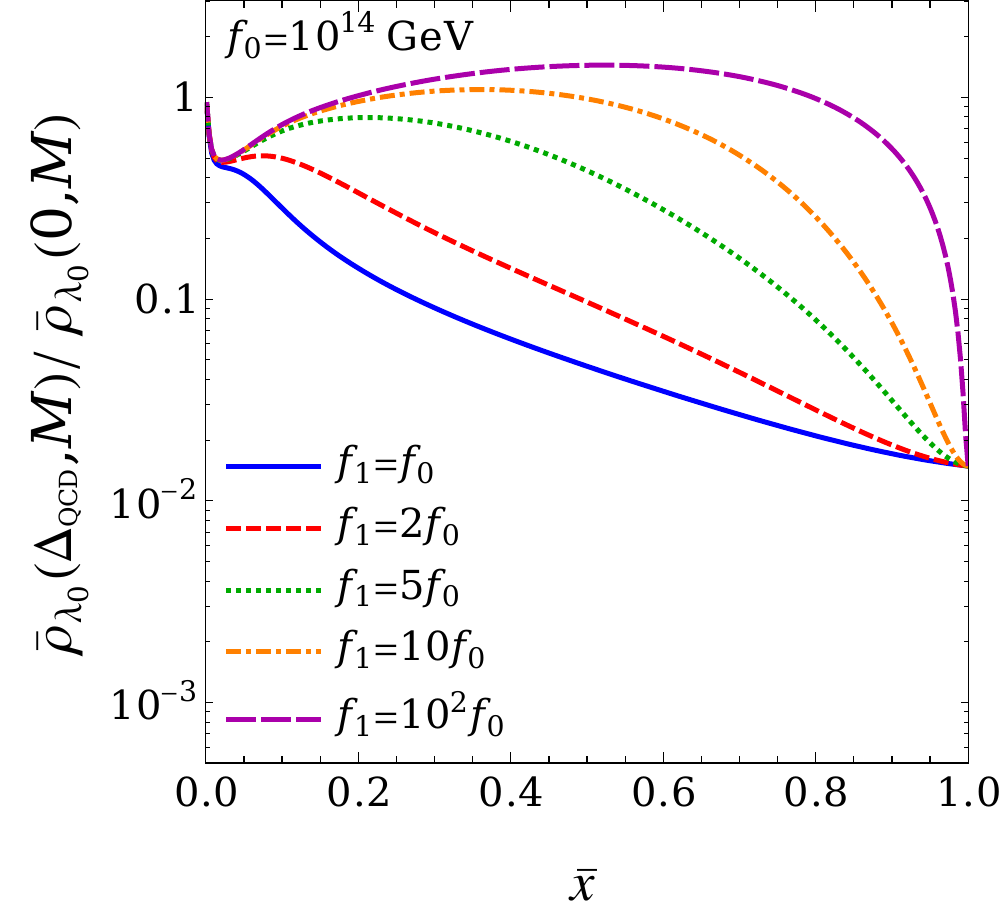}
\includegraphics[width=0.32\textwidth]{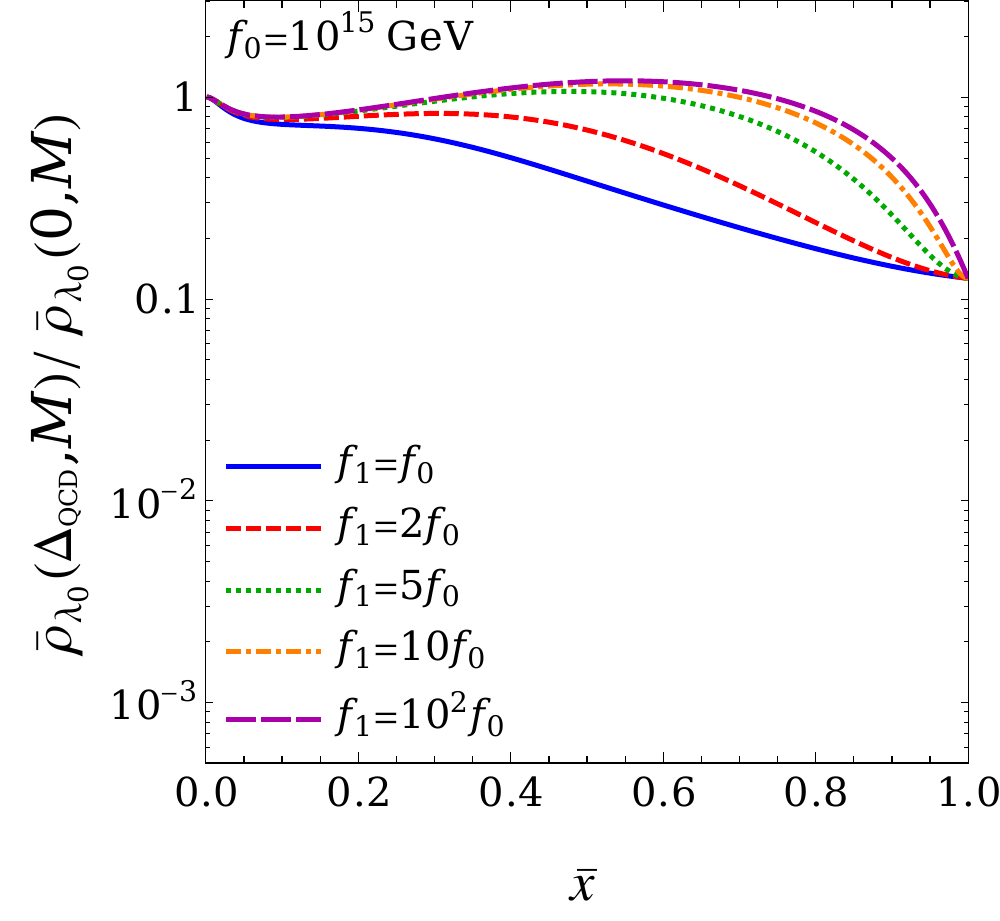}
\caption{Same as Fig.~\ref{fig:firstset} except that the late-time energy densities 
$\rhobar_{\lambda_0}(\Delta_{\rm QCD},M)$ are now normalized to the values that they would have 
had if the QCD phase transition had been treated as instantaneous.
These plots thus illustrate the effects that emerge due to the actual non-zero width of the QCD phase transition ---
effects which survive even in the \mbox{$M\to\infty$} decoupling limit (for which \mbox{$\xbar \to 1$}).
Note that unlike the plots in Fig.~\ref{fig:firstset}, here the normalization factors 
$\rhobar_{\lambda_0}(0,M)$ are different for each $f_1/f_0$ curve and
in fact even vary with $\xbar$ {\it along}\/ each curve.}
\label{fig:secondset}
\end{figure*}

Our results are shown in Fig.~\ref{fig:firstset}.
We see from Fig.~\ref{fig:firstset} that the late-time energy density associated with the QCD axion 
is suppressed relative to what would have been expected in the \mbox{$M\to\infty$} decoupling limit.
Indeed, this suppression is significant, occasionally stretching over many orders of magnitude, 
and arises regardless of the mass $\xbar$ of the QCD axion,
its corresponding Peccei-Quinn scale $f_0$, or the Peccei-Quinn scale $f_1$ associated with the
second axion.
Indeed, none of this would have occurred were it not 
for the non-zero width of the instanton-induced QCD phase transition.
This then demonstrates the phenomenological relevance of this non-zero width, and the need
to incorporate these kinds of effects in studies of multi-axion theories.

Note that this suppression of the late-time energy density
due to the presence of the second axion field persists {\it even in the limit for which \mbox{$f_1\gg f_0$}}\/.
However, this somewhat counter-intuitive result should not come as a surprise.
As we know, the limit \mbox{$f_1/f_0\to\infty$} corresponds to one in which \mbox{$\xibar\to 0$} --- \ie, one in which 
the late-time mixing vanishes.  However, this does not imply that the mixing is small
throughout the time evolution --- in fact, both $\theta$ and its time derivative can be quite sizable {\it during}\/ the 
mass-generating phase transition.
Indeed, this is particularly true for the relatively  
small values of $\xbar$ which correspond to \mbox{$\etabar <0$}.
Thus, it is {\it during}\/ the mass-generating phase transition that the bulk of the energy density 
can be transferred to the second axion --- even when \mbox{$f_1\gg f_0$}!
This is therefore an effect
that arises as a direct consequence of
the non-zero width of the 
instanton-induced QCD phase transition and its associated
timescale.

Interesting as these results are, they only address the issue 
of how our late-time energy density compares with what would arise 
in the decoupled case --- \ie, the case with no mixing (\mbox{$\xibar=0$}).
In this paper, however, we have also repeatedly studied another comparison, namely to the case of an {\it instantaneous}\/ phase transition --- \ie, 
the case in which we imagine our phase transition to have \mbox{$\Delta_{\rm QCD}=0$}.
The corresponding results are shown in Fig.~\ref{fig:secondset}.

Once again, it is not difficult to understand the features in these plots.
First, let us discuss the endpoints at \mbox{$\xbar=1$}.
Of course, these endpoints correspond to the decoupling limit (because \mbox{$\xbar=1$} implies \mbox{$\xibar=0$}, as we have seen in Fig.~\ref{fig:etavslambda0f0});
thus the physics at these endpoints includes the effects from only the lighter (QCD) axion itself.
As we see from Fig.~\ref{fig:secondset}, even in this limit the resulting late-time energy density is significantly suppressed by factors ranging from 5 to 500
as $f_0$ ranges from $10^{15}$~GeV to $10^{13}$~GeV.
Thus the effects of properly accounting for the non-zero width $\Delta_{\rm QCD}$
of our instanton-induced phase transition 
are extremely important, even for a single QCD axion! 

Moving away from the decoupling limit towards smaller values of $\xbar$,
we now begin to incorporate the additional effects of the mixing with the second axion field.
As we see from 
Fig.~\ref{fig:secondset}, 
these effects {\it enhance}\/
the resulting late-time energy density above what it would have been 
in the decoupling limit.  In so doing, they can  even possibly 
overcome the above suppression completely --- a prospect 
that depends on the value of $f_1/f_0$  ---
and thereby potentially {\it increase}\/ the late-time energy density 
above what it would have been for an instantaneous phase transition!
This enhancement ultimately reflects the same physics that we have already seen
in Figs.~\ref{fig:figname1} and \ref{fig:figname2} for small mixing saturations in the \mbox{$t_G\approx t^{(0)}_{\zeta}$} region.
Indeed, we see from Fig.~\ref{fig:etavslambda0f0} that larger values of $f_1/f_0$ correspond
to smaller mixing saturations $\xibar$.   

Finally, we can also understand why these curves all begin at $1$ for \mbox{$\xbar=0$}.
At this endpoint of these plots, 
our axion fields remain overdamped during the mass-generating phase transition.
They thus remain insensitive to the transition width $\Delta_{\rm QCD}$.

\begin{figure}[t]
\includegraphics[width=0.45\textwidth]{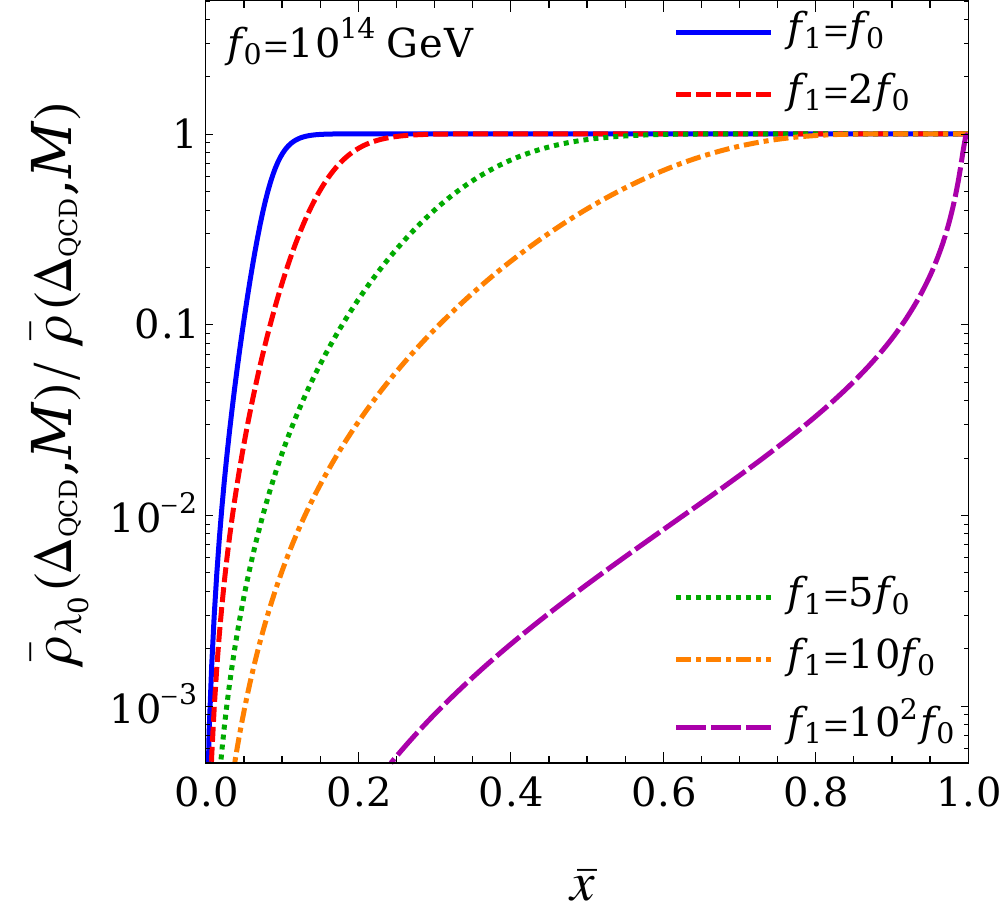}
\caption{The late-time fraction $\rhobar_{\lambda_0}/\rho$ of the total late-time energy 
density which is associated with lighter axion field, plotted as a function of $\xbar$.
We see that this fraction generally decreases as $f_1/f_0$ increases, and
also increases as a function of $\xbar$.}
\label{fig:fraction}
\end{figure}

We can also consider the fraction of the {\it total}\/ energy density 
of our two-axion system 
which is associated with the lighter axion field.
These results are shown in Fig.~\ref{fig:fraction}.
In general, we see that as $f_1/f_0$ increases,
an increasing fraction of the total energy density is associated with the heavier axion field.
However, 
as \mbox{$\xbar\to 1$},
we see that virtually all of the total energy density can be associated with the 
lighter axion field. 

\begin{figure*}[bth]
\includegraphics[width=0.40\textwidth]{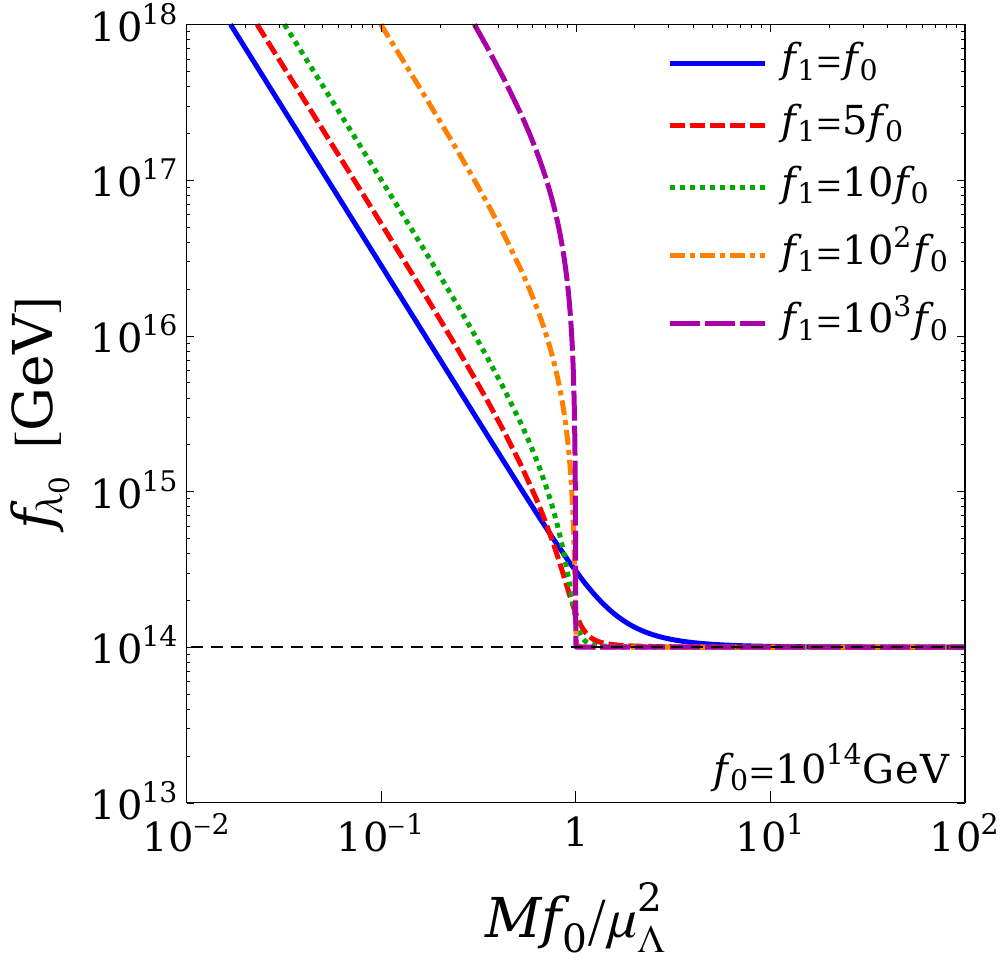}
\hskip 0.3 truein
\includegraphics[width=0.40\textwidth]{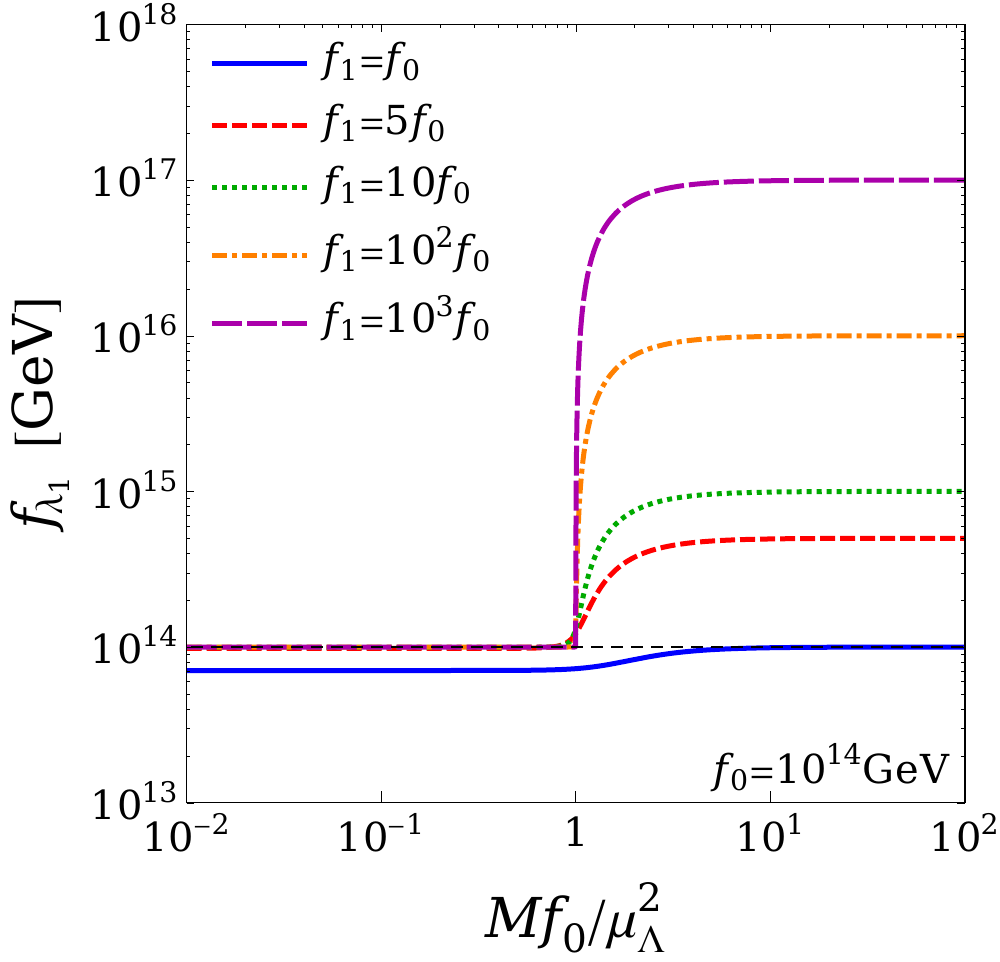}
\caption{The effective coupling scales $f_{\lambda_0}$ and $f_{\lambda_1}$ associated
with our axion mass eigenstates $\phi_{\lambda_0}$ and $\phi_{\lambda_1}$,
plotted as functions of $M f_0/\mu_\Lambda^2$.  We see that these scales
are never significantly smaller than the $f_i$ coupling scales associated with our
original unmixed states, and that either one or the other typically takes a value 
which significantly exceeds this when \mbox{$f_1/f_0\gg 1$}.}
\label{fig:couplings}
\end{figure*}

As a final comment, we remark that it is the combination \mbox{$\phi_0/f_0+\phi_1/f_1$} which not only
solves the strong-CP problem but which also couples to  
gluons.  Depending on the particular model under study, this combination may also couple
to photons and other Standard-Model (SM) particles.
We can therefore decompose this combination into our late-time mass eigenstates 
in order to determine the effective mass scales ${f}_{\lambda_0}$ and
${f}_{\lambda_1}$ 
that govern the relevant SM couplings 
of our mass eigenstates at any moment in time:
\beq
          { \phi_0 \over f_0} + 
          { \phi_1 \over f_1}  ~\equiv~
          { \phi_{\lambda_0} \over {f}_{\lambda_0}} + 
          { \phi_{\lambda_1} \over {f}_{\lambda_1}}~.
\eeq
The late-time values of these effective coupling scales are plotted in Fig.~\ref{fig:couplings}.

Several features of these plots are worthy of note. 
For example,
we observe that in all cases 
the $\overline{f}_{\lambda_i}$ coupling scales associated with our mass
eigenstates are never significantly smaller than the $f_i$ coupling scales associated with our
original unmixed states.   Thus neither of our mass eigenstates will decay too rapidly to SM 
states, a situation which might potentially have resulted in phenomenological difficulty.
Likewise,
combining the results in Fig.~\ref{fig:couplings} with those of Figs.~\ref{fig:etavslambda0f0} and \ref{fig:fraction},
we see that there 
are a variety of different phenomenological situations which can emerge, depending on the underlying
parameters in our model.

For example, let us consider the case with \mbox{$Mf_0/\mu_\Lambda^2 \approx 0.4$} and \mbox{$f_1\approx 5 f_0$}.   We immediately learn from  
Fig.~\ref{fig:couplings} that our QCD axion coupling scale is \mbox{$f_{\lambda_0}\approx 10^{15}$~GeV}, which is a full order of magnitude
higher than the value that it would have had in the decoupling limit.
This implies that the QCD axion has an enhanced invisibility  as assessed through its interactions with SM states.
However, we see from Fig.~\ref{fig:etavslambda0f0} that these parameter combinations correspond to \mbox{$\xbar \approx 0.4$},
whereupon we see from Fig.~\ref{fig:fraction} that this in turn corresponds to an energy-density fraction \mbox{$\rhobar_{\lambda_0}/\rhobar \approx 80\%$}.
This is thus a situation in which our QCD axion is invisible but nevertheless carries the bulk of the total late-time energy density.

A somewhat different situation emerges if we keep \mbox{$f_1\approx 5 f_0$} but now take \mbox{$M f_0/\mu_\Lambda^2 \approx 0.1$}.
In this case, we learn from 
Fig.~\ref{fig:couplings} that our QCD axion coupling scale is even higher --- \mbox{$f_{\lambda_0}\approx 10^{17}$~GeV} ---
which suppresses its couplings to SM states by three orders of magnitude compared with the decoupling limit.
However, we see from Fig.~\ref{fig:etavslambda0f0} that these parameter combinations now correspond to \mbox{$\xbar\approx 0.1$},
whereupon we see from Fig.~\ref{fig:fraction} that this in turn corresponds to an energy-density fraction \mbox{$\rhobar_{\lambda_0}/\rhobar \approx 3\%$}.
This is thus a situation in which our QCD axion is invisible and carries almost {\it none}\/ of the total energy density --- all because
of its mixing with the second axion in the presence of a non-zero phase transition width!
 
The opposite situation, of course, emerges for large $M$.
For example, with \mbox{$M f_0/\mu_\Lambda^2\approx 10$} we find that the physics of this model is roughly insensitive to the value of $f_1/f_0$;
we find \mbox{$\xbar \approx 1$} and 
\mbox{$\rhobar_{\lambda_0}/\rhobar \approx 100\%$}.
This, of course, signifies nothing but the approach to our decoupling limit.
 However, as long as we are not precisely {\it at}\/ the decoupling limit,
we still find from Fig.~\ref{fig:firstset} that the energy density associated with this lighter axion field remains considerably suppressed
compared with what we would have found at the full decoupling limit.
Thus in this sense our QCD axion has regular couplings to SM states
but nevertheless carries relatively little total energy 
density.

Needless to say, our goal in this section has not been to propose a complete phenomenological model of axion physics.  
Rather, it has merely been to illustrate the rich implications that can emerge for a QCD axion
in the presence of mixing with a second axion when the non-zero width of the QCD phase transition is properly taken into account.
As we have seen, the total energy density of such a two-axion system can be 
significantly suppressed 
relative to what might have been expected in the decoupling limit,  
and the individual energy densities associated with each axion can be severely distorted.
Indeed, this lesson has been one of the primary themes of this paper.

\FloatBarrier

\section{Beyond two fields\label{sec:BeyondTwoFields}}


\FloatBarrier

In Sect.~\ref{sec:AToyModel}, a toy model was constructed in order to study the effects
that arise in multi-scalar models due to the simultaneous presence of two key ingredients:
mass-generating phase transitions with finite widths, and non-zero mixing between
the different components.  For the sake of simplicity, our toy model consisted of only two scalar fields,
as this is the minimum number that could be chosen for our purposes.
However there is no reason that we are limited to two fields, and indeed the number of fields may be extended
arbitrarily.  Therefore, in order to gain a quick sense of what possibilities might emerge with an increased number of
fields, we shall now briefly consider several aspects of the case with three fields.
A more general study of the $N$-field case will be presented in Ref.~\cite{toappear}.

In analogy with Eq.~(\ref{massmatrix}), we shall begin by
assuming a rescaled (dimensionless) mass matrix $\mathcal{M}^2(t)$ which takes the form
\beq
\mathcal{M}^2(\tau) ~=~ \begin{bmatrix}[1.2] 
          m_{00}^2(\tau) & m_{01}^2(\tau) & m_{02}^2(\tau)\\ 
          m_{01}^2(\tau) & 1 + m_{11}^2(\tau) & m_{12}^2(\tau) \\ 
          m_{02}^2(\tau) & m_{12}^2(\tau) & 4 + m_{22}^2(\tau)
        \end{bmatrix} ~.
\label{N3matrix}
\eeq
Here \mbox{$\tau\equiv M t$}
(where $M$ is the overall mass parameter which has been scaled out, as in Sect.~\ref{sec:AToyModel}),
and we assume that all components
have a common time-dependence associated with a mass-generating phase transition whose properties
are exactly those of our two-component model.
Note that the diagonal components of this matrix
imply that our three fields have masses $0$, $M$, and $2M$
at early times prior to the phase transition.
These values are chosen for simplicity,
and are also motivated by
the Kaluza-Klein masses that might result in a theory with a flat extra dimension (in which case we would identify $M$
as the inverse length of this dimension).

\begin{figure}[t]
\begin{center}
\hspace*{-0.4cm}\includegraphics[width=0.45\textwidth,keepaspectratio]{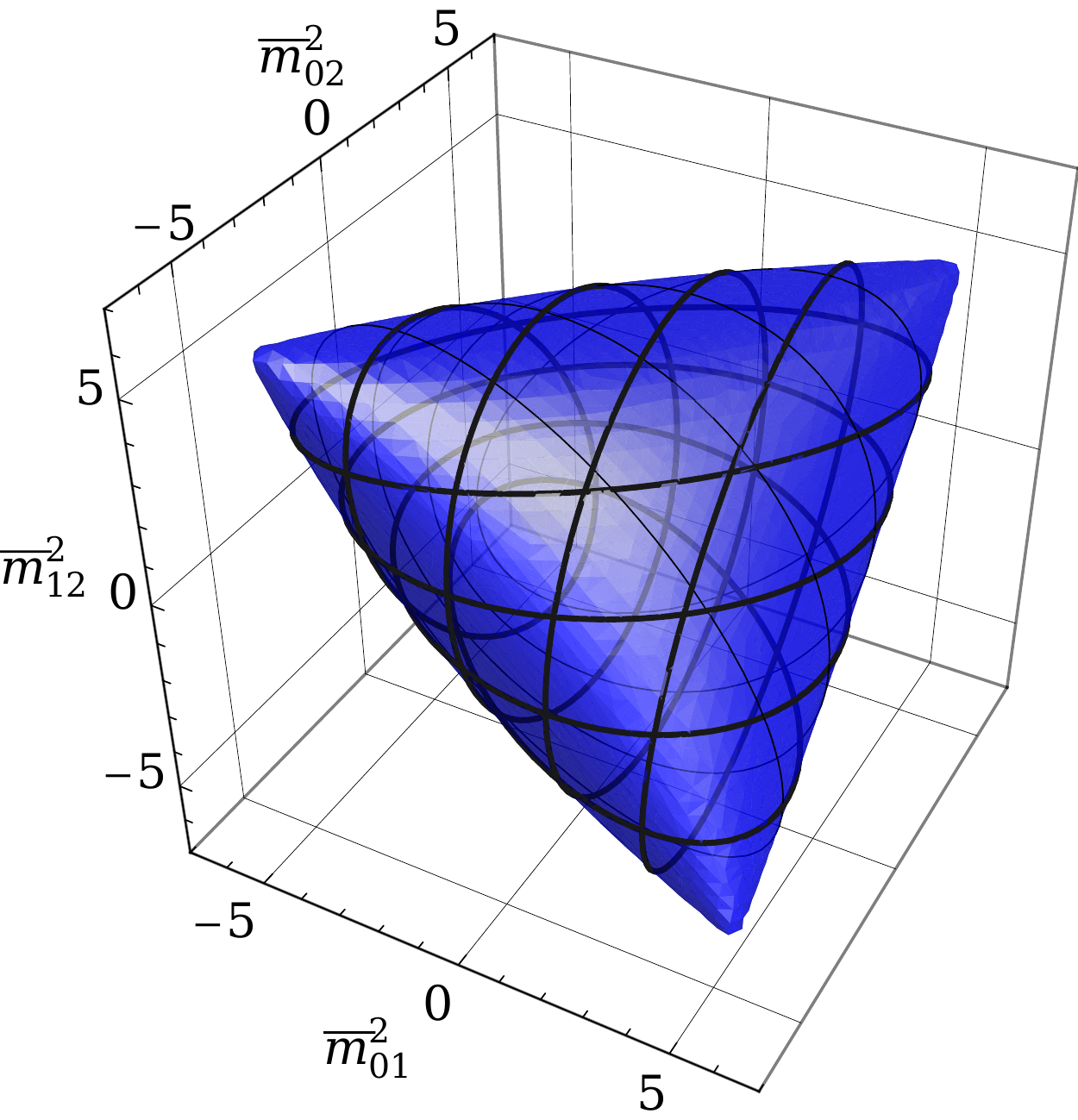}
\end{center}
\caption{
The region within the three-dimensional mixing space parametrized by
\mbox{$\lbrace \mbar^2_{01}, \mbar^2_{02}, \mbar^2_{12}\rbrace$} for which 
the mass matrix $\calM^2$ in Eq.~(\ref{N3matrix}) is positive-semidefinite.
For this plot we have taken
\mbox{$\overline{m}^2_{00}=5$}, \mbox{$\overline{m}^2_{11}=5$}, and \mbox{$\overline{m}^2_{22}=1$}.} 
\label{fig:posdefregion}
\end{figure}

\begin{figure}[t]
\centering
\hspace*{-0.4cm}\includegraphics[width=0.45\textwidth,keepaspectratio]{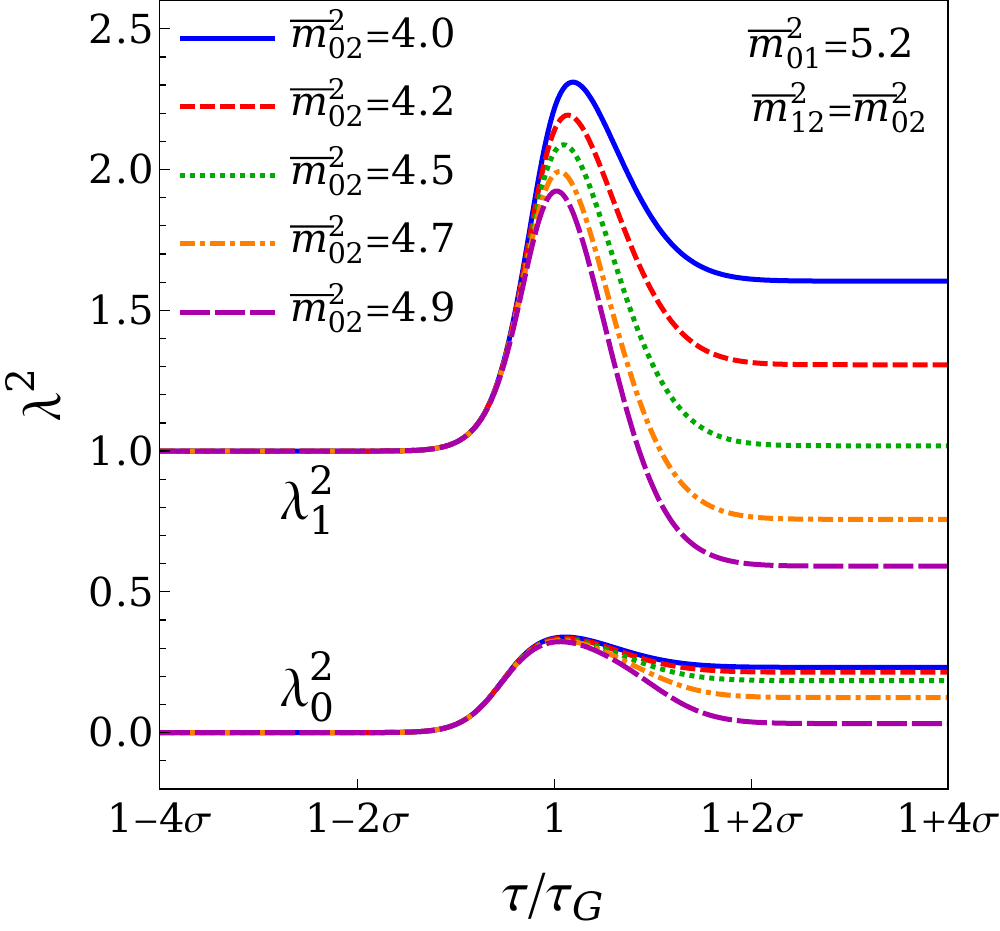}
\caption{The two smaller mass eigenvalues $\lambda_0^2$ and $\lambda_1^2$ in our three-field model, plotted
as a function of time for different values of \mbox{$\mbar^2_{02}=\mbar_{12}^2$}.    
As in Fig.~\ref{fig:posdefregion}, we have taken 
\mbox{$\overline{m}^2_{00}=5$}, \mbox{$\overline{m}^2_{11}=5$}, and \mbox{$\overline{m}^2_{22}=1$}. 
We see that each field can now separately have its own pulse with its own effective frequency, resulting 
in the possibility of {\it two}\/ parametric resonances
which can either reinforce or interfere with each other.}  
\label{fig:lambdaN3}
\end{figure}

\begin{figure*}[t]
\centering
\hspace*{-0.4cm}
\includegraphics[width=0.335\textwidth,keepaspectratio]{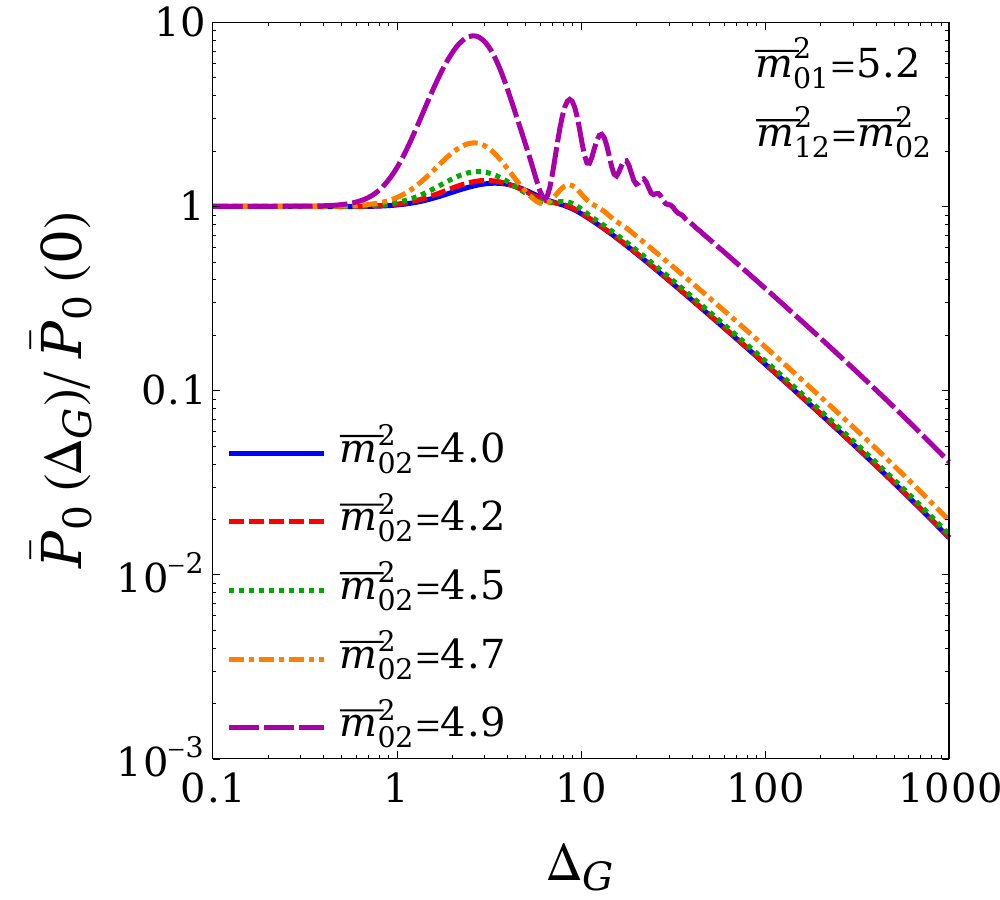}
\includegraphics[width=0.335\textwidth,keepaspectratio]{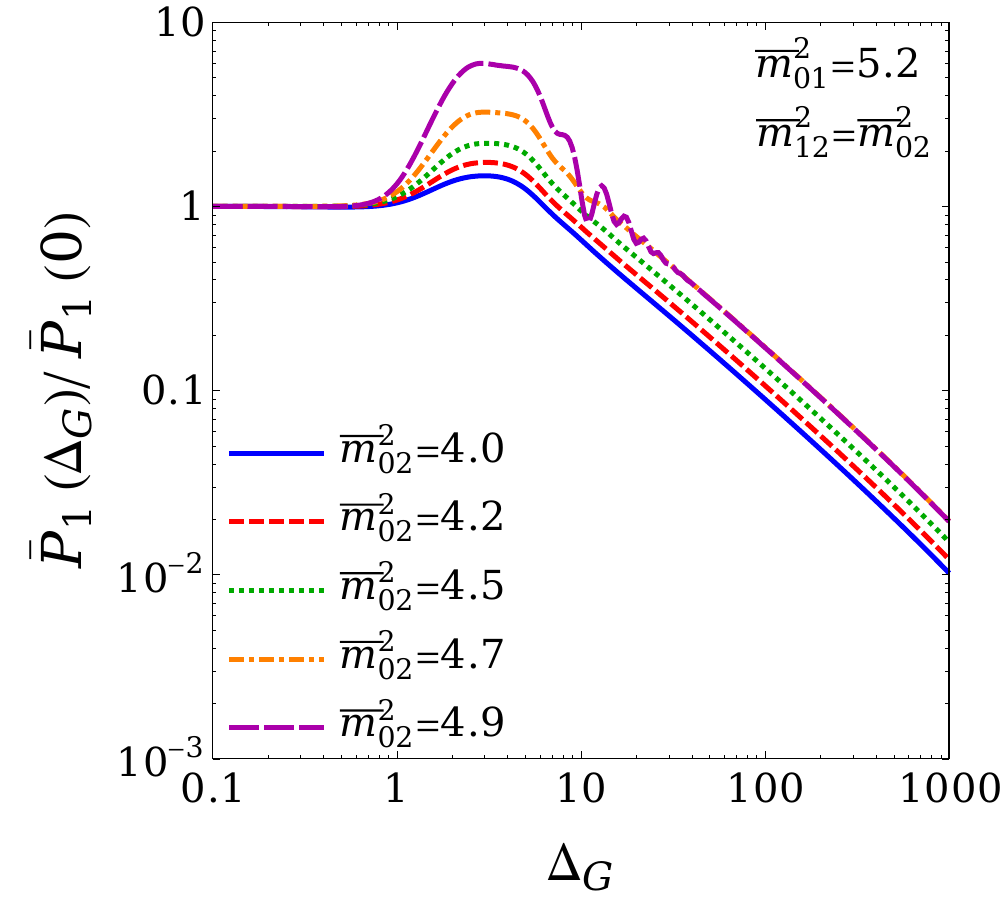} 
\includegraphics[width=0.335\textwidth,keepaspectratio]{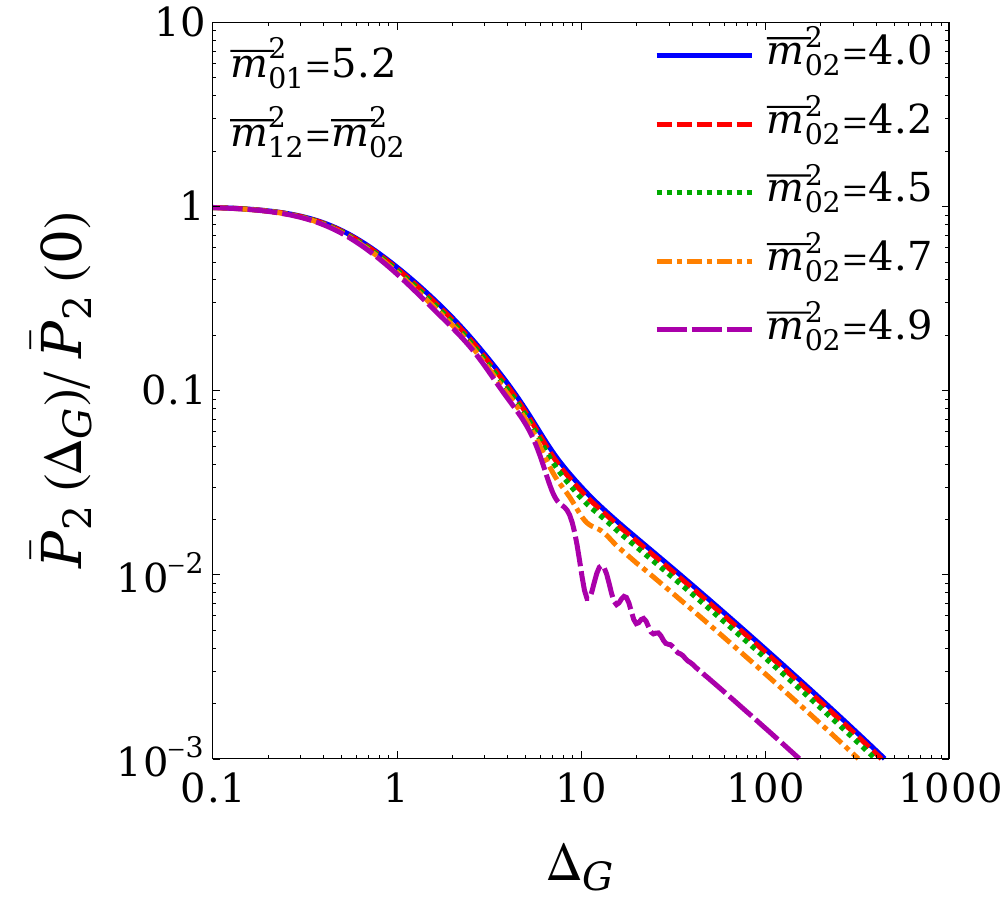}
\caption{The partial-sum energy densities $\overline{P}_\ell(\Delta_G)$, \mbox{$\ell=0,1,2$}, plotted as functions of the phase-transition width $\Delta_G$ and
normalized to the
values $\overline{P}_{\ell}(0)$
that they would have had for an instantaneous phase transition.
As in Figs.~\ref{fig:posdefregion} and \ref{fig:lambdaN3},
we have again taken \mbox{$\overline{m}^2_{00}=5$}, \mbox{$\overline{m}^2_{11}=5$}, and \mbox{$\overline{m}^2_{22}=1$}. 
We observe many features familiar from our two-component toy model, including parametric resonances which enhance the corresponding energy densities
as well as suppressions by many orders of magnitude that develop for sufficiently large $\Delta_G$. 
However, there are also new features which arise due to the fact that we now have more than two fields in our system. 
The center panel, in particular, illustrates the manner in which the separate resonances from the two lighter fields reinforce each other and produce a {\it plateau}\/
along which the partial-sum  energy density $\overline{P}_1$ 
is enhanced by nearly an order of magnitude over a significant $\Delta_G$-interval.}
\label{fig:rhoN3}
\end{figure*}

Inherent in this mass matrix are now {\it three}\/ independent mixing angles which correspond to
$m_{01}^2$, $m_{02}^2$, and $m_{12}^2$.   However, just as in the two-field case in Sect.~\ref{sec:AToyModel},
there are ultimately constraints on these mixing parameters which ensure that our mass matrix 
$\calM^2$ remains positive-semidefinite.
Such constraints are illustrated in Fig.~\ref{fig:posdefregion} for the case with
\mbox{$\overline{m}^2_{00}=5$}, \mbox{$\overline{m}^2_{11}=5$}, and \mbox{$\overline{m}^2_{22}=1$} --- a fixed parameter choice
we shall adopt for all of the plots in this section.

In our two-component toy model, we have seen that the lighter field can occasionally develop a ``pulse'' which is
ultimately responsible for the parametric-resonance and re-overdamping phenomena in Sects.~\ref{sec:TheParametricResonance} 
and \ref{sec:ReOverDamping} respectively.
However, in a {\it three}\/-component model, it turns out that the {\it two}\/ lightest fields can each develop a pulse.
This is illustrated in Fig.~\ref{fig:lambdaN3}.
Indeed these two pulses need not have the same effective frequencies or correspond to the same values of $\Delta_G$.
As we shall see, this can then give rise to multiple resonances which can reinforce or interfere with each other
and thereby produce new effects which transcend those realizable with only a single pulse.
We also note that the pulse of our second mass eigenvalue $\lambda^2_1$
can leave this eigenvalue with a smaller magnitude at late times than at early times --- a phenomenon which was not possible
for $\lambda^2_0$ in the two-field case.    This has the effect of allowing a further  elongation   
of the time interval during which this field can remain in a re-overdamped or parametrically resonant state. 

To illustrate the interplay between the different resonances,
we can consider the late-time energy densities of this system.
Since we now have three individual contributions to the total energy density,
we shall highlight their separate effects by considering the partial sums 
\beq
   P_\ell  ~\equiv~ \sum_{i=0}^{\ell}\rho_{\lambda_i} ~.
\label{eq:partialsumdef}
\eeq
In Fig.~\ref{fig:rhoN3} we have plotted these partial sums, with each normalized to the value that it would have had
if the mass-generating phase transition had been instantaneous.
This demonstrates the combined effects of the mixing and the finite phase-transition width.
The left panel of Fig.~\ref{fig:rhoN3} 
illustrates the behavior of \mbox{$\overline{P}_0\equiv \rhobar_{\lambda_0}$} alone:
this behavior is similar to what we have already seen in our two-field toy model,
for which the pulse in the evolution of $\lambda_0^2$ produces resonant peaks in the corresponding energy-density component at late times. 
These peaks grow stronger for larger $\mbar_{02}^2$, 
with the energy density in this field dissipating inversely with $\Delta_G$
once \mbox{$\Delta_G \gtrsim 2\pi/\lambda_0$}. 
However,  the center panel shows the combined contribution to the total energy density from the two lighter fields
and thus exhibits something new: 
the contributions of the two separate parametric resonances 
combine to reinforce each other and even produce an effective enhancement {\it plateau}\/ which stretches over an extended interval in $\Delta_G$.
Indeed, all along this plateau, the energy densities of our system can be enhanced by as much as nearly an order of magnitude!
Finally, the right panel of Fig.~\ref{fig:rhoN3} illustrates the behavior of the total energy density $\rhobar$.
Note that in this case, the heaviest field effectively dominates the energy density for \mbox{$\Delta_G\lesssim 5$}, thereby 
washing out those parametric resonances that appear in that range, while leaving those that appear for \mbox{$\Delta_G\gtrsim 5$}.  

The purpose of this section has merely been to provide several short examples of new phenomena that can arise when more than two scalar fields
are involved.  We shall defer a more detailed study of these general cases to Ref.~\cite{toappear}.

\FloatBarrier

\section{Conclusions\label{sec:Conclusions}}


In this paper we have examined some of the novel effects that emerge 
when multiple scalars mix with each other and experience a time-dependent mass-generating phase transition
within the context of a time-dependent background cosmology.
As we have explicitly demonstrated within a relatively simple two-field toy model,
these effects include both enhancements and 
suppressions of the total late-time energy density of the system as well
as the late-time energy densities of the individual scalar components.
These enhancements and suppressions need not be mere ${\cal O}(1)$ effects;   indeed, we have seen
that the resulting late-time energy densities can often be altered by many orders of magnitude.
These alterations are even sufficient to produce
wholesale shifts in the identity of the field carrying the bulk of the late-time cosmological
abundance. 

We also found that our late-time energy densities can exhibit large parametric resonances which
make them extremely sensitive to relatively small changes in the degree of mixing or in 
the width of the phase transition.   Thus, any effects which might change --- even slightly ---
the rate at which the phase transition unfolds 
can have a significant impact on the late-time abundances of our fields.
We also demonstrated that our system can enter 
a so-called ``re-overdamped'' phase in which the field values and energy densities behave 
in a manner quite different from what would normally be associated
with pure dark matter or vacuum energy.  Indeed, we saw that the system can remain in 
this re-overdamped phase for a considerable length of time,
suggesting that this phase be considered alongside ``vacuum energy'' and ``matter''
as an equally valid new behavioral phase that a scalar field can
experience during a significant portion of cosmological evolution.
Finally, we illustrated the importance of some of these effects  
within the context of a simple model in which a second axion is incorporated 
into the standard QCD axion theory.

As indicated in the Introduction, 
our results can be taken as an explicit demonstration
of how critical it can be to properly incorporate the non-zero but finite widths of mass-generating phase transitions
when discussing late-time energy densities in cosmological settings involving multiple mixed scalars. 
Indeed, treating these widths as being either zero or infinite (the latter corresponding to a so-called ``adiabatic'' approximation in the axion case) can lead to results 
for late-time quantities such as cosmological abundances
which are significantly distorted from their true values, often by many orders of magnitude!
Given the sensitivity with which the different slices of the so-called ``cosmic pie'' 
have already been measured, such effects will inevitably play critical roles in determining
whether a particular cosmological scenario is viable or ruled out.

Needless to say,
the features we have discussed in this paper give rise to many new possibilities for 
the phenomenology of the early universe and cosmological evolution.
As such, new approaches to model-building can readily be imagined.
For example, 
many extensions to the Standard Model involve scalar fields which are 
massless (or effectively
massless) at high scales, but for which masses are generated dynamically at lower scales.
Examples include string moduli, string axions, and other axion-like particles.  
However, these particles can be cosmologically 
problematic if these dynamically-generated masses are sufficiently large because
they can overclose the universe once they begin oscillating and act as massive matter.  
Indeed, this is the crux of the cosmological moduli 
problem~\cite{Coughlan:1983ci,Banks:1993en,deCarlos:1993jw}.  
Moreover, particles
of this sort are often produced through mechanisms such as vacuum misalignment which
are completely decoupled from the dynamics of inflation.  In such cases, spatial 
fluctuations in the energy density of these particles therefore give rise to 
isocurvature perturbations, the observational limits on which are quite 
severe~\cite{Planck:2015xua}.  

Fortunately, mechanisms such as those we have discussed in this paper can suppress 
the resulting late-time energy densities by many orders of magnitude.
This enhanced energy dissipation can therefore provide a novel way of alleviating such
constraints on models of new physics, and may represent an alternative solution
to the moduli and isocurvature problems.   They may also allow a general weakening 
of the cosmological bounds on the energy scales associated with the new physics
(such as the Peccei-Quinn scale associated with axions).

As we have discussed,
the features explored in this paper can lead
not only to a reduction in the
total late-time energy density of our system, but also to changes 
in the late-time {\it partitioning}\/ of that total energy density amongst the fields 
in our theory.
In new physics scenarios which involve multiple scalar fields,
this too can have observable phenomenological consequences.  For example,
certain classes of string models are generically expected to give rise to
large numbers of light axion-like 
particles~\cite{Witten:1984dg,Conlon:2006tq,Svrcek:2006yi,Arvanitaki:2009fg},
the masses of which must be generated dynamically via non-perturbative effects.
In situations in which the same dynamics contributes to the generation 
of mass terms for multiple such particles, the relative cosmological abundances 
of those particles can be significantly altered by the effects we have 
studied here.

There are many ways in which the analysis in this paper might be extended and generalized.
For example, in this paper we have focused primarily on scalar fields
which have a negligible back-reaction on the background cosmology, at least
during the epoch of dynamical mass generation.  In other words, we have assumed that our
scalars do not contribute significantly to the total energy density of the 
universe until well after their masses have settled into their late-time values.
However, in principle, we could relax this assumption and broaden this study to include
scalars which {\it do}\/ have a significant impact on the background cosmology.
This would then establish a non-linear back-reaction which could possibly enrich
the dynamics of this system even further, 
with the evolution of the Hubble parameter $H(t)$ itself depending on the behaviors
of the scalar energy densities as functions of time.
 
This possible extension of our work could be particularly important if the scalars 
involved are those which are directly responsible for triggering
periods of rapid cosmological expansion, including 
cosmic inflation.
Indeed, a wide variety of inflationary models 
involving multiple scalar fields
exist in the literature;  these include 
hybrid inflation~\cite{Linde:1993cn,Copeland:1994vg}, 
assisted inflation~\cite{Liddle:1998jc,Kanti:1999vt,Kanti:1999ie,Kaloper:1999gm},
$N$-flation~\cite{Dimopoulos:2005ac}, and multi-field stochastic 
inflation~\cite{Adshead:2008gk}.
Likewise, several constructions have been proposed through which a potential for the 
fields responsible for cosmic inflation is generated 
dynamically~\cite{Dimopoulos:1997fv,Izawa:1997df,Izawa:1997jc,Harigaya:2012pg}.
Such constructions are typically motivated by the possibility of realizing
a viable inflationary potential without the introduction of extremely small 
couplings or arbitrary mass scales.  It would therefore be extremely interesting to explore 
the cosmological consequences of a time-dependent, dynamically-generated inflaton
potential in the context of multi-field inflation. 

In this connection, it is perhaps worth emphasizing that 
the parametric resonance we have discussed in 
Sect.~\ref{sec:TheParametricResonance} is entirely unrelated to the 
parametric resonance which gives rise to an epoch of explosive 
production of light scalars in inflationary models with a preheating 
phase~\cite{Kofman:1994rk,Kofman:1997yn}.
Indeed, the effect we have discussed in this paper involves a resonance between the 
oscillation frequency of the light scalar, as given by its mass $\lambda_0$,
and the effective modulation frequency $\omega_{\mathrm{eff}}$ of that mass.  
As such, this parametric resonance 
is ultimately driven by the dynamics of the mass-generation mechanism and 
leads to an enhancement in the amplitude for coherent oscillations of the
zero-momentum mode of $\phi_{\lambda_0}$.     
By contrast, the effect discussed in preheating scenarios involves a resonance 
between the frequency of the coherently-oscillating inflaton field and the frequency
associated with a given momentum mode of the lighter scalar field into which it 
decays.  Such a resonance is ultimately driven by inflaton dynamics and results in particle-like
excitations of this lighter scalar.

Other extensions and generalizations of our work are also possible.
For example, our original dynamical equations in Eq.~(\ref{eq:singlephieqofmotion})
assume that there are no significant spatial variations in our field values,
but in general it is possible to allow our field values to vary not only in time but also in space.
This could then give rise to distinctly different physics in different spatial regions,
with domain walls potentially separating different regions experiencing different phases
and different energy densities.   Such spatial variations and domain walls
would inevitably introduce additional length scales into the problem, 
and thereby likely give rise to new complex dynamical behaviors for
our system as a whole.

In a similar vein, we could likewise imagine that our fundamental dynamical equations in 
Eq.~(\ref{eq:singlephieqofmotion})
are modified by the inclusion of {\it source terms}\/ reflecting possible interactions
with more complex cosmological environments (\eg, with other fields).
Such source terms could then give rise to additional features
which effectively modify our mass eigenvalues and eigenstates,
trigger enhancements or suppressions of late-time energy densities, and produce
additional kinds of resonances.
Indeed, such effects would be analogous to the well-known 
``matter effects'' (\ie, MSW effects) that 
arise when neutrinos propagate through matter rather than through a pure vacuum.

Another possible generalization of our 
toy model is to consider more than two scalar fields.
Indeed, as briefly discussed 
in Sect.~\ref{sec:BeyondTwoFields},
there is no limit to the number of scalar fields which may be considered.
These models can therefore grow in both complexity and sophistication compared
with the two-component model we have studied here.

One natural possibility along these lines is to consider the case in which
the different fields $\phi_i$ are nothing but the Kaluza-Klein (KK) modes of a single
higher-dimensional field $\Phi$.   
This is indeed a well-motivated possibility, as string theory naturally gives rise to many scalars
(axions, geometric/gravitational moduli, gauge-neutral singlets, and so forth)
that populate the ``bulk'' volume transverse to the brane  
on which gauge interactions reside.  As such, the different KK modes 
of such bulk fields are natural dark-matter candidates, even though they might be unstable ---
an observation which has motivated the Dynamical Dark Matter framework of 
Refs.~\cite{Dienes:2011ja,Dienes:2011sa,Dienes:2012jb}.
In this case, the masses that we have assumed to exist prior to the
phase transition would be nothing but the corresponding KK masses,
while the extra time-dependent contributions $\mbar_{ij} h(\tau)$ 
are presumably generated by a phase transition 
which breaks the higher-dimensional symmetries that would otherwise have
aligned our mass eigenstates with KK eigenstates.
The resulting mixing structure is therefore determined according to the 
intrinsically higher-dimensional geometric symmetries of the theory and the manner
in which these symmetries might be broken by the phase transition.
(For example, in the five-dimensional axion scenario of Refs.~\cite{Dienes:1999gw,Dienes:2011sa,Dienes:2012jb}
the instanton-induced phase transition takes place purely on the brane
and thereby breaks a translation symmetry
that otherwise existed along the extra dimension.)
In general, the cosmological properties of such KK systems can be extremely rich,
and will be discussed in more detail in Ref.~\cite{toappear}.

Given these observations, 
it might at first glance seem that the two-component toy model we have studied in this paper --- while illustrative
for pedagogical purposes --- might not have been worthy of the detailed attention 
we have afforded it.
However, this model can always be viewed as the low-energy limit of a more complete
model (such as the model of Kaluza-Klein cosmology discussed above)
in which only the contributions from the lightest two modes are considered.
Indeed, in many cosmological settings, the contributions of the lightest fields are likely
to dominate the resulting phenomenology and/or carry the largest abundances.
Thus, in this sense, our two-component toy model 
can be viewed as the common root
underlying 
a variety of much more general models --- a ``kernel'' which gives rise to
rather universal features and behaviors that will continue to appear (perhaps with further embellishments) in more
complete or realistic settings.
As such, then, this toy model can be viewed as representing a universal low-energy 
limit of a wide class of models of this type
whose leading-order phenomenologies will be exactly those we have 
investigated in this paper.

\begin{acknowledgments}
We would like to thank D.~Chung, S.~Su, and S.~Watson for useful discussions.
The research activities of KRD and JK were supported in part by the Department of Energy under 
Grant DE-FG02-13ER-41976, while the research activities of BT were supported
in part by internal funding from Reed College.
The research activities of KRD were also supported in part by
the National Science Foundation through its employee IR/D program.  
The opinions and conclusions expressed herein are those of the authors,
and do not represent any funding agencies.
\end{acknowledgments}

\bigskip

\appendix

\FloatBarrier

\section{~~Narrow-width behavior of the late-time energy density:  Analytical results}
\label{notation}


\setcounter{footnote}{0}

As we have seen, the presence of a non-zero width $\Delta_G$ for the 
mass-generating phase transition prevents us from obtaining analytical results
for most of our quantities of interest.
However, in the case with \mbox{$\Delta_G=0$} (corresponding to an instantaneous phase
transition), the resulting dynamics can be treated analytically.
In this Appendix we provide analytical results for the late-time energy
density $\overline{\rho}$.

For \mbox{$\Delta_G=0$}, the behavior of each of our two fields is relatively simple.
Prior to the phase transition at $\tau_G$, only the massless field has a non-zero value $A_0$,
as in Eq.~(\ref{eq:initialconditions});
this field is necessarily overdamped, however, and thus remains effectively constant 
until the phase transition at $\tau_G$.   At that point, the mass matrix receives the 
extra contribution which mixes the two fields with an angle $\theta$;
the initial massless field value 
is thus redistributed at \mbox{$\tau=\tau_G$} such that
\beq
           \phi_{\lambda_0}(\tau_G) = A_0  \cos\theta~,~~~~ 
           \phi_{\lambda_1}(\tau_G) = A_0  \sin\theta~.
\label{redistributed}
\eeq
Each field then evolves forward in time independently.  

At first glance, motivated by the behavior illustrated in 
Fig.~\ref{fig:rhosinglefield}, we might try to estimate the total late-time energy density $\rhobar$
using a virial approximation in which we assume that the energy density associated with each field $\phi_{\lambda_i}$ 
remains constant until \mbox{$\tau^{(i)}_{\zeta}\equiv \kappa/2\lambda_i$} [where \mbox{$\kappa\equiv 2$} (or $3/2$) for a 
matter- (radiation-) dominated universe], after which it scales as \mbox{$\sim \tau^{-\kappa}$}.
Defining \mbox{$\tau_\lambda\equiv {\rm max}\, \lbrace \tau_G,\kappa/2\lambda\rbrace$},
we would then write
\beq
   \overline{\rho} ~\approx~ 
      \sum_{\lambda}\rho_{\lambda}(\tau_{\lambda})
       \left( \frac{\tau}{\tau_\lambda}\right)^{-\kappa}  
  =  \frac{1}{2}\sum_{\lambda} \,\lambda^2 \left[\phi_{\lambda}(\tau_G)\right]^2
             \left( \frac{\tau}{\tau_\lambda}\right)^{-\kappa}~.  
\label{eq:virialapproximation}
\eeq
Unfortunately, this approximation breaks down for any fields $\phi_i$ 
for which the threshold of critical damping \mbox{$\tau^{(i)}_{\zeta}= \kappa/2\lambda_i$} is
close to the phase transition time $\tau_G$. 

For this reason, we shall instead evaluate our field values and corresponding energy densities exactly.
In the limit of an instantaneous phase transition occurring at $\tau_G$, 
the exact solution for each field is given by
\beqn
    \phi_\lambda(\tau) &=& \frac{\pi}{2} \,[\phi_\lambda(\tau_G)]\,
            \, \lambda \tau_G \, \left( \frac{\tau}{\tau_G}\right)^{-\kappa_-} \,\times\nonumber\\
    && \left[
             J_{\kappa_+}(\lambda\tau_G) 
             Y_{\kappa_-}(\lambda\tau) -
             Y_{\kappa_+}(\lambda\tau_G) 
             J_{\kappa_-}(\lambda\tau) 
      \right]~~\nonumber\\
\eeqn
where \mbox{$\kappa_\pm \equiv (\kappa\pm 1)/2$},
where $\lambda$ refer to the mass eigenvalues for \mbox{$\tau \geq \tau_G$},  and where
$J_\nu(z)$ and $Y_\nu(z)$ are Bessel functions of the first 
and second kind, respectively. 
From this result we may directly calculate the corresponding energy densities for
all \mbox{$\tau\geq \tau_G$}, obtaining
\beq
    \rho_\lambda(\tau) ~=~ \frac{\pi^2}{8} \,[\phi_\lambda(\tau_G)]^2\,
            \, \lambda^4 \, (\tau_G\tau) \, 
        \left( \frac{\tau}{\tau_G}\right)^{-\kappa} 
          {\cal J}_\lambda(\tau)~~~ \\
\eeq
where 
\beqn
    {\cal J}_\lambda &=& \phantom{+}  
           \left[ 
             J_{\kappa_+}(\lambda\tau_G) 
             Y_{\kappa_-}(\lambda\tau) -
             Y_{\kappa_+}(\lambda\tau_G) 
             J_{\kappa_-}(\lambda\tau) \right]^2 \nonumber\\
          && +
           \left[ 
             J_{\kappa_+}(\lambda\tau_G) 
             Y_{\kappa_+}(\lambda\tau) -
             Y_{\kappa_+}(\lambda\tau_G) 
             J_{\kappa_+}(\lambda\tau) \right]^2 ~.\nonumber\\
\label{eq:A}
\eeqn
Given these exact results, we can now extract the late-time behavior of $\rho_\lambda$
through the asymptotic expansions
\beqn
   J_\alpha(z) &\sim&  \sqrt{\frac{2}{\pi z}}\,\cos\left(z - \frac{\alpha\pi}{2} - \frac{\pi}{4}\right) ~+~ ...\nn\\
   Y_\alpha(z) &\sim&  \sqrt{\frac{2}{\pi z}}\,\sin\left(z - \frac{\alpha \pi}{2} - \frac{\pi}{4}\right) ~+~ ...~~~~ 
\label{eq:Bessellarge}
\eeqn
which hold for \mbox{$z\gg 1$}.
With this approximation we find that
\beq
        \overline{\cal J}_\lambda ~\approx~   \frac{2}{\pi \lambda \tau}\,
             \left[
                J^2_{\kappa_+} (\lambda\tau_G)  
                + Y^2_{\kappa_+} (\lambda\tau_G) \right] ~+~...~,
\label{latetimeJ}
\eeq
whereupon use of Eq.~(\ref{redistributed})
yields the total late-time energy density
\beqn
    \rhobar &=& \frac{\pi^2}{8} \, A_0^2 \,(\tau_G \tau) \,\left( \frac{\tau}{\tau_G}\right)^{-\kappa}~\times\nonumber\\
           && ~~~~~\left(
               \lambda_0^4  \, \overline{\cal J}_{\lambda_0} \cos^2\theta 
             + \lambda_1^4  \, \overline{\cal J}_{\lambda_1} \sin^2\theta \right)~.
\label{eq:asymptoticrho}
\eeqn
 
The result in Eq.~(\ref{eq:asymptoticrho}) is an exact expression for the late-time energy density $\rhobar$ in the 
\mbox{$\Delta_G\to 0$} limit.
Given this result, several things are immediately clear.
First, we observe that in the case of matter-dominated universe (corresponding to \mbox{$\kappa=2$}),
the total late-time energy density simplifies to
\beqn
   \rhobar &=& \frac{ A_0^2}{4\tau^2} \, \biggl[
         2 \, +\, (\lambda_0^2 + \lambda_1^2 ) \tau_G^2 \, +\,
             (\lambda_0^2 - \lambda_1^2 ) \tau_G^2  \cos(2\theta)\biggr] \nonumber\\ 
           &=& 
   \frac{ A_0^2}{2\tau^2} \, (
         1 \, +\, m^2_{00}  \tau_G^2 )~ ,
\label{true}
\eeqn
where in passing to the second line we have used the results in Eq.~(\ref{lambdabarvalues}).
As a result, all $\theta$-dependence drops out of the final result for the total late-time energy density.
In other words, {\it for a matter-dominated universe, the corresponding late-time energy 
density is independent of mixing}\/!

Second, we see that the virial approximation in Eq.~(\ref{eq:virialapproximation})
corresponds to the limit in which \mbox{$\tau_G\lambda_i\gg 1$} for both fields.
Indeed, in this limit we see that \mbox{$\tau_\lambda = \tau_G$} for each field
(since both fields are already significantly underdamped
when the phase transition occurs),
and we can likewise apply the asymptotic expansions in Eq.~(\ref{eq:Bessellarge})
to the remaining Bessel functions in Eq.~(\ref{latetimeJ}), thereby reproducing the result in Eq.~(\ref{eq:virialapproximation}).
Further use of 
the results in Eq.~(\ref{lambdabarvalues})
then gives the total late-time energy density
\beq
        \rhobar ~\approx~ 
   \frac{ A_0^2}{2} \, m_{00}^2 \, \left( \frac{\tau}{\tau_G}\right)^{-\kappa}~,
\eeq
which is again independent of the mixing --- now regardless of the value of $\kappa$!
However, comparing this result with the exact result in Eq.~(\ref{true}) for the case of 
a matter-dominated universe illustrates the imperfect 
nature of the virial approximation.
 
\FloatBarrier

\section{~~Late-time energy densities:  An alternative approach }
\label{alternative}


\setcounter{footnote}{0}

In the main body of this paper, we have consistently viewed 
the matrix elements $m_{ij}^2$
as our independent variables.   These have been reparametrized 
in other forms, giving rise to variables such as $m_{\rm sum}^2$, $\Delta m^2$, 
$\eta$, $\theta$, and $\xi$,
but in all cases these independent variables have served as inputs.
We then calculated the values of dependent quantities such as the eigenvalues $\lambda_i$,
energy densities $\rho$, and so forth.
Thus, when we have studied how a certain dependent variable such as a late-time energy density
$\rhobar$ depends on the mixing, we have implicitly 
held the diagonal matrix elements $\mbar^2_{00}$ and $\mbar^2_{11}$ 
(or equivalently $\mbar^2_{\rm sum}$ and $\etabar$) fixed, and varied 
the off-diagonal matrix element $\mbar_{01}^2$ (or equivalently
the mixing angle $\thetabar$ or mixing saturation $\xibar$).

However, another possibility is to study the mixing-dependence of quantities such as $\rhobar$ 
when we hold the late-time {\it eigenvalues}\/ $\lambdabar_i$ fixed instead.
This approach can be useful in situations where the masses of our states are
known and it is only the mixing between such states that we wish to vary.

In order to implement this approach, we 
must first invert our usual algebraic relations and 
express \mbox{$\lbrace m^2_{00}, m^2_{11}\rbrace$}
in terms of \mbox{$\lbrace \lambda_0^2, \lambda_1^2\rbrace$}.
However, we must also recognize that our previous assumption of holding 
 \mbox{$\lbrace m^2_{00}, m^2_{11}\rbrace$} fixed is also implicitly buried
within the definition of another variable:  the mixing saturation $\xi$.   
Although $m_{01}^2$ or $\theta$ parametrize the mixing in absolute terms and are thus 
independent of such assumptions,
the mixing {\it saturation}\/ $\xi$ defined in Eq.~(\ref{eq:xidefinition}) describes
the mixing as a fraction of the {\it maximum allowed}\/ mixing,
and this maximum allowed mixing has always been determined relative to
a fixed \mbox{$\lbrace m^2_{00}, m^2_{11}\rbrace$}.
What we now need, 
by contrast, is a new variable $\xi_\lambda$ which describes the mixing
as a fraction of the maximum mixing allowed for a given fixed  
\mbox{$\lbrace \lambda_0^2, \lambda_1^2\rbrace$}.
Indeed, this is an entirely different quantity.

Thus, we seek to express 
\mbox{$\lbrace m^2_{00}, m^2_{11}, \xi\rbrace$}
in terms of \mbox{$\lbrace \lambda_0^2, \lambda_1^2,\xi_\lambda\rbrace$}.
Equivalently, defining
\beq
         \lambda^2_{\rm sum} ~\equiv ~ \lambda_0^2 + \lambda_1^2~
\eeq
and
\beq
         \eta_\lambda ~\equiv ~ { \Delta \lambda^2 \over \lambda^2_{\rm sum}}  
          ~\equiv~ {\lambda_1^2-\lambda_0^2\over \lambda_0^2 + \lambda_1^2}~
\eeq
in analogy with $m_{\rm sum}^2$ and $\eta$,
we would like to express 
\mbox{$\lbrace m^2_{{\rm sum}}, \eta, \xi\rbrace$}
in terms of 
\mbox{$\lbrace \lambda^2_{{\rm sum}}, \eta_\lambda, \xi_\lambda\rbrace$}.
Surprisingly, these relations ultimately take a relatively simple form:
\beqn
         m_{\rm sum}^2 &=& \lambda_{\rm sum}^2      \nonumber\\
      \eta^2 &=&  \eta_\lambda^2 \left({ 1-\xi_\lambda^2\over 1-\eta_\lambda^2 \xi_\lambda^2}\right)         \nonumber\\
         \xi &=&   \eta_\lambda \xi_\lambda~.
\label{inverses}
\eeqn
These relations in turn imply that
\beq
      m_{01}^2 ~=~ 
\half \, (\Delta \lambda^2) \, \xi_\lambda  \left( {1-\eta_\lambda^2\over 1-\eta_\lambda^2 \xi_\lambda^2}\right)~.
\eeq
Likewise, following Eq.~(\ref{tanrelation}),
we can also determine the absolute mixing angle $\theta$ via
\beq
  \tan^2 (2\theta) ~=~ {\xi^2\over \eta^2} (1-\eta^2) 
               ~=~ \xi_\lambda^2 \left(  {1-\eta_\lambda^2\over 1-\xi_\lambda^2} \right)~. 
\label{thetaangle}
\eeq
Note that \mbox{$0 \leq \eta_\lambda \leq 1$}, whereas
\mbox{$-1 \leq \eta \leq 1$}.
Also note that the eigenvalues $\lambda_i^2$ are independent of the 
sign of $\eta$ (as evident from Fig.~\ref{fig:lambdares2}).
It is for this reason that our inverse relations in Eq.~(\ref{inverses}) 
determine $\eta$ [and likewise $\tan (2\theta)$] only up to an overall sign.
Thus, strictly speaking, our variable map really trades
\mbox{$\lbrace \eta,\xi\rbrace$} for \mbox{$\lbrace \eta_\lambda,\xi_\lambda,\epsilon\rbrace$}
where \mbox{$\epsilon \equiv {\rm sign}(\eta)= \pm 1$} is a discrete $\mathbb{Z}_2$ phase.
Having \mbox{$\epsilon>0$}
corresponds to \mbox{$\eta>0$} and \mbox{$\tan (2\theta)>0$}.
However, we shall avoid writing $\epsilon$ in what follows, understanding the proper
phase choice to be implicit.

In principle, 
given this variable map, we could now work our way through this entire paper
again using the variables
\mbox{$\lbrace \lambda^2_{{\rm sum}}, \eta_\lambda, \xi_\lambda\rbrace$}
rather than the variables 
\mbox{$\lbrace m^2_{{\rm sum}}, \eta, \xi\rbrace$}.
Indeed, every curve in every plot which has been labeled in terms of
certain numerical values for the first set of parameters could now equivalently be relabeled
in terms of different numerical values for the second set of parameters.
In cases where these curves describe the behavior of 
a certain quantity such as an energy density as a function of time, 
or the behavior of a quantity such 
as a late-time energy density as a function of a phase-transition width $\Delta_G$, 
these curves would themselves remain intact
and continue to take the same shapes as before because our time and $\Delta_G$
parameters are independent of our variable change.
All that would change are the labels associated with each such curve.
Indeed, every feature
that appears in our plots 
for certain fixed values of
\mbox{$\lbrace \mbar_{\rm sum}^2,\etabar\rbrace$}
(such as the oscillations due to parametric resonances or the re-overdamping of certain fields)
would continue to exist when
\mbox{$\lbrace \lambdabar_{\rm sum}^2,\etabar_\lambda\rbrace$}
are held fixed instead. 
As a result, nothing we have said in this paper concerning the parametric-resonance
or the re-overdamping phenomena would be affected by this change in variables.

The only changes that could potentially occur due to this variable exchange
concern the effects of mixing 
as discussed in Sects.~\ref{sec:LateTimeEnergyDensity} and \ref{sec:indiv}.
Of course, each individual curve within the plots in Figs.~\ref{fig:figname1},
\ref{fig:figname2}, \ref{fig:figname3},
\ref{fig:rhocomponents}, and \ref{fig:rhocomponentsratio}
would remain exactly as before and  would merely be relabeled using 
\mbox{$\lbrace \lambda^2_{{\rm sum}}, \eta_\lambda, \xi_\lambda\rbrace$}
rather than \mbox{$\lbrace m^2_{{\rm sum}}, \eta, \xi\rbrace$}.
However, what {\it would}\/ change is the resulting 
 {\it grouping}\/ of these curves into distinct panels
in which all curves share common values of, say, $\eta_\lambda$ rather than $\eta$.
Indeed, although the space of all possible curves is not changed, these curves would
experience a reshuffling into differently-grouped subsets.
As a result, the only physics questions whose answers might change
in this process are those which rely on comparisons  between the different curves 
 {\it within}\/ a given grouping.
Clearly, questions pertaining to the effects of mixing (\ie, the effects of changing the values of $\xi$ or $\xi_\lambda$
while holding $\eta$ or $\eta_\lambda$ fixed)  are in this category.

\begin{figure*}[thb]
\includegraphics[width=0.40\textwidth]{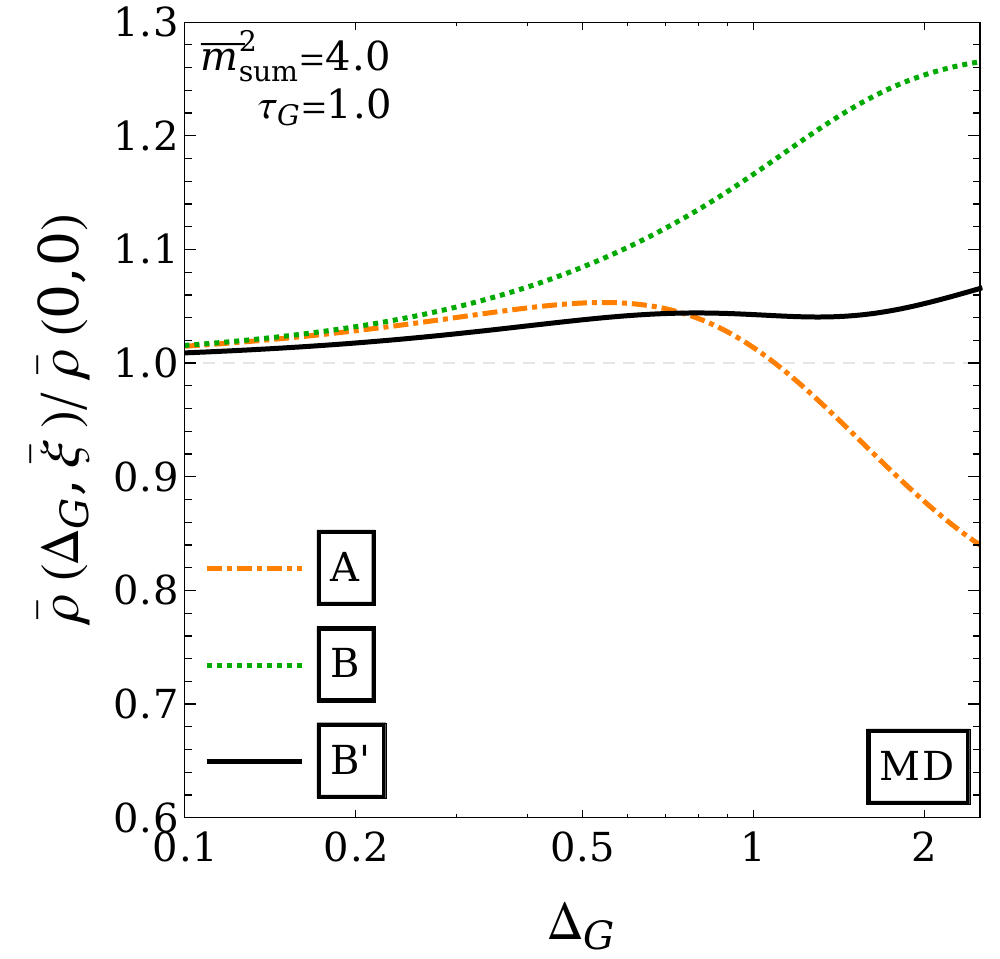}
\hskip 0.3 truein
\includegraphics[width=0.40\textwidth]{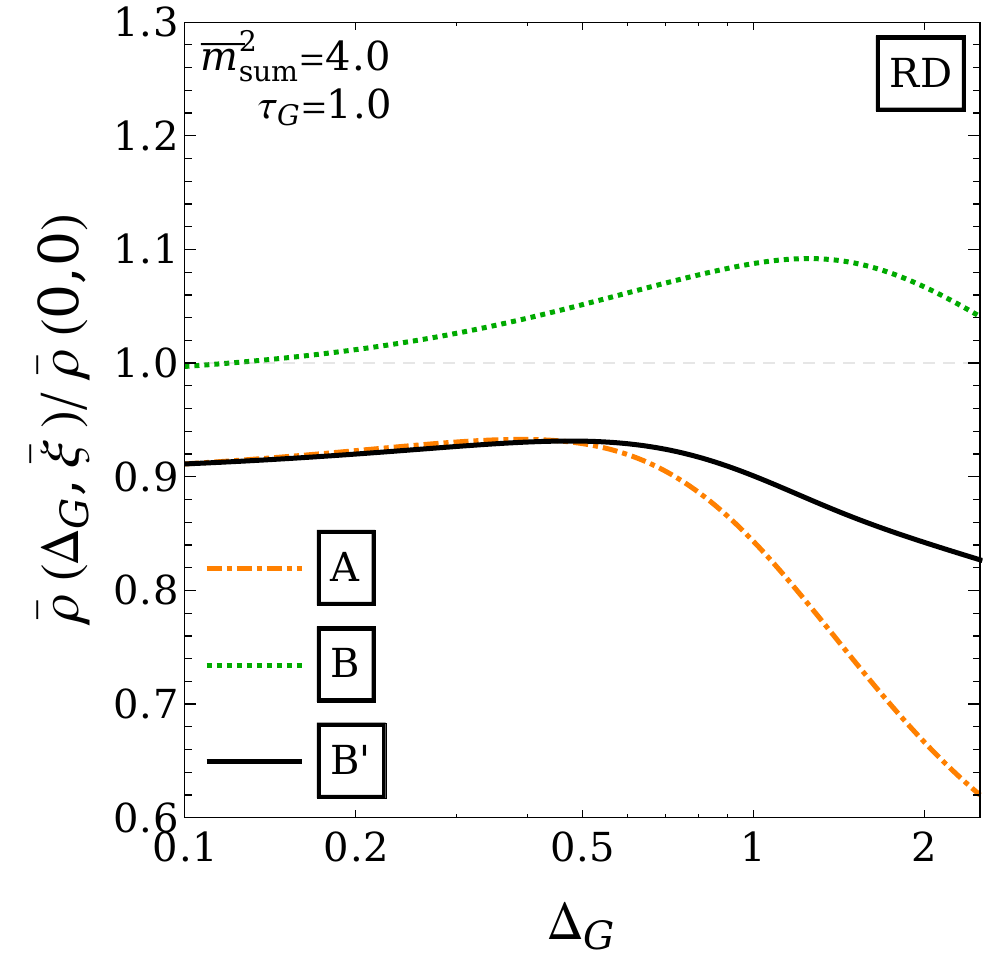}
\caption{The late-time energy density $\rhobar$,  plotted as 
a function of $\Delta_G$ in a matter-dominated universe (left panel) 
or radiation-dominated universe (right panel) and normalized to its value for an instantaneous phase transition
with zero mixing.   
In each case, the curves~$A$, $B$, and $B'$ correspond to the parameter 
choices in Table~\ref{paramtable}: 
curves~$A$ and $B$ differ by a fixed \mbox{$\Delta\thetabar = \thetabar_B-\thetabar_A$} while holding
$\etabar$ fixed, while curves~$A$ and $B'$
differ by the same $\Delta\thetabar$  but
instead hold $\etabar_\lambda$ fixed. 
Thus, the difference between curves~$B$ and $B'$ in each case illustrates the
impact of holding $\etabar$ fixed versus holding $\etabar_\lambda$ fixed when performing
a shift of the mixing angle $\Delta\theta$ relative to the common curve~$A$.     }
\label{fig:holdingfixed}
\end{figure*}

Rather than provide an exhaustive study of such regroupings and their outcomes, we shall here content ourselves
with providing a concrete example of the effects that can emerge by focusing on a single pair of curves
which we shall call $A$ and $B$.
These curves can be taken to show, for example, a late-time energy density $\rhobar$  as a function of $\Delta_G$, 
but they will differ in their absolute mixing by an angle $\Delta \thetabar$ when \mbox{$\lbrace \mbar_{\rm sum}^2, \etabar\rbrace$} are held fixed.
We will then calculate a third curve $B'$ which also differs from the curve $A$ by the same absolute angle $\Delta\thetabar$
when  \mbox{$\lbrace \lambdabar_{\rm sum}^2, \etabar_\lambda\rbrace$} are held fixed instead.
In this way, comparing curves $B$ and $B'$ will then give us an idea of the difference that can come from the choice of which variables
to hold fixed.

For concreteness, we select curves~$A$ and $B$ to be the \mbox{$\xibar=0.9$} and \mbox{$\xibar=0.5$} curves 
within the right panel of Fig.~\ref{fig:figname2}.    
Within this panel, \mbox{$\Delta \mbar^2=4$} and \mbox{$\etabar=0.75$} for both curves.
Use of Eq.~(\ref{inverses}) then provides us with the corresponding
\mbox{$\lbrace \lambdabar_{\rm sum}^2, \etabar_\lambda,\xibar_\lambda \rbrace$} values 
for each of these curves:  we find 
\mbox{$\lbrace 4.0, 0.96, 0.94\rbrace$} for curve~$A$
and  
\mbox{$ \lbrace 4.0, 0.82, 0.61\rbrace$} for curve~$B$.
Likewise, use of Eq.~(\ref{thetaangle}) with either the
\mbox{$\lbrace \etabar,\xibar\rbrace $} or \mbox{$\lbrace \etabar_\lambda,\xibar_\lambda\rbrace$} variables
then tells us that \mbox{$\thetabar_A\approx 19.22^\circ$} 
and \mbox{$\thetabar_B\approx 11.90^\circ$}.   [The fact that these values emerge in each case using
either    
set of variables is a useful cross-check that the absolute mixing angle $\thetabar$ is
indeed invariant across the mapping in Eq.~(\ref{inverses}).]
We now wish to define a third curve~$B'$
such that $\etabar_\lambda$ is held fixed relative to curve~$A$
(\ie, \mbox{$\etabar_\lambda^{(B')}= \etabar_\lambda^{(A)} \approx 0.96$})
while the mixing-angle shift 
\mbox{$\Delta \thetabar_{AB'} \equiv \thetabar_{B'}-\thetabar_A$}
matches \mbox{$\Delta \thetabar_{AB} \equiv \thetabar_B-\thetabar_A$}.  
This latter condition implies that 
\mbox{$\thetabar_{B'}=\thetabar_B\approx 11.90^\circ$},
whereupon use of Eq.~(\ref{thetaangle}) tells us that \mbox{$\xibar_\lambda^{(B')}\approx 0.84$}.
Use of Eq.~(\ref{inverses}) then tells us that 
\mbox{$\lbrace \mbar_{\rm sum}^2, \etabar,\xibar\rbrace  \approx 
\lbrace 4, 0.88, 0.80\rbrace$} for curve~$B'$.
These parameter values for curves~$A$, $B$, and $B'$ are summarized in 
Table~\ref{paramtable}.

\begin{table}[h!]
\begin{tabular}{||c|| c|c|| c | c||c||}
\hline
\hline
    ~ & $\etabar$ & ~~$\xibar^{\phantom{2^2}}_{\phantom{2_2}}  $ &  ~$\etabar_\lambda$ & $\xibar_\lambda$ & $\thetabar$ \\

\hline
 A$\phantom{'}$    &    0.75 &  0.90 &  0.96 &  0.94 &  19.22$^\circ$ \\      
 B$\phantom{'}$    &    0.75 &  0.50 &  0.82 &  0.61 &  11.90$^\circ$ \\   
 ~~B$'$~~ &    ~~0.88~~ &  ~~0.80~~ &  ~~0.96~~ &  ~~0.84~~ &  ~~11.90$^\circ$~~ \\   
\hline
\hline
\end{tabular}
\caption{Parameters defining the curves $A$, $B$, and $B'$ 
shown in Fig.~\protect\ref{fig:holdingfixed}.   All curves correspond to
\mbox{$ \mbar_{\rm sum}^2 = \lambdabar_{\rm sum}^2 = 4$} and \mbox{$\epsilon = + $}.}
\label{paramtable}
\end{table}

In Fig.~\ref{fig:holdingfixed} we plot Curves~$A$, $B$, and $B'$.
The left panel of 
Fig.~\ref{fig:holdingfixed} 
shows these curves for a matter-dominated universe, while the right panel shows these same
curves for a radiation-dominated universe.
Thus, while curves~$A$ and $B$ in the left panel of Fig.~\ref{fig:holdingfixed} respectively correspond to the \mbox{$\xibar=0.9$} and \mbox{$\xibar=0.5$} curves
in the right panel of Fig.~\ref{fig:figname2}, curves~$A$ and $B$ in the right panel of Fig.~\ref{fig:holdingfixed} 
respectively correspond to the \mbox{$\xibar=0.9$} and \mbox{$\xibar=0.5$} curves in the right panel of Fig.~\ref{fig:figname1}.

As evident from Fig.~\ref{fig:holdingfixed}, there are significant differences between curves~$B$ and $B'$.
Indeed, these curves are sufficiently distinct in the radiation-dominated case (right panel)
to turn what would have been
an overall {\it enhancement}\/ of the normalized late-time energy density 
(curve~$B$)
into an overall {\it suppression}\/ (curve~$B'$)!
We thus conclude that the question of which variables to hold fixed can indeed be a critical  
one for certain types of physics questions within certain regions of parameter space.
This also highlights the need to be extremely careful about specifying which parameters are held
fixed when making statements concerning the effects of varying some parameters relative to others.
In this paper, in order to study the effects of mixing, 
we have consistently held our diagonal mass-matrix elements $m^2_{00}$  and $m_{11}^2$ fixed 
while varying $m_{01}^2$ (or equivalently varying $\theta$ or $\xi$).   This ultimately represents a choice which 
stems from our desire to perturb the potential generated by our underlying mass-generating phase transition
as minimally as possible. 
However, this issue will be explored 
further in Ref.~\cite{toappear}.

\clearpage


\bibliography{sources}


\end{document}